\documentclass[aps,10pt,twocolumn,groupedaddress,superscriptaddress,floatfix,pra]{revtex4-1}
\usepackage{amsmath}
\usepackage{amssymb}
\usepackage{graphicx}
\usepackage{textcomp}
\usepackage{natbib}
\usepackage{bm}

\newcommand{\blu}[1]{\textcolor{black}{#1}}

\usepackage[usenames,dvipsnames]{color}	
\usepackage{ulem}

\usepackage{tikz}
\usepackage{pgfplots}
\pgfplotsset{compat=newest}
\usetikzlibrary{spy}

\begin{document}

\title{Quantum computers as universal quantum simulators: state-of-art and perspectives} 

\author{Francesco Tacchino} %\email{francesco.tacchino01@ateneopv.it}    
\affiliation{Dipartimento di Fisica, Universit\`a di Pavia, via Bassi 6, I-27100, Pavia, Italy}
\author{Alessandro Chiesa} %\email{}    
\affiliation{Dipartimento di Scienze Matematiche, Fisiche, e Informatiche, Universit\`a di Parma, I-43124, Parma, Italy}
\author{Stefano Carretta} %\email{}    
\affiliation{Dipartimento di Scienze Matematiche, Fisiche, e Informatiche, Universit\`a di Parma, I-43124, Parma, Italy}
\author{Dario Gerace}\email{corresponding author: dario.gerace@unipv.it}
\affiliation{Dipartimento di Fisica, Universit\`a di Pavia, via Bassi 6, I-27100, Pavia, Italy}

%\keywords{Quantum simulation,Quantum circuits,Superconducting qubits,Trapped ions}
%\pacs{xxxx, xxxx, xxxx}                                         
%\date{\today}
\begin{abstract}
The past few years have witnessed the concrete and fast spreading of quantum technologies for practical computation and simulation. In particular, quantum computing platforms based on either trapped ions or superconducting qubits have become available for simulations and benchmarking, with up to few tens of qubits that can be reliably initialized, controlled, and measured. The present review aims at giving a comprehensive outlook on the state of art capabilities offered from these near-term noisy devices as universal quantum simulators, i.e.\ programmable quantum computers potentially able to calculate the time evolution of many physical models. First, we give a \blu{pedagogic} overview on the basic theoretical background pertaining digital quantum simulations, with a focus on hardware-dependent mapping of spin-type Hamiltonians into the corresponding quantum circuit model \blu{as a key initial step towards simulating more complex models}. Then, we review the main experimental achievements obtained in the last decade \blu{regarding the digital quantum simulation of such spin models}, mostly employing the two leading quantum architectures. We compare their performances and outline future challenges, also in view of prospective hybrid technologies, towards the ultimate goal of reaching the long sought quantum advantage for the simulation of complex many body models in the physical sciences.
\end{abstract}

\maketitle

%{\bf Short biographies of corresponding authors}

%{\bf Francesco Tacchino} is a PhD student in quantum technologies at the University of Pavia, where he got  his M.Sc.\ in Theoretical Physics in 2016. His research interests include quantum computing and quantum simulations on near-term devices, quantum thermodynamics, and the intersections between machine learning and the physical sciences.

%{\bf Dario Gerace} has been associate professor of theoretical condensed matter physics at the  University of Pavia since 2015. He graduated from the same University in 2005, he was then post-doctoral researcher in quantum photonics at ETH Zurich, and came back to Pavia as an assistant professor in 2008, with tenure in 2009. His current research interests span from quantum technologies and quantum simulations of complex many body systems, to quantum optics and thermodynamics, nanophotonics and radiation-matter interaction in nanostructured systems, and more recently quantum machine learning.

\section{Introduction}

% SIMULATORS
When trying to accurately describe the dynamical behavior of physical systems made of several interacting fundamental constituents, and from these explain the complexity of natural aggregates following a bottom up approach, the well established classical laws of physics fail to give an accurate picture of reality, as it is now accepted and understood. In fact, quantum mechanics is arguably the most complete and successful theory we currently have to effectively describe the dynamics of the elementary constituents of our universe. A great deal of methods and simulation tools have been developed in the last century, such as quantum Monte-Carlo \cite{gubernatis_quantum_2016}, molecular dynamics \cite{haile_molecular_1992}, and tensor networks \cite{montangero_introduction_2018} to name a few examples, which allow solving some of the theoretical models formulated in quantum mechanical terms and correctly describe a large variety of quantum phenomena. 
The very concept of ``simulation'' has a broadly understood meaning in Science, Technology, Engineering, and Mathematics (STEM) applications. In fact, simulating any natural phenomenon is equivalent to artificially reproduce its properties and its dynamical evolution in time. This is primarily carried out through an accurate mathematical modeling, i.e.\ a mapping of the information we know about a system of interest onto a certain set of variables and equations, followed by an analytic or most often numerical solution. The resulting set of mathematical identities (or the computer with its numerical program aimed at solving them) can then be named a \textit{simulator}. Such a simulator is used to study the behavior of the real system under fairly general conditions, to make predictions and to test new hypotheses, the only limitations being the validity of the initial modeling and the available computational power.  \\
\begin{figure}[ht]
\centering
\includegraphics[width=\columnwidth]{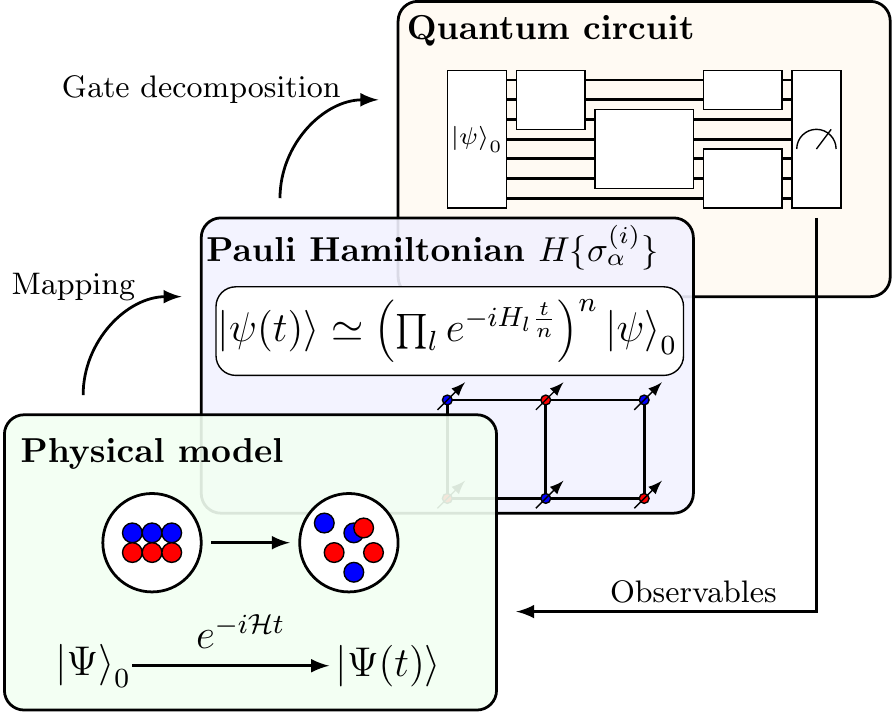}
\caption{Conceptual illustration of a universal simulator implemented on a digital quantum computing device. A physical model describes the quantum state evolution in a physical space, $\Psi(t)$; this evolution can be approximated to arbitrary precision by mapping the given model on a spin-type model (which can be easily encoded, e.g., onto a qubits-based register), and slicing the time evolution according to the Trotter-Suzuki formula (see text); the sequence of unitary operations can then be programmed through a quantum circuit model to be directly run on a quantum computer, giving the approximated evolved state as an output, $\psi(t)$. }
\label{fig:schemeUQS}
\end{figure}
% QUANTUM SIMULATORS
It is generally accepted that most of the models we currently deal with cannot be solved exactly with classical computing machines, such as modern supercomputers obeying the laws of classical physics. The main reason  lies in the exponential scaling of time and memory resources needed to correctly capture the dynamics of the relevant physical variables with increasing system size. This is especially true when strong correlations between the system parties play a dominant role, which is the case in most interesting situations. In such cases, even the most elaborate but inevitably approximate classical simulation approaches so far developed fail in giving the correct answers.
Hence, \textit{quantum simulators} have long been proposed as a possible solution, building on the general idea that since Nature ultimately behaves quantum mechanically, only a computing machine obeying quantum mechanical laws would be able to accurately simulate it \cite{manin_computable_1980,benioff_computer_1980,feynman_simulating_1982}. A quantum simulator is a system under high control of the experimenter, which is able to reproduce the dynamical behavior of a given physical model, irrespective of the degree of internal correlations or entanglement between the model's degrees of freedom. Following this route, a plethora of \textit{analog} quantum simulators have been proposed~\cite{fisher_boson_1989,jaksch_cold_1998,hartmann_strongly_2006,greentree_quantum_2006,angelakis_photon-blockade-induced_2007,carusotto_numerical_2008,carusotto_fermionized_2009,tian_circuit_2010,gerace_analog_2012,viehmann_observing_2013,du_superconducting_2015,reiner_emulating_2016} and developed~\cite{greiner_quantum_2002,friedenauer_simulating_2008,gerritsma_quantum_2010,kim_quantum_2010,islam_onset_2011,nguyen_acoustic_2015,steinhauer_observation_2016,labuhn_tunable_2016,bernien_probing_2017,roushan_spectroscopic_2017,zhang_observation_2017,zhang_experimental_2018}, in which the physical properties of a targeted model are reproduced on a physical set-up under \blu{externally controlled} conditions. On the other hand, \textit{digital} quantum simulators are programmable and general purpose quantum devices, which promise a larger flexibility on the models to be solved~\cite{lloyd_universal_1996,ortiz_quantum_2001,somma_simulating_2002,verstraete_quantum_2009,weimer_rydberg_2010,santini_molecular_2011,casanova_quantum_2012,mezzacapo_digital_2012,jordan_quantum_2012,raeisi_quantum-circuit_2012,hauke_quantum_2013,las_heras_digital_2014,chiesa_digital_2015,garcia-alvarez_digital_2017,jiang_quantum_2018,kivlichan_quantum_2018}. In this respect, digital quantum simulators are quantum computing machines not restricted to emulate the dynamics of targeted models, but satisfying DiVincenzo criteria \cite{divincenzo_physical_2000} for quantum computation. Here we will consider such digital quantum computers as universal quantum simulators (UQS)~\cite{lloyd_universal_1996}, meaning that they are able, in principle, to reproduce with arbitrary precision the dynamics of any Hamiltonian model that can be suitably encoded on a given quantum register and translated into a sequence of gate operations, as schematically illustrated in Fig.~\ref{fig:schemeUQS}. The time evolution of the physical model is mapped onto an effective model defined on the quantum hardware degrees of freedom, in which the time evolution can be programmed in digital steps through a sequence of unitary operations defined by a quantum circuit \cite{nielsen_quantum_2000}. This mapping will be the focus of the present review. \blu{As it will be recalled, the concept of \textit{universality} requires that the model Hamiltonian be the sum of locally interacting terms, which in turn implies that the number of required operations does not grow exponentially with the system size. This, in practice, restricts the class of Hamiltonian models that can be usefully simulated (although to the ones that are most physically relevant).}  \\
On a more refined level, it is worth mentioning that hybrid digital-analog quantum simulators have also been proposed, aimed at combining the easier scalability of analog approaches with the intrinsic universality of digital quantum simulations~\cite{mezzacapo_digital_2015}. 
Here, analog blocks allow for the direct simulation of the time dynamics on a large number of variables, thus reducing the number of digital operations and errors, while digital blocks are included to introduce a variety of possible interaction models. This paradigm is hailed as a promising route leading to universal digital-analog quantum computation.

% THIS REVIEW
Several excellent reviews have been published in the last few years, giving a broad account of quantum simulators, either general purpose~\cite{buluta_quantum_2009,georgescu_quantum_2014,sanders_efficient_2013,preskill_quantum_2018} or more focused on specific categories and/or quantum hardware \cite{blatt_quantum_2012,bloch_quantum_2012,aspuru-guzik_photonic_2012,houck_-chip_2012,wendin_quantum_2017,lamata_digital-analog_2018}. Here we give a targeted overview on near term digital quantum computers as devices able to perform universal quantum simulations. This goes in line with the fast pace of advancement in different quantum computing technologies that have made programmable devices available, thus attracting widespread interest worldwide. 
In fact, current quantum processors \blu{promise to overcome} the intrinsic limitations of simulating complex many body physics with classical computing machines, \blu{although it is still  difficult to predict when this will happen}. A targeted goal would be to reach the long-sought ``quantum advantage'', i.e.\ a certified gain in either memory or temporal efficiency obtained for the solution of a quantum problem with respect to the equivalent simulation being performed on a classical supercomputer. Without entering into the subtleties related to a rigorous definition of quantum advantage, here we just consider that a quantum computer with fully operational $N=50$ qubits is able to store something like $8\times 2^N\sim 9\cdot 10^{15}$ bytes of information (i.e.\ 9 Pb, assuming 8 bytes to store a complex number in single-precision), which roughly corresponds to the random access memory of state-of-art supercomputers \cite{pednault_breaking_2017,boixo_characterizing_2018}. \blu{Should this threshold be met with an actual quantum simulation involving the whole quantum hardware} in the current Noisy Intermediate Scale Quantum devices (NISQ) era~\cite{preskill_quantum_2018}, \blu{it would represent} the quantum advantage turning point. \blu{On a longer and still unpredictable timescale, even} farther-reaching consequences are expected should fully fault tolerant and scalable quantum hardware become available~\cite{schindler_experimental_2011,you_simulating_2013,barends_superconducting_2014,corcoles_demonstration_2015}, in which $N>100$ logical qubits have to be complemented with a much larger number of auxiliary quantum bits aimed at correcting errors. \\ 
Our aim is to give an overview of the field that could be useful to the beginning researcher or student, trying to keep a pedagogic approach over the elementary theoretical background throughout the manuscript, and then summarizing the main experimental achievements and prospective developments. Since most Hamiltonian models can be mapped onto spin-type ones, being able to efficiently simulate spin models on actual quantum computing devices is crucial, not only because they possess interesting many body dynamics themselves but also to open the door to the universal quantum simulation of a large class of quantum models (typically interacting fermionic particles) that are intractable by classical computation means \cite{troyer_computational_2005}. Paradigmatic examples are the Hubbard model in condensed matter \cite{casanova_quantum_2012,barends_superconducting_2014}, or the Schwinger model in lattice gauge field theory \cite{hauke_quantum_2013,martinez_real-time_2016}. In particular, we emphasize the role of specific quantities that are known to be difficult to compute but extremely important in the description of the dynamical properties of many body systems, such as quantum correlations. \\
In terms of actual quantum hardware, we will focus on reviewing the main experimental achievements obtained in the last decade, specifically dealing with the simulation accuracy of targeted spin Hamiltonians on different quantum platforms. While several alternatives are currently being pursued to realize actual non-error corrected quantum processors~\cite{ladd_quantum_2010}, from photonic integrated circuits \cite{aspuru-guzik_photonic_2012} to spins in semiconductors \cite{awschalom_quantum_2013}, we concentrate upon the two leading architectures that have been dominating the scene: trapped ions manipulated through external microwave or optical fields \cite{monroe_scaling_2013,schindler_quantum_2013,bruzewicz_review-trap-ion_2019}, and superconducting circuits working at microwave frequencies \cite{clarke_superconducting_2008,schoelkopf_wiring_2008,devoret_superconducting_2013,Gu_review_SCcircuits_2017}.  
We \blu{anticipate} that interesting results might already be within reach in NISQ processors, despite the relatively small number of useful operations and non-error corrected qubits currently available on such devices.
On a parallel sight, while the main object is restricted to quantum simulations of physical models and STEM applications in general, actual quantum processors might eventually turn to solve complex problems in other fields as well. As examples, classification and scheduling tasks, stock market pricing \cite{woerner_quantum_2019,martin_towards_2019} and machine learning \cite{biamonte_quantum_2017} might benefit from speedup advantages over classical computers. The basics of quantum circuit programming reported in this review may be a useful starting point. Last but not least, these topics settle within the quantum technologies roadmap promoted at the European level through the recently funded Quantum Flagship \cite{acin_quantum_2018}.

\section{Theory of Digital Quantum Simulations}

When the main object of a physical theory is to determine the evolution in time of a given system, most problems are formulated in terms of a set of differential equations. Their solution is at the heart of many simulation protocols nowadays, from molecular dynamics to aircraft design. A very common situation is, for example, a linear set of equations such as
\begin{equation}
\frac{d\vec{x}}{dt} = M\vec{x}
\end{equation}
where $M$ is a matrix and $\vec{x}$ represents a vector of dynamical variables. Once an initial condition $\vec{x}(0)$ is given, the formal solution to the above equation is simply
\begin{equation}
\vec{x}(t) = e^{Mt}\vec{x}(0)
\end{equation}
Implementing such a solution on a computer routine gives a useful tool to fully solve the system dynamics, provided that the size of the numerical problem is within reach of the available computational resources. 
In quantum mechanics, the paradigmatic example is the Schr\"{o}dinger equation (here and in the following, we take $\hbar = 1$)
\begin{equation}
\frac{d\left|\Psi\right\rangle}{dt} = -i\mathcal{H}\left|\Psi\right\rangle
\end{equation}
where $\mathcal{H}$ is known as the Hamiltonian operator. This complex-valued differential equation is solved by computing the unitary time-evolution operator $\mathrm{U}(t) = e^{-i\mathcal{H}t}$. Indeed, once the latter is known, any initial condition can be evolved linearly as
\begin{equation}
\left|\Psi (t)\right\rangle = \mathrm{U}(t)\left|\Psi(0)\right\rangle
\end{equation}
Matrix exponentiation is a very common numerical task arising in many interesting simulation scenarios, and crucially in the field of quantum mechanical systems. On classical computers, this task turns out to be provably difficult in terms of the matrix size, most notably for quantum mechanical simulations, where the exponential increase of the size of the Hilbert space of a composite system with the number of sub-systems leads to an exponential demand of time and memory resources.

In 1982, Richard Feynman conjectured that using a controllable quantum mechanical system as a computing resource, instead of a classical object, would provide significant advantages in the simulation of quantum systems~\cite{feynman_simulating_1982}. Indeed, just about fifteen years later, in 1996, Seth Lloyd proved that idea to be essentially correct~\cite{lloyd_universal_1996}, with the sole limitation that the systems to be simulated only carry local interactions between their constituent subsystems. In these review, we will thus concentrate on system Hamiltonians of the form\begin{equation}\label{eq:local}
\mathcal{H} = \sum_l^L \mathcal{H}_l
\end{equation}
where $\mathcal{H}_l$ acts locally only on a portion of the total system. \blu{As a matter of fact,  many Hamiltonian models of physical interest can be formulated as in Eq.~(\ref{eq:local}).}

\subsection{The quantum computer as a universal quantum simulator}

Given \blu{a Hamiltonian $\mathcal{H}$} that models the physical system under investigation, the problem of computing the corresponding time evolution operator $\mathrm{U}(t) = \exp\left(-i\mathcal{H}t\right)$ is equivalent to the task of implementing a well defined unitary matrix. A quantum computer endowed with a universal set of quantum gates is by definition able to perform any unitary transformation, albeit not necessarily in an efficient number of elementary operations~\cite{nielsen_quantum_2000}. What Lloyd actually proved is that universal quantum computers can calculate $\mathrm{U}(t)$ efficiently (i.e. with polynomial time and memory resources in the size of the target system) when $\mathcal{H}$ is a sum of local terms. The proof is based on two fundamental facts: first, \blu{in the circuit model for UQS we can implement generic} transformations by successively performing elementary unitary operations (quantum gates), and appending one unitary $\mathrm{U}_A$ after another $\mathrm{U}_B$ in the circuit results in a total unitary, which is mathematically the product $\mathrm{U}_A\mathrm{U}_B$ being applied to the state of the qubit register. Second, any unitary operation $\mathrm{U}$ acting on $N$ qubits can be implemented with $O(2^{2N})$ elementary operations (we recall that the dimension of the Hilbert space associated to $N$ qubits is $d=2^N$)~\cite{barenco_elementary_1995,nielsen_quantum_2000}. \blu{Following Ref.~\onlinecite{lloyd_universal_1996}, we now suppose that we are given a Hamiltonian expressed as the sum of $L$ local terms, as in Eq. (\ref{eq:local}), where $L \propto p\cdot N$ such that $p$ measures some degree of locality (e.g., the number of nearest neighbors or second-to-nearest neighbors in a lattice), and $N$ is the total number of qubits required to encode the computation. Hence, the number of local terms, $L$, scales polynomial with $N$}. In general, according to the rules above, computing directly $\mathrm{U}(t) = \exp\left(-i\mathcal{H}t\right)$ requires $O(2^{2N})$ operations, and it is therefore exponentially inefficient. However, let us call $m_l$ the dimension of the subsystem over which the action of $\mathcal{H}_l$ is restricted. Typically, we will have $m_l \ll 2^N$, since local terms only involve few-body interactions. In this case, the unitary $\mathrm{U}_l(t) = \exp\left(-i\mathcal{H}_lt\right)$ can be computed with $O(m_l^{2})$ operations. The overall product
\begin{equation}
\tilde{\mathrm{U}} = \prod_l \mathrm{U}_l(t)
\end{equation}
can therefore be obtained on a universal quantum computer by juxtaposing the circuit implementations of the single $\mathrm{U}_l(t)$ unitaries, and it would take at most $O(L m_{max}^2)$ elementary operations, where $m_{max} = \operatorname{max}_l {m_l}$. The final step of the reasoning lies in the following mathematical identity, which is known as the Suzuki-Trotter (ST) decomposition:
\begin{equation}
e^{-i\sum_l \mathcal{H}_l t} = \lim_{n\to\infty} \left(\prod_l e^{-i \mathcal{H}_l t/n}\right)^n
\label{eq:ST}
\end{equation}
Unless all the $\mathcal{H}_l$ operators commute, in which case the ST identity is exact already for $n=1$, the product of local unitaries will not be exactly equal to the total target unitary $\mathrm{U}(t) = \exp\left(-i\mathcal{H}t\right)$. However, it can be shown that $\forall n$
\begin{equation}
\mathrm{U}(t) = e^{-i\sum_l \mathcal{H}_l t} = \left(\prod_l e^{-i \mathcal{H}_l t/n}\right)^n + O\left(\frac{t^2}{n}\right)
\end{equation}
which means that we can approximate arbitrarily well the desired unitary operator by repeating $n$ times the sequence of gates corresponding to the product of local terms for time slices $t/n$. All in all, we were able to break our original problem into smaller pieces, $e^{-i \mathcal{H}_l t/n}$, which can now be implemented efficiently using only a limited set of elementary gates and give the correct answer up to an arbitrarily small digital error $O(t^{2}/n)$. Indeed, for any $\epsilon>0$ and $t$, there exists a $n_\epsilon$ such that $\mathrm{U}(t)$ can be computed within an approximation $\epsilon$ in at most $n_\epsilon Lm_{max}^2$ operations. This is polynomial in $N$ whenever $L = \text{poly}(N)$, as for example in the case of nearest neighbors interactions.

\subsection{Quantum simulations cookbook}
\label{sec:cookbook}

From now on, we will assume to work with a universal quantum computer, described in the standard \blu{circuit} model as a (quantum) digital device, i.e. qubit-based, obeying the algebra of Pauli matrices and operating with a universal set of quantum gates~\cite{nielsen_quantum_2000}. The problem of quantum simulation can then be formulated and solved on such a machine by taking a few simple steps, which we are going to outline in the following. 

First, define a model Hamiltonian of interest $\mathcal{H}$. This should contain all the dynamical information necessary to describe and characterize the physical quantum system under investigation. The most appropriate set of variables and operators will appear in the mathematical structure of $\mathcal{H}$.

Second, map the target Hamiltonian $\mathcal{H}$ onto its representation on the qubit Pauli algebra
\begin{equation}
\mathcal{H}\rightarrow H(\{\sigma_\alpha\})
\end{equation}
In simpler terms, this means finding a suitable encoding of the degrees of freedom of the target system into a number $N$ of qubits. The resulting mapped Hamiltonian $H$ will then be written in terms of Pauli matrices. Notice that this mapping is straightforward for physical systems consisting of collections of spin-$1/2$ objects, as they also obey Pauli algebra, but it is possible in principle for a large class of physical system, \blu{as it will be shown in the following with some specific example.}
The quantum simulation will be efficient whenever such $H$ is the sum of local terms. Notice that this is usually not a limitation in many practical cases, as most physical processes are inherently local in nature. \blu{However, local Hamiltonian models get mapped into non-local ones, as, e.g., when the well known Jordan-Wigner transformation is applied \cite{jordan_uber_1928} to encode fermionic degrees of freedom. These models might be efficiently simulated on quantum hardware implementing multi-qubit gates, as it will be detailed in the following.}

Third, assuming the target Hamiltonian is mapped onto a sum of local contributions
\begin{equation}
H = \sum_l H_l
\end{equation}
check whether $[H_l,H_{l'}] = 0 \,\, \forall l,l'$. If that is the case, then
\begin{equation}
e^{-iHt} = \prod_l e^{-iH_l t}
\end{equation}
with no digital error. Otherwise, choose the number of ST steps (sometimes referred to as Trotter steps), $n$, that is appropriate for the required degree of precision, in such a way that
\begin{equation}
e^{-iHt} \simeq \left(\prod_l e^{-iH_l t/n}\right)^n
\end{equation}
This application of the ST formula is sometimes called \textit{Trotterization} in quantum simulations jargon.

Fourth, translate each local unitary $e^{-iH_l t}$ (or $e^{-iH_l t/n}$) into a sequence of quantum gates. This is always possible in at most $O(m_l^2)$ operations and with any universal set of single- and two-qubits operations available on a general purpose quantum computer~\cite{nielsen_quantum_2000}. The total quantum circuit encoding the time evolution will be the juxtaposition of all the sequences corresponding to the factors in the ST decomposition, repeated $n$ times.

Finally, add initial state preparation at the beginning of the circuit and an appropriate set of measurements at the end to recover expectation values of the relevant observable quantities on the evolved quantum state.

The points above represent a quite general set of instructions towards the design of a quantum simulation algorithm. In the following, we will give some explicit examples to show how this is done in practical cases. Of course, such techniques are not limited to actual simulations of real physical systems, but can become a tool for a larger class of computational tasks whenever the problem of interest can be encoded in a Hamiltonian quantum dynamics.

\section{Quantum Circuits}

Among the steps that must be undertaken in order to practically design and realize a digital quantum simulation, the translation of unitary operators into elementary quantum gates is the one that is most typically hardware-dependent. It is also critical in terms of results and performance, particularly in the present era of noisy and intermediate-scale prototypes of quantum processors, where the interplay between hardware properties and target features is stronger.

Several universal sets of single- and two-qubit gates are known~\cite{nielsen_quantum_2000}, all in principle equally valid as a primitive set to realize any quantum simulation. However, every real hardware platform usually comes with a native set of operations that, due to the physical characteristics of the device, are readily implemented in practice. The platform is in itself capable of implementing universal quantum computation, and is thus a potential UQS, if and only if the native set is a universal set in the usual quantum computing sense. If that is the case, any target unitary evolution can be translated in a combination of the native operations without unnecessary overhead. Processors based on different technological platforms may also exhibit distinct topological properties, i.e.\ different qubit-qubit inter-connectivity and limitations in gate directionality. While these do not pose hard limitations to the computational power of the platform, since they can always be compensated via, e.g., $\mathrm{SWAP}$ operations, they may results in some overhead in the total length of the simulations. Hence, in this NISQ era some platforms are more suitable for the simulation of certain physical models (e.g.\ trapped ions, featuring built-in all-to-all connectivity, can more easily simulate long-range interactions), thus making a fair comparison of performances less straightforward~\cite{cross_validating_2018}. Of course, it should be reminded that, as a general rule, only systems described by local interaction terms are somehow guaranteed to be efficiently mapped on a quantum computing register. 

\subsection{Pauli algebra and spin Hamiltonians}

The mathematical properties of qubits are those of spin-$1/2$ systems, thus obeying the algebraic properties of Pauli matrices. The latter can be written in the computational basis representation as
\begin{equation}
\begin{split}
\sigma_x = \begin{pmatrix}
0 & 1 \\ 
1 & 0
\end{pmatrix} \quad \sigma_y
= \begin{pmatrix}
0 & -i \\ 
i & 0
\end{pmatrix} \quad \sigma_z
= \begin{pmatrix}
1 & 0 \\ 
0 & -1
\end{pmatrix} \end{split}
\end{equation}
and satisfy the following commutation and anti-commutation rules
\begin{equation}
[\sigma_\alpha,\sigma_\beta] = 2i\epsilon_{\alpha\beta\gamma}\sigma_\gamma\, , \quad \{\sigma_\alpha,\sigma_\beta\} = 2\delta_{\alpha\beta}\mathbb{I} 
\end{equation}
where $\alpha,\beta,\gamma \in \{x,y,z\}$, $\epsilon_{\alpha\beta\gamma}$ is the Levi-Civita tensor, $\delta_{\alpha\beta}$ is the Kronecker delta and $\mathbb{I}$ is the identity matrix.

In order to be simulated on a qubit-based architecture, any target Hamiltonian, $\mathcal{H}$, has to be mapped into an equivalent Hamiltonian, $H$, of interacting spin-$1/2$ operators. As already mentioned, this step is straightforward for paradigmatic spin-$1/2$ Hamiltonians (e.g., implementing Heisenberg or Ising models), but effective mappings are known for a large variety of cases, ranging from spin $S>1/2$~\cite{santini_molecular_2011,chiesa_digital_2015,tacchino_electromechanical_2018} to fermionic and fermionic-bosonic systems~\cite{ortiz_quantum_2001,santini_molecular_2011,casanova_quantum_2012,mezzacapo_digital_2012,chiesa_digital_2015,mezzacapo_digital_2015,barends_digital_2015,garcia-alvarez_digital_2017,kandala_hardware-efficient_2017,dumitrescu_cloud_2018,kivlichan_quantum_2018}, including lattice models related to gauge theories~\cite{martinez_real-time_2016,klco_quantum-classical_2018}. The generator of time evolution in a $N$-qubit digital quantum simulation therefore takes the general form
\begin{equation}
H = \sum_{\substack{i = 1\\ \alpha=x,y,z}}^N h^{(1)}_{\alpha,i}\sigma_\alpha^{(i)} + \sum_{\substack{i,j = 1\\ \alpha,\beta=x,y,z}}^N h^{(2)}_{\alpha\beta,ij}\sigma_\alpha^{(i)}\sigma_\beta^{(j)}
\label{eq:su2ham}
\end{equation}
containing in general both single- and two-spin terms, to which any other many body term time evolution can, in principle, be reduced (see Sec.~\ref{sec:multiQint}). Whenever the overall structure of $H$ retains a local nature, as it is the case for many physically relevant examples, its translation into elementary gate operations can be done efficiently. In the following, we will provide a dictionary of useful decomposition rules in terms of different universal sets of gates. Most of them are derived from real use-case scenarios and can therefore be straightforwardly applied to well known physical models. 

\subsection{Single-qubit rotations}

Once the target Hamiltonian $\mathcal{H}$ is reduced to its counterpart $H$ on $N$ spin-$1/2$ systems, a register of $N$ qubits can be used to encode and carry out the quantum simulation via the identification of each qubit with a single spin-$1/2$ element. All currently proposed and realized quantum computing platforms allow addressing single qubits with tailored control pulses to perform single qubit gates. The most general single qubit $\mathrm{SU}(2)$ operation has the form
\begin{equation}
\mathrm{U}(\theta,\phi,\lambda) = \begin{pmatrix}
\cos(\theta/2) & -e^{i\lambda}\sin(\theta/2) \\ 
e^{i\phi}\sin(\theta/2) & e^{i(\lambda+\phi)}\cos(\theta/2)
\end{pmatrix}
\end{equation}
and can be obtained, for example, by combining well known single qubit quantum gates such as the Hadamard gate
\begin{equation}
\mathrm{H} =  \frac{1}{\sqrt{2}}\begin{pmatrix}
1 & 1 \\ 
1 & -1
\end{pmatrix}
\end{equation}
and the phase gate
\begin{equation}
\Phi(\delta) = \begin{pmatrix}
1 & 0 \\ 
0 & e^{i\delta}
\end{pmatrix}
\end{equation}
Indeed, the following identity holds:
\begin{equation}
\mathrm{U}(\theta,\phi,\lambda) = e^{-i\theta/2}\Phi\left(\frac{\pi}{2}+\phi\right)\mathrm{H}\Phi(\theta)\mathrm{H}\Phi\left(-\frac{\pi}{2}+\lambda\right)
\end{equation}
Rotations around the coordinate axes
\begin{equation}
\mathrm{R}_\alpha (\theta) = \exp\left(-i\frac{\theta}{2}\sigma_\alpha\right) \quad \alpha = x,y,z
\end{equation}
can be implemented, up to global phase factors, by choosing particular parameters in $\mathrm{U}(\theta,\phi,\lambda)$. For example, $\mathrm{R}_z(\lambda) = e^{-i\lambda/2}\Phi(\lambda) = \mathrm{U}(0,0,\lambda)$, $\mathrm{R}_x(\theta) = \mathrm{U}(\theta,-\pi/2,\pi/2)$ and $\mathrm{R}_y(\theta) = \mathrm{U}(\theta,0,0)$. Vice-versa, any platform capable of implementing single-qubit rotations around the coordinate axes can in principle realize an arbitrary $\mathrm{U}(\theta,\phi,\lambda)$ via the following identity
\begin{equation}
\mathrm{U}(\theta,\phi,\lambda) = \mathrm{R}_z (\phi)\mathrm{R}_x (\theta)\mathrm{R}_z (\lambda)
\end{equation}
In Eq.~\eqref{eq:su2ham}, any single-spin term 
\begin{equation}
H_1^{(i)} = \sum_{\alpha=x,y,z} h^{(1)}_{\alpha,i}\sigma_\alpha^{(i)}
\end{equation}
essentially represents a magnetic field applied to the $i$-th qubit along the direction identified by the vector $\vec{h} = (h^{(1)}_x,h^{(1)}_y,h^{(1)}_z)$. The induced time evolution
\begin{equation}
\mathrm{U}_1^{(i)}(t) = e^{-iH_1^{(i)}t}
\end{equation}
is a precession around the $\vec{h}$ axis, with the corresponding action on a qubit being a rotation of the Bloch vector. This can always be expressed in the $\mathrm{U}(\theta,\phi,\lambda)$ form, and  therefore as a combination of rotations around the coordinate axes or of Hadamard and
phase gates. Other decompositions of general $\mathrm{SU}(2)$ transformations, as well as approximate results employing only a finite set of fixed-phase single qubit operations instead of continuous-valued ones, are also known.~\cite{nielsen_quantum_2000,chow_universal_2012}
% Examples in trapped ions, SC qubits (with ideal Z?)

\subsection{Two-qubits gates}
\label{sec:2qInteractions}

Two-spin interactions appearing in the general Pauli Hamiltonian, Eq.~\eqref{eq:su2ham}, are usually implemented in digital quantum simulation protocols as combinations of single- and two-qubits gates. The typical evolution operator has the form
\begin{equation}
\mathrm{U}_{\alpha\beta}^{(i,j)}(t) = e^{-iH_{\alpha\beta}^{(i,j)}t} = e^{-i\delta\sigma_\alpha^{(i)}\otimes\sigma_\beta^{(j)}} 
\end{equation}
where $\delta$ is a dimensionless phase factor. These terms arise naturally in the simulation of many renown spin models such as the Heisenberg model
\begin{equation}
H = J \sum_{\langle i,j \rangle} \left( \sigma_x^{(i)}\sigma_x^{(j)} + \sigma_y^{(i)}\sigma_y^{(j)} + \sigma_z^{(i)}\sigma_z^{(j)} \right),
\end{equation}
the $\mathrm{XYZ}$ model
\begin{equation}
H = \sum_{\langle i,j \rangle} \left( J_{xx}\sigma_x^{(i)}\sigma_x^{(j)} + J_{yy}\sigma_y^{(i)}\sigma_y^{(j)} + J_{zz}\sigma_z^{(i)}\sigma_z^{(j)} \right),
\end{equation}
which reduces to the so called $\mathrm{XY}$ model if $J_{zz} = 0$, or the transverse field Ising model 
\begin{equation}
H = \sum_i h_i \sigma_x^{(i)} + \sum_{\langle i,j \rangle} J_{zz}\sigma_z^{(i)}\sigma_z^{(j)}
\label{eq:TIMgeneralH}
\end{equation}
Here $\langle i,j \rangle$ denote nearest neighbors spin pairs. 

The exact and most effective decomposition of $\mathrm{U}_{\alpha\beta}^{(i,j)}(t)$ terms into elementary quantum gates varies from platform to platform, depending on the available set of native operations. One common situation, typical of, e.g., superconducting qubit technology with cross-resonance interactions~\cite{rigetti_fully_2010,chow_simple_2011,sheldon_procedure_2016}, is a native universal set 
\begin{equation}
\mathcal{S}_1 = \{\mathrm{R}_\alpha(\theta), \mathrm{CNOT}\}
\end{equation}
containing single qubit rotations and the two-qubit $\mathrm{CNOT}$ entangling gate
\begin{equation}
\mathrm{CNOT} = \begin{pmatrix}
1 & 0 & 0 & 0 \\
0 & 1 & 0 & 0 \\
0 & 0 & 0 & 1 \\
0 & 0 & 1 & 0
\end{pmatrix}
\end{equation}
Let $\mathrm{ZZ}(\delta)$ be the unitary operation
\begin{equation}
\mathrm{ZZ}(\delta) = e^{-i\delta\sigma_z\otimes\sigma_z}
\label{eq:ZZdef}
\end{equation}
This can be realized using the elementary quantum gates belonging to $\mathcal{S}_1$ with the following quantum circuit:
\begin{equation}
\includegraphics[scale=1.2]{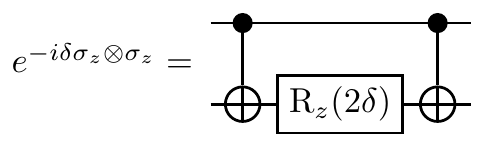}
\end{equation}
Other terms generated by $\sigma_\alpha\otimes\sigma_\beta$ can be obtained from the construction above by suitable changes of reference frames, implemented with single qubit rotations. Indeed, remembering the following identities 
\begin{equation}
\begin{split}
\mathrm{R}_y \left(\frac{\pi}{2}\right) \sigma_z \mathrm{R}_y \left(-\frac{\pi}{2}\right) = & \, \sigma_x \\
\mathrm{R}_x \left(\frac{\pi}{2}\right) \sigma_z \mathrm{R}_x \left(-\frac{\pi}{2}\right) = & \, - \sigma_y
\end{split}
\end{equation}
it is straightforward to verify that
\begin{equation}
\includegraphics[scale=0.9]{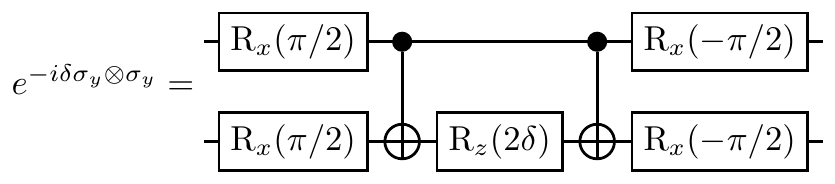}
\end{equation}
and
\begin{equation}
\includegraphics[scale=0.9]{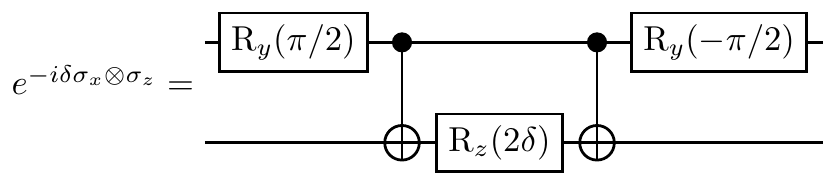}
\end{equation}
These gate sequences can be combined to simulate all of the paradigmatic spin models mentioned above. For example, it is straightforward to prove that for the two-qubit Heisenberg model we have
\begin{equation}
e^{-i\delta\left(\sigma^{(1)}_x\sigma^{(2)}_x+\sigma^{(1)}_y\sigma^{(2)}_y+\sigma^{(1)}_z\sigma^{(2)}_z \right)} = \mathrm{XX}(\delta) \mathrm{YY}(\delta) \mathrm{ZZ}(\delta)
\label{eq:2qHeisenbergDeco}
\end{equation}
where $\mathrm{AB}(\delta) = e^{-i\delta\sigma_a\otimes\sigma_b}$, \blu{and all of the terms on the right hand side commute with each other}. More detailed examples will be given in Sec.~\ref{sec:examples}.

Another universal set, defined $\mathcal{S}_2$, that often arises in superconducting realizations and proposals of quantum simulators replaces the $\mathrm{CNOT}$ gate with a parametric $\mathrm{XX}+\mathrm{YY}$ interaction~\cite{las_heras_digital_2014,salathe_digital_2015,mckay_universal_2016,tacchino_electromechanical_2018}
\begin{equation}
\mathrm{U}_{xy}(\delta) = e^{-i\delta(\sigma_x\otimes\sigma_x + \sigma_y\otimes\sigma_y)}
\end{equation}
In this case, we can take as the fundamental building block $\mathrm{XX}(\delta) = e^{-i\delta\sigma_x\otimes\sigma_x}$, to which all other unitary evolution terms generated by $\sigma_\alpha\otimes\sigma_\beta$ can be reduced with single-qubit changes of reference frame. The $\mathrm{XX}(\delta)$ gate is realized in $\mathcal{S}_2$ as
\begin{equation}
\includegraphics[scale=0.9]{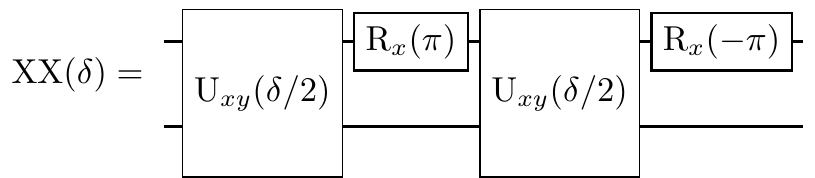}
\label{eq:XXfromXY}
\end{equation}

Finally, let us call $\mathcal{S}_3 = \{\mathrm{R}_\alpha(\theta), \mathrm{C}\Phi(\delta)\}$ the universal set of quantum gates containing all single qubit rotations and the controlled phase gate
\begin{equation}
\mathrm{C}\Phi(\delta) = \begin{pmatrix}
1 & 0 & 0 & 0 \\
0 & 1 & 0 & 0 \\
0 & 0 & 1 & 0 \\
0 & 0 & 0 & e^{i\delta}
\end{pmatrix}
\end{equation}
The latter is natively implemented on superconducting platforms with state dependent frequency shifts~\cite{carretta_quantum_2013,barends_superconducting_2014,chiesa_robustness_2014,barends_digital_2015,reagor_demonstration_2018}, and is closely related to the Ising interaction generated by $H_{\text{Ising}}\propto \sigma_z\otimes\sigma_z$~\cite{jones_robust_2003}. In view of the latter property, it is not surprising that the $\mathrm{ZZ}(\delta)$ building block can be obtained directly from a single $\mathrm{C}\Phi(\delta)$ just with single qubit corrections and apart from an overall phase:
\begin{equation}
\includegraphics[scale=0.9]{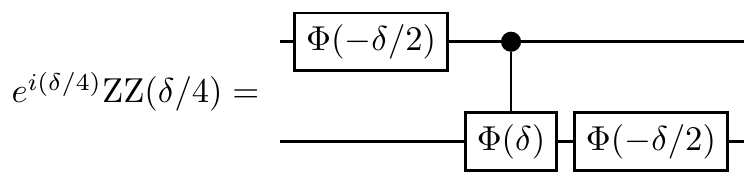}
\label{eq:ZZfromCP}
\end{equation}
An equivalent construction with two $\mathrm{C}\Phi(\delta)$ is the following
\begin{equation}
\includegraphics[scale=0.9]{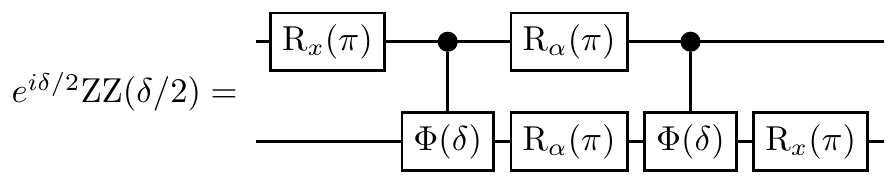}
\label{eq:ZZfrom2CP}
\end{equation}
where rotations around $\alpha = x,y$ enable the range of negative and small angles in those real experimental setups where the achievable phases $\delta$ in a single $\mathrm{C}\Phi(\delta)$ gate might be limited due to hardware constraints~\cite{barends_digital_2015}.

In quantum simulators based on trapped ions technology~\cite{schindler_quantum_2013,schindler_quantum_2013-1,lanyon_universal_2011}, the fundamental set of operations, which we will call $\mathcal{S}_4$, typically includes individual single qubit $z$ rotations
\begin{equation}\label{eq:T1gate}
\mathrm{T}_1^{(j)}(\theta) = e^{-i\theta\sigma_z^{(j)}},
\end{equation}
collective non-entangling operations
\begin{equation}\label{eq:T2gate}
\mathrm{T}_2(\theta) = e^{-i\theta\sum_j\sigma_z^{(j)}} ,\quad \mathrm{T}_3(\theta,\phi) = e^{-i\theta\sum_j\sigma_\phi^{(j)}}
\end{equation}
where $\sigma_\phi = \cos\phi\sigma_x + \sin\phi\sigma_y$, and M{\o}lmer-S{\o}rensen collective entangling gates~\cite{molmer_multiparticle_1999}
\begin{equation}\label{eq:T4gate}
\mathrm{T}_4(\theta,\phi) = e^{-i\theta\sum_{i<j}\sigma_\phi^{(i)}\sigma_\phi^{(j)}}
\end{equation}
Any subset of qubits can in principle be addressed with the collective gates, while leaving the others untouched. On a 2-qubit quantum register, $\mathrm{T}_4(\delta,0)$ can for example be used to obtain $\mathrm{XX}(\delta)$. Of course, the naturally collective character of trapped ions quantum gates is best exploited for the quantum simulation of long range and multiple-body interactions.

It is worth pointing out that while the elementary decomposition of typical two-qubits interaction terms reported here can be used to perform the digital quantum simulation of generic spin Hamiltonians, this is not necessarily the optimal strategy in general. Indeed, further optimization of, e.g.,\ combined two qubit operations can lead to an overall reduction of the total number of gates for particular target Hamiltonian models~\cite{vidal_universal_2004,las_heras_digital_2014,klco_quantum-classical_2018,chiesa_quantum_2019}. Examples of these techniques applied to the Heisenberg model simulated with $\mathcal{S}_1$ and $\mathcal{S}_2$ universal sets are discussed in Sec.~\ref{sec:examples} below. 

\subsection{Multiple-qubit interactions}
\label{sec:multiQint}

The generalization of $\mathrm{U}_{\alpha\beta}^{(i,j)}(\delta)$ building blocks to $N$-qubit interactions leads to unitary evolution terms of the form
\begin{equation}
\mathrm{U}_{\alpha_1\dots\alpha_N}(\delta) = e^{-i\delta\bigotimes_{i}\sigma^{(i)}_{\alpha_i}}
\end{equation}
These can be in principle always decomposed into single- and two-qubit operations. An example within the $\mathcal{S}_1$ universal set is the following:
\begin{equation}
\includegraphics[scale=0.9]{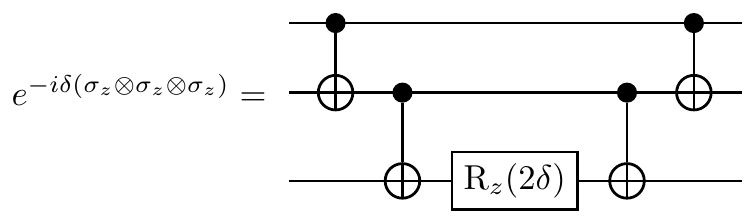}
\end{equation}
The pattern can be generalized to any $N>3$, and changes of reference frames can be applied to individual qubits as done for the $N=2$ case. 

In trapped ions processors, whose universal set $\mathcal{S}_4$ natively contains many body interactions, the decomposition of $N$-body terms can usually be done very efficiently using M{\o}lmer-S{\o}rensen gates~\cite{muller_simulating_2011} and the limits on $N$ are in principle dictated only by the scalability of the hardware set-up itself.

\subsection{Suzuki-Trotter decomposition and digital error}
% Figure fixed digital error vs fixed ntrot

\begin{figure}
\centering
\includegraphics[width=\columnwidth]{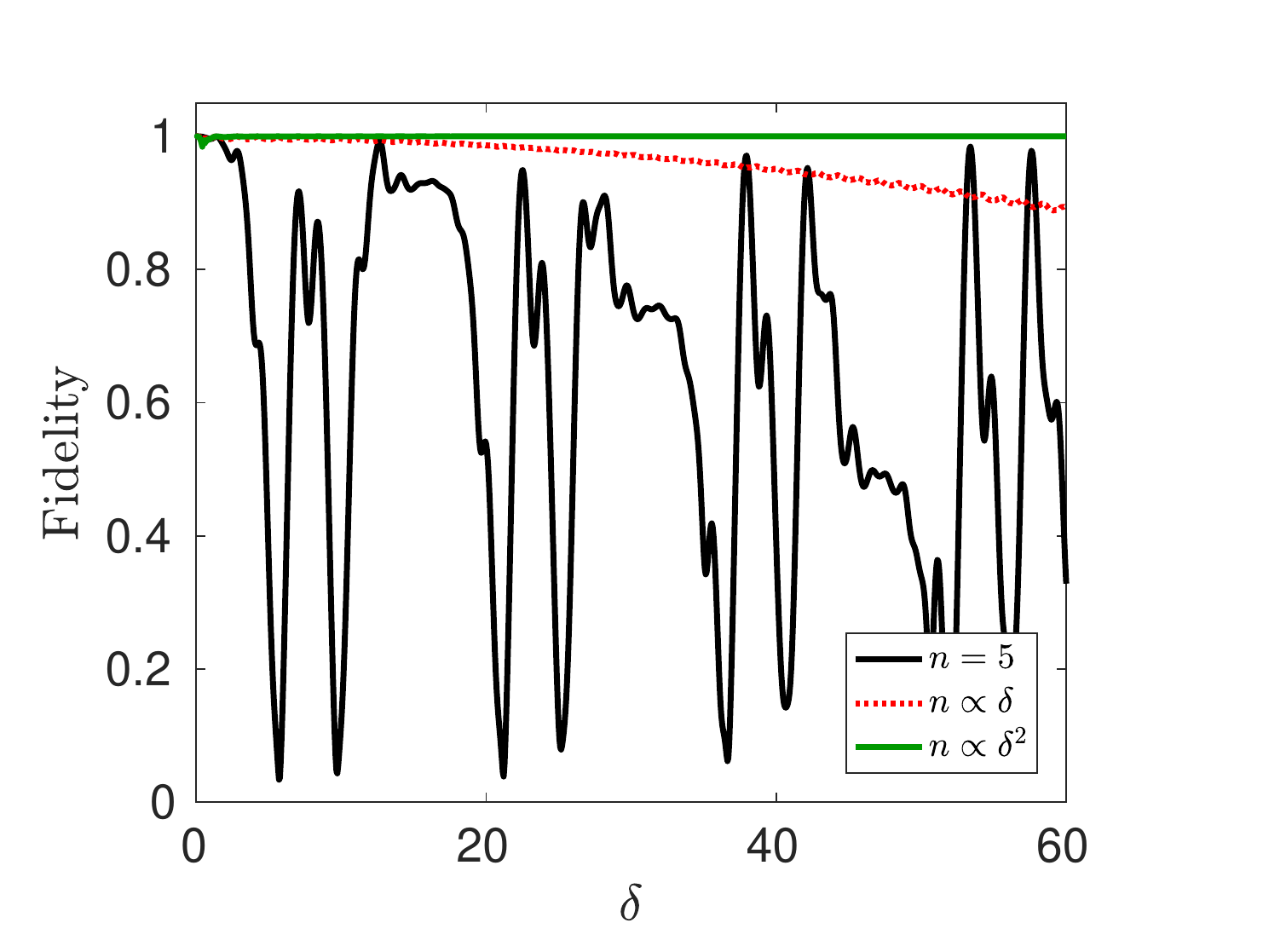}
\caption{Fidelity of the digital evolution for $|\psi_0\rangle = |00\rangle$, $\mathrm{O}_1 = -i(\sigma_x^{(1)}+\sigma_x^{(2)})$ and $\mathrm{O}_2 = -i\sigma_z^{(1)}\sigma_z^{(2)}$. The solid black line shows the fixed $n=5$ approach, which fails after a very short phase evolution. The dotted red line shows the case in which $n$ increases linearly with $\delta$ according to $n=\delta/2\epsilon$, while the solid green line shows the case in which the increase in $n=\delta^2/2\epsilon$ keeps the digital error fully under control. In the plot, $\epsilon=0.1$ and $n$ goes up to $n\simeq 10^4$ when the scaling is quadratic with the phase.}
\label{fig:DigitalError}
\end{figure}

When designing a quantum simulation which requires the non-trivial application of Suzuki-Trotter approximation formula, Eq.~\eqref{eq:ST}, the degree of acceptable digital error must be carefully assessed. This is critical for intermediate-scale non error-corrected quantum processors, where the increase in the number of gates which comes with the increase in the number $n$ of Trotter steps cannot proceed indefinitely without affecting the quality of the results. In practical cases, it is usually sufficient for the digital error to be just smaller than the hardware noise. If $\mathrm{O}_1$ and $\mathrm{O}_2$ are two operators such that $[\mathrm{O}_1,\mathrm{O}_2]\neq 0$, the so called first-order Suzuki-Trotter formula gives~{\cite{hatano_finding_2005}}
\begin{equation}
e^{(\mathrm{O}_1 + \mathrm{O}_2)\delta} \simeq \left(e^{\mathrm{O}_1\frac{\delta}{n}}e^{\mathrm{O}_2 \frac{\delta}{n}}\right)^n -  \frac{\delta^2}{2n}[\mathrm{O}_1,\mathrm{O}_2]
\label{eq:STfirstorder}
\end{equation}
A better scaling of the digital error can be obtained at the cost of an additional factor per iteration using the second-order formula~{\cite{hatano_finding_2005}}
\begin{equation}
e^{(\mathrm{O}_1 + \mathrm{O}_2)\delta} = \left(e^{\mathrm{O}_2 \frac{\delta}{2n}}e^{\mathrm{O}_1\frac{\delta}{n}}e^{\mathrm{O}_2 \frac{\delta}{2n}}\right)^n +  \mathcal{O}\left(\frac{\delta^3}{n^2}\right)
\end{equation}
In both cases, a ratio $r_\epsilon = \delta^p/n^q$ controls the digital error as a function of the target evolution phase and  the number of Trotter steps. Two different strategies can therefore be envisioned. 

On the one hand, one could aim at a fixed digital precision $\epsilon$ over the whole range of the dynamical simulation. This requires to increase the number of Trotter steps, and consequently the total length of the quantum circuit to be computed, keeping the ratio $r_\epsilon$ fixed. As an example, for the first-order formula in Eq.~\eqref{eq:STfirstorder} we get
\begin{equation}
n_\epsilon(\delta) \propto \frac{\delta^2}{2\epsilon}
\end{equation}
Notice that while the number of digital steps increases, the phase evolution $\delta_n = \delta/n$ required in each step decreases as $1/n$, thus keeping the overall computation time on the physical hardware linear in the total phase provided that each digital step can be implemented with a coherent operation of duration $t \propto 1/\delta_n$~\cite{lloyd_universal_1996}.

On the other hand, when the maximum length of quantum circuits that can be faithfully realized is \textit{de facto} limited, such as in state-of-the-art noisy quantum processors, it might be convenient to follow a different approach, namely to keep fixed the length of the quantum circuit (i.e.\ the number of steps $n$). This produces a phase-dependent digital error scaling e.g.\ with $\delta^2$ in the first-order case. The fixed computational complexity, and consequently the uniform effect of hardware noise over the whole simulation, comes at the cost of a limited range of phases (and therefore of physical times) in which the results of the simulation agree with the target model. Hybrid solutions are also possible, e.g.\ by selecting reasonable number of steps $n$ in different intervals of phases $\delta$, always with the primary goal of balancing the total error arising both from the hardware noise and software-level approximations. In Fig.~\ref{fig:DigitalError} we compare the two different approaches (fixed $\epsilon$ or fixed $n$) by showing how the fidelity $\left|\langle\psi_0|\psi_n(\delta)\rangle\right|$ of the digitally evolved state $|\psi_n(\delta)\rangle =  \left(\exp\left(\mathrm{O}_1\frac{\delta}{n}\right)\exp\left(\mathrm{O}_2 \frac{\delta}{n}\right)\right)^n|\psi_0\rangle$ with respect to the exact evolution $|\psi_{ex}\rangle = \exp\left(\left(\mathrm{O}_1 + \mathrm{O}_2\right)\delta\right)|\psi_0\rangle$ decreases at long evolution times $t\propto \delta$ when $n$ is fixed or increases only linearly with the phase. In this simple 2-qubit case, we choose $\mathrm{O}_1 = -i(\sigma_x^{(1)}+\sigma_x^{(2)})$ and $\mathrm{O}_2 = -i\sigma_z^{(1)}\sigma_z^{(2)}$, corresponding to interaction terms \blu{typical of an Ising model in a transverse field}.

\subsection{Extracting physical observables}
\label{sec:physObs}

At the end of a quantum simulation, the final state $|\psi (t)\rangle$ of the quantum register is measured to retrieve information about the physical properties of the system under study. With an appropriate mapping of the generic observable of interest, $\mathbb{O}$, onto a combination spin-$1/2$ operators, the expectation value $\langle\mathbb{O}(t)\rangle = \langle \psi(t) | \mathbb{O}|\psi(t) \rangle$ can be reconstructed by a readout procedure combining, e.g.,\ appropriate unitary operations $\mathrm{U}_{\text{meas}}$ and measurements in the computational basis. The reason why $\mathrm{U}_{\text{meas}}$ might be needed is that the eigenstates of $\mathbb{O}$ are, in general, different from the computational basis states: for example, if $\mathbb{O}= \sigma_x$ for a single qubit, the readout of $\langle\sigma_x(t)\rangle$ can be done by performing a Hadamard gate (i.e., mapping $\sigma_x\mapsto\sigma_z$) followed by a standard measurement in the computational basis. Joint qubit measurements are also possible, in general, as a way of characterizing the output quantum state~\cite{salathe_digital_2015}.

More refined strategies allow the extraction of complex physical quantities and to optimize the efficiency of the measurement process. Here we will review in particular ancilla-assisted observation of dynamical correlation functions and of the spectrum of an Hermitian operator. This topic is discussed in detail in Ref.~\onlinecite{somma_simulating_2002}.

\begin{figure}
\centering
\includegraphics[width=\columnwidth]{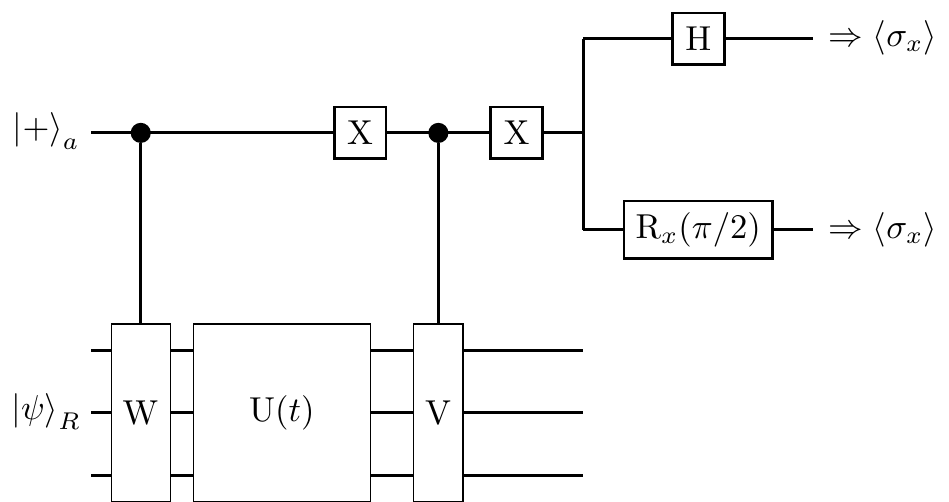}
\caption{Ancilla-based algorithm to compute dynamical correlation functions. The two alternative paths at the end of the circuit show a possible choice of unitaries $\mathrm{U_{meas}}$ which, followed by a measurement in the computational basis, give access to the real and imaginary parts of $\mathcal{C}_{\mathrm{VW}}(t)$, proportional to $\langle\sigma_x\rangle$ and $\langle\sigma_y\rangle$ respectively.}
\label{fig:VWcorr}
\end{figure}

\begin{figure*}[t]
\centering
\includegraphics[width=\textwidth]{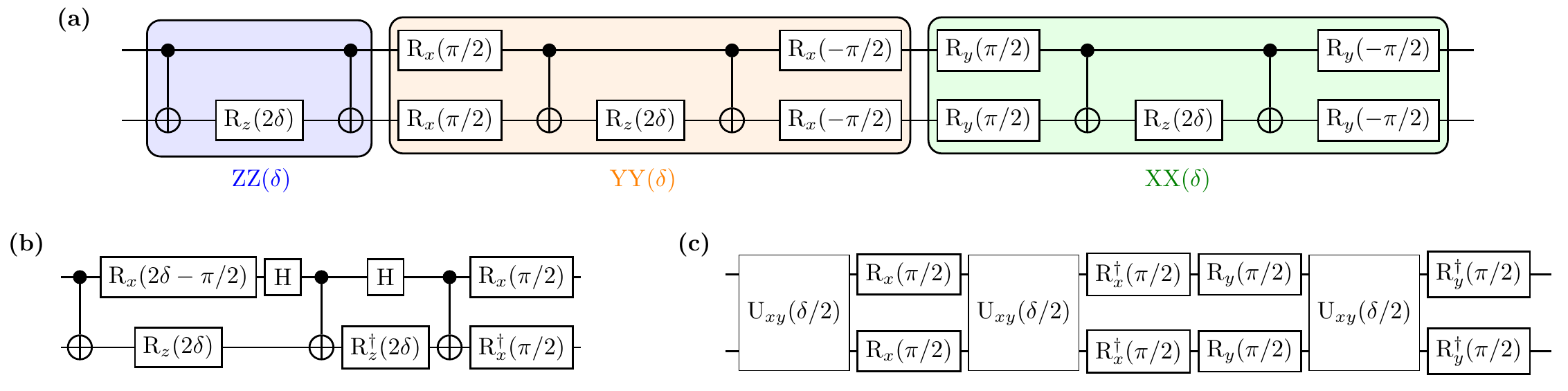}
\caption{Quantum circuits for the digital quantum simulation of the 2-qubit Heisenberg model with $\mathcal{S}_1$ and $\mathcal{S}_2$ universal sets. (a) $6$-$\mathrm{CNOT}$ decomposition. (b) $3$-$\mathrm{CNOT}$ decomposition.~\cite{vidal_universal_2004,chiesa_quantum_2019} (c) $3$-$\mathrm{U}_{xy}$ decomposition.~\cite{las_heras_digital_2014,ferrando-soria_switchable_2016}}
\label{fig:circuitsHeis2}
\end{figure*}

Given a $N$-qubit state $|\psi\rangle$, a Hamiltonian $H$ generating time evolution and two unitary operators $\mathrm{V}$ and $\mathrm{W}$, we define the dynamical correlation $\mathcal{C}_{\mathrm{VW}}(t)$ function as the quantity
\begin{equation}
\mathcal{C}_{\mathrm{VW}}(t) = \langle \mathrm{V}^\dagger (t) \mathrm{W}\rangle = \langle\psi|e^{iHt} \mathrm{V}^\dagger e^{-iHt} \mathrm{W}|\psi\rangle
\end{equation} 
The quantum circuit in Fig.\ \ref{fig:VWcorr} describes how to compute $\mathcal{C}_{\mathrm{VW}}(t)$ using a quantum register and an ancilla qubit $a$. Here we assume that the quantum register is already prepared in the desired state $|\psi\rangle$, e.g.\ the ground state of the target physical system, and that the ancilla starts in the quantum superposition $\sqrt{2}|+\rangle = |0\rangle + |1\rangle$. The joint initial state of the quantum register $R$ and the ancilla is therefore $|\phi\rangle_{aR} = |+\rangle_a|\psi\rangle_R$. The first step is a $\mathrm{W}$ unitary performed on $R$ and controlled by the ancilla: 
\begin{equation}
|\phi\rangle_{aR} \rightarrow \frac{1}{\sqrt{2}}\left(|0\rangle_a|\psi\rangle_R + |1\rangle_a \mathrm{W}|\psi\rangle_R \right)
\end{equation}
A quantum circuit implementing the digital simulation of the time evolution $\mathrm{U}(t) = e^{iHt}$ is then applied to the quantum register to evolve the state $|\psi\rangle$, thus leading to
\begin{equation}
\frac{1}{\sqrt{2}}\left(|0\rangle_a \mathrm{U}(t)|\psi\rangle_R + |1\rangle_a \mathrm{U}(t) \mathrm{W}|\psi\rangle_R \right)
\end{equation}
Finally, a $\mathrm{V}$ unitary is applied to $R$, controlled by the state $|0\rangle$ of the ancilla (this can be obtained by adding $\mathrm{X} \equiv \sigma_x$ quantum gates on $a$ before and after the standard controlled operation). The output state is:
\begin{equation}
|\phi_{out}\rangle = \frac{1}{\sqrt{2}}\left(|0\rangle_a \mathrm{V}\mathrm{U}(t)|\psi\rangle_R + |1\rangle_a \mathrm{U}(t) \mathrm{W}|\psi\rangle_R \right)
\end{equation}
A measure of the observable $\sigma_x$ on the ancilla gives
\begin{equation}
\begin{split}
\langle\sigma^{(a)}_x\rangle = & \, \operatorname{Tr}\left[\left(\sigma^{(a)}_x\otimes\mathbb{I}\right)|\phi_{out}\rangle\langle \phi_{out}|\right] \\ = & \, \operatorname{Re}\left[\mathcal{C}_{\mathrm{VW}}(t)\right]
\end{split}
\end{equation}
In a similar way, $\operatorname{Im}\left[\mathcal{C}_{\mathrm{VW}}(t)\right]$ can be obtained by measuring $\langle\sigma^{(a)}_y\rangle$, in a second run of the algorithm. In total
\begin{equation}
\langle2\sigma^{(a)}_+\rangle = \mathcal{C}_{\mathrm{VW}}(t)
\end{equation}
where $2\sigma_+ = \sigma_x + i\sigma_y$. The same scheme can be applied to equal-time correlations by removing the unitary evolution or by moving it at the beginning of the circuit to evolve some initial state. %In addition to that, if the initial state $|\psi\rangle$ of the quantum register is some $|\psi (t)\rangle$, e.g.\ the output of a previous digital simulation, the algorithm can be applied to extract the time evolution of equal-time correlations $\mathbb{C} (t) = \langle\psi(t)|\mathrm{V}^\dagger \mathrm{W}|\psi(t)\rangle$. 
It is worth noting explicitly that the useful information at the end of the proposed procedure is accessible through the ancilla $a$ alone, while the larger quantum register $R$ needs not to be measured at the end. The algorithm can also be generalized efficiently to the extraction of $n$-point time-correlation functions~\cite{pedernales_efficient_2014} and of the expectation value of any operator which can be expressed as $\mathbb{O} = \sum_j c_j \mathrm{V}_j^\dagger\mathrm{W}_j$ where $\mathrm{V}_j,\mathrm{W}_j$ are unitary operators~\cite{somma_simulating_2002}.

With the addition of a classical Fast Fourier Transform (FFT), the strategy described above for time correlation functions can be used to extract the spectrum of a Hermitian operator $\mathrm{Q}$. The most relevant example in physical problems is certainly $\mathrm{Q} = H$, for some Hamiltonian of interest $ H$. The hybrid quantum-classical approach, first proposed in Ref.~\onlinecite{somma_simulating_2002} and then further developed and applied (see e.g.\ Ref.~\onlinecite{chiesa_quantum_2019}), requires the quantum register $R$ to be initialized in a state $|\psi\rangle$ with some overlap with the eigenstates $|Q_l\rangle$ of $\mathrm{Q}$ 
\begin{equation}
|\psi\rangle = \sum_l \lambda_l |Q_l\rangle
\end{equation}
Since by hypothesis the target operator is Hermitian, its exponential $\mathrm{U}_\mathrm{Q}(\theta) = e^{-i\mathrm{Q}\theta}$ is a unitary operator. This can be realized on the quantum register in exactly the same way as any standard time-evolution operator $\mathrm{U}_H(\theta) = e^{-iH\theta}$. We can then compute the expectation value $\langle\psi|\mathrm{U}_\mathrm{Q}(\theta)|\psi\rangle$ with the ancilla-based protocol described in the prevoius paragraph, setting e.g.\ $|\psi\rangle_R = |\psi\rangle$, $t = 0$ (i.e.\ removing the time evolution $\mathrm{U}(t)$ part in Fig.~\ref{fig:VWcorr}), $\mathrm{W}=\mathrm{U}_\mathrm{Q}(\theta)$ and $\mathrm{V}=\mathbb{I}$. In general, the result will be of the form
\begin{equation}
\langle\mathrm{U}_\mathrm{Q}(\theta)\rangle = \sum_l |\lambda_l|^2 e^{-iq_l\theta}
\end{equation}
where $q_l$ are the eigenvalues of $\mathrm{Q}$. Applying FFT  to the variable $\theta$ then yields
\begin{equation}
\mathrm{FFT}\left(\langle\mathrm{U}_\mathrm{Q}(\theta)\rangle\right) = \sum_l 2\pi|\lambda_l|^2\delta(q-q_l)
\end{equation}

\subsection{Examples}
\label{sec:examples}

\begin{figure}[t]
\centering
\includegraphics[width=\columnwidth]{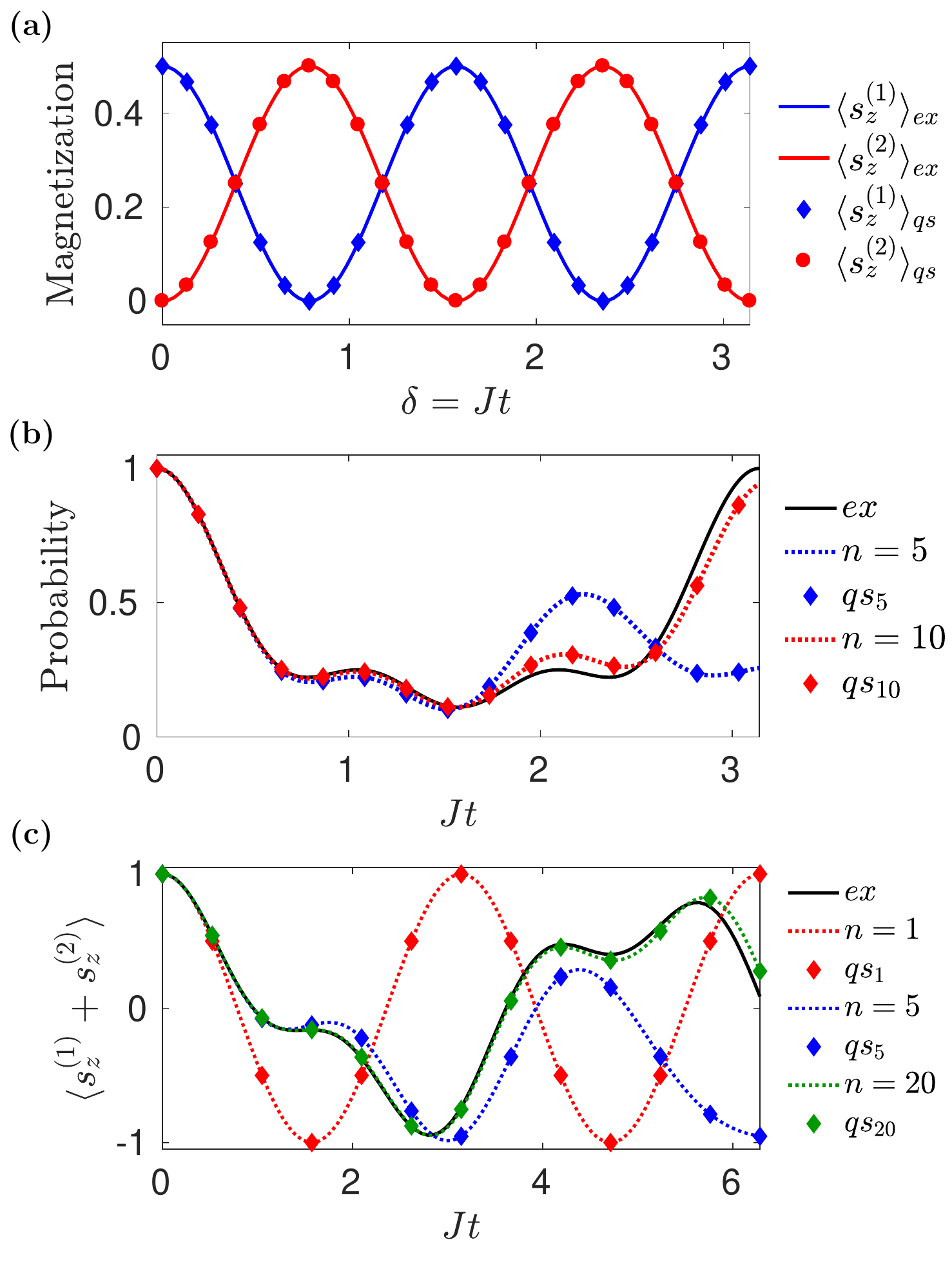}
\caption{Digital quantum simulation of spin models. The exact (`$ex$') curves are obtained with full Hamiltonian exponentiation, the dotted lines represent ideal digital approximation computed with the Suzuki-Trotter formula, and the quantum simulations data points (`$qs$') are computed numerically by matrix multiplication using the decomposition in elementary quantum gates. (a) Individual spin magnetization for the 2-qubit Heisenberg model, using the decomposition in Fig.~\ref{fig:circuitsHeis2}b for the digital quantum simulation. The initial state of the two spins is $\sqrt{2}|\psi_0\rangle = \left|\uparrow\right\rangle\left(\left|\uparrow\right\rangle+\left|\downarrow\right\rangle\right)$. (b) Time evolution of the occupation probability of the initial state $|\psi_0\rangle = |100\rangle$ of 3 qubits interacting as a linear Heisenberg chain with open ends in an external field. Here $J_{12} = J_{23} = J $ and $Bg = 20 J$. (c) Total magnetization of a pair of qubits interacting according to the transverse field Ising model, with $J_{zz}=J$ and $Bg = 2J$. The digital quantum simulation is performed using the $\mathcal{S}_1$ fundamental set of operations. }
\label{fig:EvoExamples}
\end{figure}

The Hamiltonian for a 2-qubit isotropic Heisenberg model is
\begin{equation}
H_{\text{Heis},2} = J\left(\sigma_x^{(1)}\sigma_x^{(2)}+\sigma_y^{(1)}\sigma_y^{(2)}+\sigma_z^{(1)}\sigma_z^{(2)}\right).
\end{equation}
The induced time evolution then reads
\begin{equation}
\begin{split}
\mathrm{U}_{\text{Heis},2}(\delta) = & \, e^{-i\delta\left(\sigma_x^{(1)}\sigma_x^{(2)}+\sigma_y^{(1)}\sigma_y^{(2)}+\sigma_z^{(1)}\sigma_z^{(2)}\right)} \\ = & \, e^{-i\delta\sigma_x^{(1)}\sigma_x^{(2)}}e^{-i\delta\sigma_y^{(1)}\sigma_y^{(2)}}e^{-i\delta\sigma_z^{(1)}\sigma_z^{(2)}}
\end{split}
\end{equation}
where $\delta = Jt$ and the second equality, which is essentially the ST formula for $n=1$, follows from $[\sigma_\alpha^{(1)}\sigma_\alpha^{(2)},\sigma_\beta^{(1)}\sigma_\beta^{(2)}] = 0 \, \forall \alpha,\beta$. Recalling Eq.~\eqref{eq:2qHeisenbergDeco} and the results in Sec.~\ref{sec:2qInteractions}, a 6-$\mathrm{CNOT}$ decomposition for arbitrary $\delta$ can be given using the universal set $\mathcal{S}_1$, as shown in Fig.~\ref{fig:circuitsHeis2}a. An equivalent and more efficient circuit in terms of number of two-qubit operations can be designed, according to the results discussed in Ref.~\cite{vidal_universal_2004}, if we consider the time evolution operator globally as a single two-qubit transformation, see Fig~\ref{fig:circuitsHeis2}b. Within $\mathcal{S}_2$, besides juxtaposing gate sequences of the form shown in Eq.~\eqref{eq:XXfromXY}, an optimal decomposition, using again only three 2-qubit gates instead of six, is reported in Fig.~\ref{fig:circuitsHeis2}c, based on the identity~\cite{las_heras_digital_2014,ferrando-soria_switchable_2016}
\begin{equation}
H_{\text{Heis},2} = \frac{J}{2}\left(H_{xxyy} + H_{xxzz}+ H_{zzyy}\right)
\end{equation}
where $H_{\alpha\alpha\beta\beta} = \sigma_\alpha^{(1)}\sigma_\alpha^{(2)}+\sigma_\beta^{(1)}\sigma_\beta^{(2)}$. In $\mathcal{S}_3$ a decomposition with three $\mathrm{C}\Phi(\delta)$ follows immediately from Eq.~\eqref{eq:ZZfromCP} and single qubit changes of reference frame. Finally, in $\mathcal{S}_4$ a possible realization of the Heisenberg interaction can be obtained for some digital resolution $\delta$ as %~\cite{lanyon_universal_2011}
\begin{equation}
\begin{split}
\mathrm{U}_{\text{Heis},2}(\delta) = \mathrm{A}\mathrm{B}\mathrm{C}\mathrm{A}\mathrm{C}^\dagger
\end{split}
\end{equation} %%%%%%%%% TO BE CHECKED... I could not understand the one in Lanyon fig. 2C!!!!!
where $\mathrm{A} = \mathrm{T}_4(\delta,0)$, $\mathrm{B} = \mathrm{T}_4(\delta,\pi/2)$ and $\mathrm{C} = \mathrm{T}_3(\pi/4,\pi/2)$. With any of the above elementary decomposition in quantum gates, the digital quantum simulation of the 2-qubits Heisenberg model can be performed and physical information can be extracted by using the methods discussed in Sec.~\ref{sec:physObs}. In a numerical example reported in Fig.~\ref{fig:EvoExamples}a we show the digital quantum simulation of the individual magnetization of the two spins, which can be extracted by measuring the observable $\sigma_z^{(i)}$ and using the definition  $\langle s_z^{(i)}\rangle =(1/2)\langle \sigma_z^{(i)}\rangle$. No digital error is present in this case.

\begin{figure}
\centering
\includegraphics[width=\columnwidth]{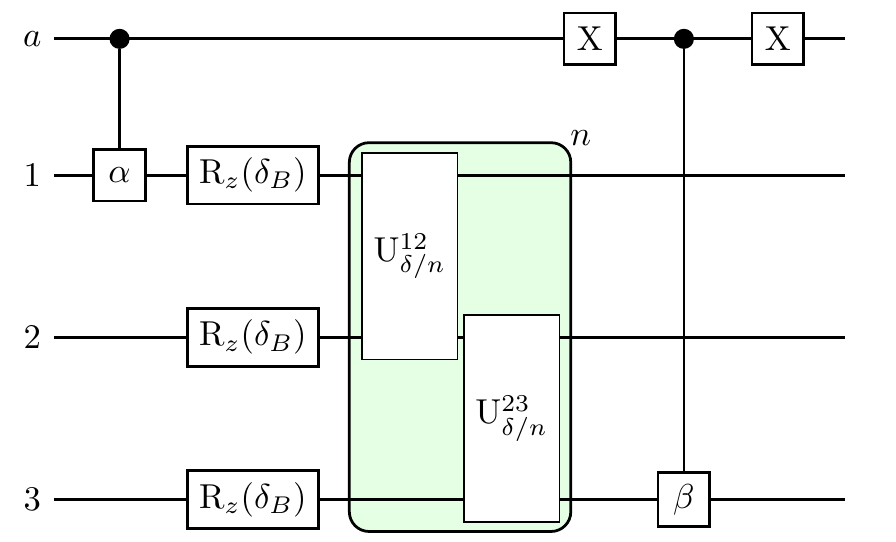}
\caption{Quantum circuit allowing to compute time-correlation functions of the 3-qubit Heisenberg model between next-to-nearest neighbors qubits, namely $\langle\sigma_\beta^{(3)}(t)\sigma_\alpha^{(1)}\rangle$. The operators $\alpha$ and $\beta$ represent (controlled) $\sigma_\alpha$ and $\sigma_\beta$ unitary transformations, $\delta_B = Bgt$ and $\mathrm{U}^{ij}_{\delta/n}$ is a shorthand notation for $\mathrm{U}^{ij}_{\text{Heis},2}(\delta_{ij}/n)$. The part inside the green box must be repeated $n$ times.}
\label{fig:HeisCirc}
\end{figure}

\begin{figure}
\centering
\includegraphics[width=\columnwidth]{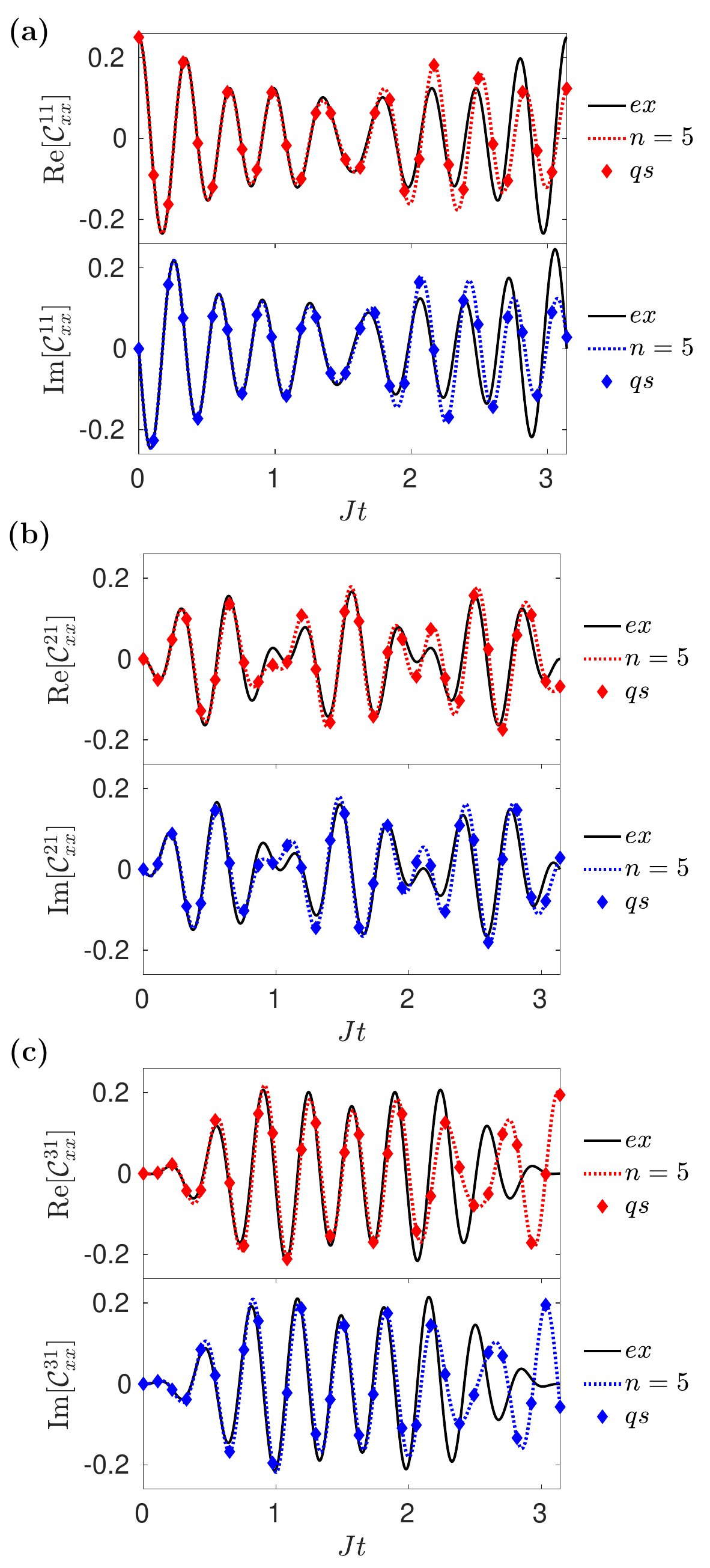}
\caption{Digital quantum simulation of dynamical correlation functions for the three spin Heisenberg model, computed using the circuit in Fig.~\ref{fig:HeisCirc}. The digital quantum simulation is shown for $n=5$ Trotter steps, with the dotted line representing the expected result for continuous phase and the data points showing the result of the corresponding quantum circuit for a selection of phase values. Here the quantum register is initialized in the quantum state corresponding to $|\psi\rangle = \left|\downarrow\downarrow\downarrow\right\rangle$ and we set $J_{12} = J_{23} = J $ and $Bg = 20 J$. (a) Autocorrelation $\langle s_x^{(1)}(t)s^{(1)}_x\rangle$.  (b) Nearest neighbors $\langle s_x^{(2)}(t)s^{(1)}_x\rangle$ cross correlation. (c) Next-to-nearest neighbors $\langle s_x^{(3)}(t)s^{(1)}_x\rangle$ cross correlation. }
\label{fig:CorrExamples}
\end{figure}

The decomposition of the 2-spin Heisenberg model into elementary quantum gates presented above can be used as a building block, in combination with single qubit rotations, to perform more complex digital quantum simulations. A 3-spin Heisenberg chain with open ends and $N_{b} = 2$ bonds, put in an external field, has a Hamiltonian of the form
\begin{equation}
H_{\text{Heis},3} = H_{B} + H_{\text{Heis},2}^{12} + H_{\text{Heis},2}^{23}
\end{equation}  
where 
\begin{equation}
H_{B} = \frac{Bg}{2}\left(\sigma_z^{(1)} + \sigma_z^{(2)} + \sigma_z^{(3)} \right)
\end{equation} 
describes a magnetic field oriented along the $z$-direction and each of the spin-spin bonds corresponds to a term
\begin{equation}
H_{\text{Heis},2}^{ij} = J_{ij}\left(\sigma_x^{(i)}\sigma_x^{(j)}+\sigma_y^{(i)}\sigma_y^{(j)}+\sigma_z^{(i)}\sigma_z^{(j)}\right)
\end{equation}
In general, the two bonds can be nonequivalent, i.e.\ $J_{12}\neq J_{23}$. Since $[H_{\text{Heis},2}^{12},H_{\text{Heis},2}^{23}]\neq 0$ (independently from the coupling constants $J_{ij}$), the quantum simulation must be carried out using the ST digital procedure, alternating the application of the results presented for the 2-spin case on the two bonds
\begin{equation}
\mathrm{U}_{\text{Heis},3}(\delta) = \left(\mathrm{U}^{12}_{\text{Heis},2}(\delta_{12}/n)\mathrm{U}^{23}_{\text{Heis},2}(\delta_{23}/n)\right)^n e^{-iH_Bt}
\end{equation}
where $\delta_{ij} = J_{ij}t$. The part describing the magnetic field on equivalent spins (we set the gyromagnetic ratio $g_1 = g_2 = g_3 = g$) corresponds to single qubit rotations around the $z$ axis. Since this part commutes with the rest, it can be performed at the beginning of the circuit without any phase discretization. In Fig.~\ref{fig:EvoExamples}b we show how these results can be used to compute the time evolution of the occupation probability of an initial state $|\psi_0\rangle = |100\rangle$.

As a third example, we recall that the Hamiltonian of the transverse field Ising model (TIM), introduced in Eq.~\eqref{eq:TIMgeneralH}, in the two-qubit case can be written as
\begin{equation}
H_{\text{TIM},2} = H_{B,x} + H_{zz}
\end{equation}
where
\begin{equation}
H_{B,x} = \frac{Bg}{2}\left(\sigma_x^{(1)}+\sigma_x^{(2)}\right) \quad H_{zz} = J_{zz}\sigma_z^{(1)}\sigma_z^{(2)}
\end{equation}
The quantum simulation of the \blu{transverse field Ising model} corresponds to the following digital process
\begin{equation}
\mathrm{U}_{\text{TIM},2}(t) = \left(\mathrm{ZZ}(J_{zz}t/n)e^{-i\left(\sigma_x^{(1)}+\sigma_x^{(2)}\right)\frac{Bgt}{2n}}\right)^n
\end{equation}
Apart from straightforward single qubit rotations around the $x$ axis, the required quantum circuit contains only $\mathrm{ZZ}$ operations, which can easily be translated into elementary quantum gates as shown in Sec.~\ref{sec:2qInteractions}. The time evolution of the total magnetization of the spin dimer along $z$ can then be extracted by measuring the expectation values of $\sigma_z^{(i)}$, see Fig.~\ref{fig:EvoExamples}c.

Finally, we also report the example of a 3-spin open Heisenberg chain with an application of the ancilla-based algorithm discussed in Sec.~\ref{sec:physObs} to the extraction of spin-spin dynamical correlations $\mathcal{C}_{ij}^{\alpha\beta} (t)=\langle s_\alpha^{(i)}(t)s^{(j)}_\beta\rangle = (1/4)\langle\sigma_\alpha^{(i)}(t)\sigma^{(j)}_\beta\rangle$ on the system ground state. The latter, for the model under study and for a sufficiently strong external field, $B$, is well approximated by $|\psi\rangle = \left|\downarrow\downarrow\downarrow\right\rangle$, which is then assumed as the initial state on the quantum register. The structure of the required quantum circuit is shown in Fig.~\ref{fig:HeisCirc} for the case of next-to-nearest neighbors cross correlations. Autocorrelations and nearest neighbors correlations can be computed in a similar way by changing the target qubit involved in the operations controlled by the ancilla. Numerical results based on $\mathcal{S}_1$ decompositions are presented in Fig.~\ref{fig:CorrExamples}. 

Despite the relatively small size of the systems presented in the previous examples, all the elements introduced in this section can be used as basic modules to extend the quantum simulation to an arbitrary number of spins with pairwise interactions. \blu{When scaling up any spin chain to a larger number of interacting elements, i.e., with $N_{b} > 2$ and possibly to different inter-qubits connectivity, one may also consider that all the terms involving disjoint sets of locally interacting spins generate independent time-evolution terms commuting with each other. These can then be simulated in parallel by using the elementary quantum circuit building blocks defined above, thus reducing the overall complexity of the quantum simulation (i.e., depth of the quantum circuit). As an example, let's consider a chain of 4 spins labelled from $1$ to $4$ with pairwise $1-2$, $2-3$, and $3-4$ interactions: the $1-2$ and $3-4$ blocks can be run in parallel \cite{santini_molecular_2011,chiesa_digital_2015}.} It is also worth mentioning that, concerning the simulation of dynamical correlation functions, in any $N$-spin system there are $\mathcal{O}(N^2)$ two-body sigma correlations of the form $\langle s_\alpha^{(i)}(t)s^{(j)}_\beta\rangle$. These quantities, which are often of great physical interest~\cite{chiesa_quantum_2019}, can then be extracted efficiently, in principle, with the ancilla-based methods discussed in Sec.~\ref{sec:physObs}, e.g.\ by repeating a polynomial number of times the calculation with slightly modified circuits for each spin pair.

\blu{As mentioned in Sec.~\ref{sec:cookbook}, many problems in digital quantum simulations can be reduced to an equivalent spin Hamiltonian, which can in turn be directly mapped onto a $N$-qubit quantum register. A significant example is provided by fermionic systems, which play a central role in fields such as quantum chemistry, solid state physics and material sciences and, at the same time, are typically hard to treat with classical computational methods. To understand how the mapping $\mathcal{H}\mapsto H(\{\sigma_\alpha\})$ is carried out in such a case, we consider an elementary two-sites Fermi-Hubbard model:
\begin{equation}
\begin{split}
\mathcal{H} = & \, - V \left(c^\dagger_{1,\downarrow}c_{2,\downarrow} + c^\dagger_{1,\uparrow}c_{2,\uparrow} + h.c. \right) \\ + & \, U \left(c^\dagger_{1,\downarrow}c_{1,\downarrow}c^\dagger_{1,\uparrow}c_{1,\uparrow} +  c^\dagger_{2,\downarrow}c_{2,\downarrow}c^\dagger_{2,\uparrow}c_{2,\uparrow}\right)
\end{split}
\end{equation}
where $h.c.$ means the Hermitian conjugate and the operator $c^\dagger_{i,s}$ ($c_{i,s}$) creates (annihilates) a fermion with spin $s \in \{\uparrow,\downarrow\}$ on site $i \in \{1,2\}$. In $\mathcal{H}$, the coefficients $V$ and $U$ are hopping and on-site repulsion energies.  Fermionic operators obey the canonical anticommutation rules
\begin{equation}
\{c_{i,s},c_{i',s'}^\dagger\} = \delta_{ii',ss'}\mathbb{I} \qquad \{c_{i,s},c_{i',s'}\} = 0
\end{equation}
The mapping is obtained by applying the Jordan-Wigner transformation~\cite{jordan_uber_1928,bari_classical_1973,ortiz_quantum_2001,somma_simulating_2002}, which in this case takes the explicit form~\cite{barends_digital_2015,lamata_digital-analog_2018}
\begin{equation}
\begin{split}
c^\dagger_{1,\downarrow} = & \, \mathbb{I} \otimes \mathbb{I} \otimes \mathbb{I} \otimes \sigma_+ \equiv \sigma_+^{(4)}\\
c^\dagger_{2,\downarrow} = & \, \mathbb{I} \otimes \mathbb{I} \otimes \sigma_+ \otimes \sigma_z  \equiv \sigma_+^{(3)} \sigma_z^{(4)}\\
c^\dagger_{1,\uparrow} = & \, \mathbb{I} \otimes \sigma_+ \otimes \sigma_z \otimes \sigma_z \equiv \sigma_+^{(2)}\sigma_z^{(3)} \sigma_z^{(4)}\\
c^\dagger_{2,\uparrow} = & \, \sigma_+ \otimes \sigma_z \otimes \sigma_z \otimes \sigma_z \equiv \sigma_+^{(1)}\sigma_z^{(2)}\sigma_z^{(3)} \sigma_z^{(4)}\\
\end{split}
\label{eq:JW}
\end{equation}
Here, as usual, we use $2\sigma_+ = \sigma_x + i\sigma_y$ and the anticommutation relations are preserved due to the properties of Pauli matrices. The resulting Pauli Hamiltonian reads
\begin{equation}
\begin{split}
H = & \,\frac{V}{2} \Big(\sigma_x^{(1)}\sigma_x^{(2)} + \sigma_y^{(1)} \sigma_y^{(2)} \\ & \, + \sigma_x^{(3)} \sigma_x^{(4)} + \sigma_y^{(3)} \sigma_y^{(4)}\Big) \\ + & \, \frac{U}{4} \Big(\sigma_z^{(1)}\sigma_z^{(4)} + \sigma_z^{(1)} + \sigma_z^{(4)} \\ & \, + \sigma_z^{(2)}\sigma_z^{(3)} + \sigma_z^{(2)} + \sigma_z^{(3)}\Big)
\end{split}
\end{equation}
Since $H$ is now the sum of local single- and two-body Pauli terms, the corresponding time evolution operator can be digitally simulated on a 4-qubit quantum register, using the techniques and the building blocks described above for paradigmatic spin-$1/2$ interactions. \\
While a one-dimensional fermionic chain with on-site and nearest-neighbor interactions can be mapped onto a spin Hamiltonian including only one- and two-spin terms \cite{santini_molecular_2011}, the application of the Jordan-Wigner transformation in more than one dimension or with long-range couplings leads to multi-spin interactions in the resulting Hamiltonian. Indeed, in the most general case, the mapping of Eq.~\eqref{eq:JW} takes the form $c^\dagger_j = \left(\prod_{l=1}^{j-1} -\sigma_z^{(l)} \right) \sigma_+^{(j)}$ \cite{somma_simulating_2002}, where we have introduced a single label $j=\{i,s\}$ for fermionic modes, addressing both site and spin variables. The resulting spin Hamiltonian contains several terms, depending on the number of occupied states existing between each pair of interacting fermions after lattice sites have been properly ordered~\cite{chiesa_digital_2015}. The quantum simulation of these terms is demanding, in terms of quantum circuit depth, for any architecture implementing only two-qubit interaction terms. Hence, in the NISQ era, the practical quantum simulation of fermionic models can strongly benefit from native many body terms in the hardware Hamiltonian, as it is the case for the trapped ion quantum processors to be briefly recalled in the following.
}

%{\bf [Other examples, larger systems?]}

\section{Experimental achievements and prospective technologies}

The last few years have represented a timeline of intense development towards the realization of quantum computing architectures. Among the plethora of possible platforms, two leading technologies have been emerging: trapped ions and superconducting quantum circuits. Here we give a brief overview of the main achievements reported to date in these two experimental set-ups. \blu{The focus will be on the perspective development of these two platforms as universal quantum simulators, rather than reviewing the technical details of the different setups, for which thorough reviews already exist.
We notice that a comparison between the two leading platforms can be done based, for instance, on relevant figures of merit such as gate fidelities or the ratio between coherence and gating times, roughly representing the number of operations that the quantum hardware can reliably perform before the quantum information is degraded. However, it is important to stress that most of the remarkably high experimental fidelities reported in the literature have been achieved on two-qubit setups that, although in principle scalable, are optimized for a specific target and are often pushed to the experimental limits. Conversely, when several qubits are connected and operated together, many new challenges emerge, such as the need to selectively address only some of them or to stem cross-talk errors. As discussed below, this yields slower and more error-prone gates. Hence, a fair study on the actual performances of a quantum simulator should be done on an architecture including several interacting qubits, in order to assess the effective scalability of the given setup. A useful concept, in this respect, is that of ``quantum volume'', recently introduced by IBM researchers with the aim of quantifying with a single paramenter the performance of a quantum computer, based on how efficiently a quantum algorithm can be run \cite{moll_quantum_2018}. The quantum volume takes into account both width (i.e., the number of qubits used) and depth (i.e., the number of reliable operations performed) of a given quantum circuit. Despite a few non trivial issues related, e.g., to individual qubit addressing and frequency crowding, adding qubits to the actual hardware mostly appears to be a technological challenge, while implementing a sequence of high-fidelity operations on several qubits could be more demanding. Indeed, this requires a considerable improvement of gate fidelities, suppression of both qubit decoherence and coherent errors due to imperfect qubit manipulations, and reduction of unwanted qubit-qubit interactions (cross-talk) whose harmful effect increases with the system size. Increasing the quantum volume truly represents the next challenge for current NISQ devices to evolve into useful quantum simulators.}\\
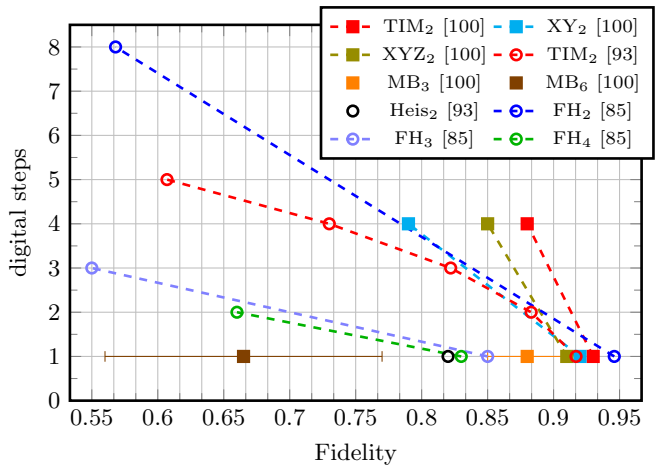
\begin{figure}
\centering
    \begin{tikzpicture}[spy using outlines={rectangle, magnification=2}] %connect spies
		\begin{axis}[%
                width=7.6cm,%\columnwidth,%8cm,
                height=5cm,
                scale only axis,
                ymin=0, ymax=8.5,
                ytick={0,1,2,3,4,5,6,7,8},
                yticklabels={0,1,2,3,4,5,6,7,8},
                xmajorgrids,
                xminorgrids,
                minor x tick num=2,
                minor y tick num=1,
                xmin=0.5334, xmax=0.9667,
                ylabel={digital steps},
                xlabel={Fidelity},
                ymajorgrids,
                %title={Stroke Ratio Comparison},
                %axis lines*=left,
                line width=1pt,
                mark size=2pt,
                legend style={at={(0.73,1.05)},anchor=north,draw=black,fill=white,align=left,legend columns=2}]

               \addplot+[%
                dashed,
                color=red,
                mark=square*,
                mark options={solid},                
                ]
                coordinates{
                 (0.93,1) (0.88,4)
                };
                \addlegendentry{\scriptsize{TIM$_2$ \cite{lanyon_universal_2011}}};

                \addplot+[%
                dashed,
                color=cyan,
                mark=square*,
                mark options={solid},                
                ]
                coordinates{
                 (0.92,1) (0.79,4)
                };
                \addlegendentry{\scriptsize{XY$_2$ \cite{lanyon_universal_2011}}};
                
                \addplot+[%
                dashed,
                color=olive,
                mark=square*,
                mark options={solid},                
                ]
                coordinates{
                 (0.91,1) (0.85,4)
                };
                \addlegendentry{\scriptsize{XYZ$_2$ \cite{lanyon_universal_2011}}};
                
                \addplot+[%
                dashed,
                color=red,
                mark=o,
                mark options={solid}             
                ]
                coordinates{
                 (0.917,1) (0.883,2) (0.822,3) (0.73,4) (0.607,5)
                };
                \addlegendentry{\scriptsize{TIM$_2$ \cite{salathe_digital_2015}}};
                
                \addplot+[%
                solid,
                only marks,
                color=orange,
                mark=square*,
                mark options={solid},
                error bars/.cd,
			    x dir=both,x explicit              
                ]
                coordinates{
                 (0.88,1) +- (0.03,0)
                };
                \addlegendentry{\scriptsize{MB$_{3}$ \cite{lanyon_universal_2011}}};
                
                \addplot+[%
                solid,
                only marks,
                color=orange!50!black,
                mark=square*,
                mark options={solid},
                error bars/.cd,
			    x dir=both,x explicit              
                ]
                coordinates{
                 (0.665,1) +- (0.105,0)
                };
                \addlegendentry{\scriptsize{MB$_{6}$ \cite{lanyon_universal_2011}}};
                
                \addplot+[%
                dashed,
                only marks,
                color=black,
                mark=o,
                mark options={solid}             
                ]
                coordinates{
                 (0.82,1)
                };
                \addlegendentry{\scriptsize{Heis$_{2}$ \cite{salathe_digital_2015}}};
                
                \addplot+[%
                dashed,
                color=blue,
                mark=o,
                mark options={solid}             
                ]
                coordinates{
                 (1-0.054,1) (1-8*0.054,8) %(2,1-2*0.054) (3,1-3*0.054) (4,1-4*0.054) (5,1-5*0.054) (6,1-6*0.054) (7,1-7*0.054) (8,1-8*0.054)
                };
                \addlegendentry{\scriptsize{FH$_2$ \cite{barends_digital_2015}}};
                
                \addplot+[%
                dashed,
                color=blue!50,
                mark=o,
                mark options={solid}             
                ]
                coordinates{
                 (1-0.15,1) (1-3*0.15,3) %(2,1-2*0.15) (3,1-3*0.15)
                };
                \addlegendentry{\scriptsize{FH$_3$ \cite{barends_digital_2015}}};
                
                \addplot+[%
                dashed,
                color=green!70!black,
                mark=o,
                mark options={solid}             
                ]
                coordinates{
                 (1-0.17,1) (1-2*0.17,2)
                };
                \addlegendentry{\scriptsize{FH$_4$ \cite{barends_digital_2015}}};

                %\coordinate (spypoint) at (axis cs:1,0.8825);
                %\coordinate (spyviewer) at (axis cs:6.5,0.9);
                %\spy [width=1.5cm,height=3cm] on (spypoint) in node [fill=white] at (spyviewer);              
		\end{axis}

    \end{tikzpicture}
\caption{Summary of state-of-art experimental digital quantum simulations. Open circles represent results obtained on superconducting circuits quantum processors, while squares correspond to experimental quantum simulations on trapped ions processors. The color code corresponds to different target models being simulated: two-spin Transverse field Ising model (TIM$_2$), two-spin XY (XY$_2$) and XYZ  (XYZ$_2$) models, 3- and 6-spin many body interactions (MB$_{3,6}$, fidelities given as estimated bounds), two-spin Heisenberg model (Heis$_2$), and 2- to 4-mode Fermi Hubbard model (FH$_x$, with $x = 2,3,4$). Although more digital steps than the ones reported here were actually performed in some of the experiments, data points are shown only when some measure of accuracy was provided in the original reference. Fidelities from Ref.~\onlinecite{salathe_digital_2015} are given with respect to the ideal evolution for a fixed phase value and initial state, while those from Ref.~\onlinecite{lanyon_universal_2011} are process fidelities given with respect to the expected digitized evolution. Finally, data from Ref.~\onlinecite{barends_digital_2015} are extrapolated linear trends of fidelity with respect to the ideal digital outcome. %The inset shows a magnified version of the region inside the smaller rectangle.
}\label{fig:ExpSummary}
\end{figure}

In Fig.~\ref{fig:ExpSummary}, we try to give a quantitative summary of the main experimental achievements reported in the recent literature. \blu{As mentioned above, it is generally difficult to directly compare experiments performed on different platforms, 
under different initial conditions}, and reporting slightly different figures of merit. \blu{However,} the graph \blu{aims at giving} a visual idea of the scenario on the digital quantum simulation of spin Hamiltonians up to date, and the capabilities offered in different quantum hardware. Evidently, there is still a considerable correlation between the number of digital steps included in the simulation and the fidelity of the final state obtained. \blu{As a general comment}, trapped ions quantum simulators allow deeper quantum circuits with better performance, i.e. a larger number of Trotter steps is possible. This is in line with recent studies comparing the two platforms when challenged with similar quantum algorithms on 5 qubits processors~\cite{linke_experimental_2017}. One may notice that 5 Trotter steps are currently a limiting value for superconducting circuit quantum simulators, where the fidelity drops to values slightly above $60\%$, \blu{still far from an acceptable result, also in view of scalability}. Finally, in terms of size of the simulated model we see that spin-models with up to 6 spins have been simulated on trapped ions processors \cite{lanyon_universal_2011}, while up to 4 spins on superconducting ones \cite{chiesa_quantum_2019}. \\
\blu{As a last comment, it is worth mentioning that great efforts have been lately devoted to design error-mitigation strategies to improve the quality of the final results from the data directly extracted from the quantum hardware. Some of them are problem specific, based, e.g., on the symmetries of the target Hamiltonian \cite{mcardle_error_2019}, or on general properties of the measured observables \cite{chiesa_quantum_2019}, which set constraints on the output of the quantum calculation. In particular, the stabilizer-like method proposed in Ref. \cite{mcardle_error_2019} could enable the detection of up to 60-80 \% of depolarizing errors, thus largely improving the results of variational algorithms, as well as of quantum simulations.
Theoretical proposals have also been developed, demonstrating that the accuracy in the expectation values of computed quantum observables can be improved by suitably interpolating to zero noise the results on a series of experiments at varying noise levels \cite{PRX_Benjamin_2017,PRX_Benjamin_2018}. This strategy has been recently applied to suppress incoherent errors in variational calculations on superconducting circuits \cite{kandala_error_2019} (see below). 
The combination of these two approaches could further improve the accuracy of the computed observables \cite{mcardle_error_2019}. } \\

We will now give a more detailed description of each of the two aforementioned experimental platforms, and the corresponding key results on the digital quantum simulation of spin models. \blu{A word of warning should be given, however, before any further discussion takes place: a number of challenges still need to be faced before either one or the other may ultimately develop into a fully operational quantum simulation technology with clear advantages over classical simulations of quantum many body models. In fact, while scalability requires that the number of qubits on a given quantum hardware be arbitrarily increased without degrading the reliability of individual preparation and read-out, as well as  quantum gating fidelities, there is no current quantum technology that is truly fulfilling this requirement. The key challenges to be faced in the forthcoming years are thus related to practically addressing these issues, even before fault tolerant quantum computation be developed. 
%While several proposals and ideas are already present in the literature, it is worth emphasizing that no real breakthrough demonstration has been published in such direction at time of writing, neither in the leading technological platforms nor in the several alternative approaches currently pursued, some of them outlined in the last part of this Section. 
}

\subsection{UQS with trapper ions }
\label{sec:TrapIons}

\blu{Atomic ions trapped in a suitable combination of static and time-varying electric fields have been representing one of the most promising routes towards realizing fully operational quantum processors since the late nineties \cite{blatt_entangled_2008,blatt_quantum_2012,monroe_scaling_2013}. As compared to neutral atoms, the trapping potential for charged atomic species can be much stronger, thus allowing to hold each single ion for several hours, and even days, with very long coherence times and a remarkably high degree of external control. This makes them a reliable solution for quantum information processing satisfying all of the required DiVincenzo criteria (for a comprehensive recent review on the state-of-art of this technology focused on quantum computing, we refer to Ref.~\onlinecite{bruzewicz_review-trap-ion_2019}). Today, digital quantum processors with up to 11 programmable ions in a linear Paul trap have been made available \cite{wright_benchmarking_2019}, also through spin-off companies such as IonQ  offering restricted cloud access (see https://ionq.com/, accessed 2019-09-01). Other companies are working on providing access to general purpose trapped ion quantum computing hardware with up to 20 qubits, such as Alpine Quantum Technologies (see, e.g., https://www.aqt.eu/, accessed 2019-09-01). In such realizations the trapping is generally created through radio frequency oscillating electric fields that generate a stable two-dimensional potential well for the charged atoms \cite{leibfried_review-trap-ion_2003}, such that ions are then confined by a further harmonic trap along a linear chain, in which the single particles are stably kept a few microns apart due to the mutual Coulomb repulsion \cite{monroe_scaling_2013}. While this is the configuration that currently allows the most performing type of apparatus as programmable NISQ devices, it is considered hardly scalable towards larger numbers of trapped ions within the same chain, as well as to higher dimensionality of the confining potential, for which new solutions will have to be realized. In particular, modularity \cite{bautista_multilayer-iontraps_2019} and 2D arrays \cite{bruzewicz_2Darray_ions_2016} are being considered as valuable approaches, for which we will likely see progress in the near future.  It is worth noticing that this type of hardware requires ultra-high vacuum and possibly laser cooling, but not necessarily cryogenic apparatus to be operated. \\
Trapped ions may be employed as reliable, coherently manipulated qubits, in which the information is encoded into the internal quantum states of the charged atomic species.  While we refer the interested reader to more specialized review works on trapped ion quantum technologies  \cite{leibfried_review-trap-ion_2003,blatt_entangled_2008,schindler_quantum_2013}, we hereby summarize its main aspects for completeness, especially concerning the ongoing developments towards programmable NISQ devices. First, we remind that different types of trapped ion qubits can be defined, depending on the frequency spacing between the relevant energy eigenstates selected to encode the logical $|0\rangle$ and $|1\rangle$. In general, \textit{hyperfine} qubits are encoded into a pair of hyperfine energy levels typically separated by GHz frequencies, while \textit{optical} qubits are defined corresponding to quadrupole active transitions in the hundreds THz range; also Zeeman qubits can be defined by application of an external static magnetic field, opening a low frequency (few MHz) and tunable gap between magnetic levels. Pros and cons characterize each specific choice, and a dedicated experimental apparatus must necessarily be developed for one qubit type or the other. Among the different possibilities, particularly advanced appear the technologies based on $^{40}$Ca$^+$ as optical qubits \cite{friis_observation_2018,hempel_quantum-chemistry_2018}, and $^{43}$Ca$^+$ or $^{171}$Yb$^+$ as hyperfine qubits \cite{ballance_fidelity-gates_2016,debnath_demonstration_2016}, respectively, although several other atomic species with a single outer electron can be successfully trapped \cite{monroe_scaling_2013,bruzewicz_review-trap-ion_2019}.  In fact, irrespective of the specific qubit realization, initialization and readout of their quantum state is performed through coherent manipulation, e.g. by using external lasers of suitable frequency to optically pump the ion state into the desired one (preparation), or by detecting resonantly scattered radiation from an optical transition (readout). The qubits' individual control (i.e., single-qubit operations) can be performed by directly coupling the $|0\rangle$ and $|1\rangle$ eigenenergy levels, for which the required tools inevitably depend on the qubits type (e.g., hyperfine qubits require microwave control or stimulated Raman coupling, while optical qubits internal states are directly coupled through a resonant laser) \cite{leibfried_review-trap-ion_2003}. Two-qubit entangling gates between ions trapped along the same chain are realized by exploiting the transverse normal vibrational modes of the whole ion string trapped in a harmonic potential \cite{cirac_quantum_1995}, which are used as a bus to transfer quantum information. Currently, the most performing multi-qubit operations  rely on a controlled-phase type of gate that was originally proposed from M{\o}lmer and S{\o}rensen \cite{molmer_multiparticle_1999}, whose formal expression is given in Eq.~(\ref{eq:T4gate}). 
The details of the specific implementation of these gates depend on the type of qubit~\cite{schindler_quantum_2013,shapira_robust_2018,webb_resilient_2018}. In general, they  require only global control lasers to entangle multiple qubits, hence avoiding beams to be independently focused on each ion (which may anyway be necessary, e.g., for single-qubit control). In hyperfine qubits-based quantum hardware, a M{\o}lmer-S{\o}rensen gate with a global laser and suitably detuned individually addressing beams allow to realize an effective two-qubit XX gate, which then becomes the native two-qubit gate on that hardware \cite{debnath_demonstration_2016}. Elementary combinations of single-qubit rotations and such XX two-qubit gate finally lead to the CNOT gate \cite{nielsen_quantum_2000}. Independently of the specific implementation, an unequivocal advantage of any trapped ion quantum hardware is that each qubit can be connected (and entangled) to any other qubit in the chain, thus practically realizing all-to-all connectivity.\\
In terms of absolute performance specifically meant for prospective applications in quantum computing, the different experimental platforms have essentially shown quite comparable figures of merit. In particular, when working with isolated or pairs of trapped ions, single-qubit operations with fidelities in the order of $99.9999\%$ have been shown \cite{harty_high-fidelity_2014}, as well as two-qubit gates reaching fidelities in excess of $99.9\%$ even in different experimental setups \cite{gaebler_fidelity-gate-set_2016,ballance_fidelity-gates_2016}. 
Typical duration for single-qubit gates varies between 100 ns and few tens of $\mu$s, while two qubit gating times range in the $\mu$s to few hundred $\mu$s interval \cite{ballance_fidelity-gates_2016,schafer_fast_2018,bruzewicz_review-trap-ion_2019}; readout is typically performed in hundreds of $\mu$s with fidelities in the $99.99\%$ range \cite{meyerson_readout_2008}. Generally speaking, a trade-off between speed and fidelity has to be found, where faster usually means more error prone.
Considering also that the reported coherence times of trapped ion qubits vary between few hundred ms \cite{bermudez_progress_2017} and hundreds of seconds (i.e., up to several minutes) \cite{harty_high-fidelity_2014,wang_memory_2017}, depending on the type of qubit, the trapped ion quantum hardware is the one currently allowing to achieve the highest coherence vs. gating time ratio, ideally in the order of $10^5$ to $10^6$ (e.g., assuming two-qubit gate times in the order of 100-200 $\mu$s, and depending on the qubit type).   \\
It is worth emphasizing, however, that the considerations above are mostly limited to few qubits quantum hardware: while analog quantum simulators of Ising chains have been shown with more than 50 trapped ions \cite{zhang_observation_2017}, digital quantum simulators are quite challenging to scale up. In fact, increasing the number of ions in the chain ultimately limits gates fidelity, mainly due to the difficulties of individually addressing single qubits while avoiding cross-talks between the different beams. Moreover, errors affecting the overall success of a quantum computation generally arise from two distinct mechanisms: decoherence, e.g. from undesired qubit-environment coupling such as spontaneous emission, frequency shift, or motional heating, and imperfect control fields, such as miscalibrated or noisy control field amplitude, frequency, or polarization, which typically result in quantum gate errors. Targeted strategies to mitigate these issues will become crucial to further improve the trapped ion quantum hardware \cite{bruzewicz_review-trap-ion_2019}.
On a quantitative level, it has recently been reported that the fidelity of multi-qubit entangling M{\o}lmer-S{\o}rensen gates degrades from $99.6\%$ with 2 optical qubits down to $86\%$ when the same quantum hardware is loaded with 10 optical qubits \cite{erhard_characterizing_2019}. On the other hand, a quantum processor based on 11 hyperfine qubits has recently been shown to achieve an all-to-all connectivity with single- and two-qubit XX gate fidelities of $99.5\%$ and $97.5\%$ on average, respectively~\cite{wright_benchmarking_2019}. It is interesting to notice that these results improve previous reports on an analogous 5-qubits quantum processor \cite{linke_experimental_2017}, as a clear signature of the ongoing development.
}
%Digital quantum processors made of atomic ions in a linear Paul trap have been representing the most promising route towards realizing fully operational quantum processors since the late nineties \cite{blatt_entangled_2008,blatt_quantum_2012,monroe_scaling_2013}.  The trapping potential is created through radio frequency oscillating electric fields that generate a stable linear potential well for the charged atoms \cite{leibfried_review-trap-ion_2003}, which are then spatially separated by a few microns due to the mutual Coulomb repulsion \cite{monroe_scaling_2013}. To date, in the order of 50 and more ions can be stably aligned within the same trap in analog quantum simulators \cite{zhang_observation_2017}, and qubits can be encoded into their internal degrees of freedom. 

%Gate operations can be realized with fidelities in excess of $99.1\%$ for single-qubit rotations and $97\%$ for two-qubit gates \cite{linke_experimental_2017}, and gate operations have duration in the range between 20 $\mu$s (single-qubit rotations) and 250 $\mu$s (two-qubits gates), against spin dephasing time that can reach the second time scale, which makes these platforms extremely advanced for quantum simulation purposes. \\

\begin{figure}[t]
\centering
\includegraphics[width=\columnwidth]{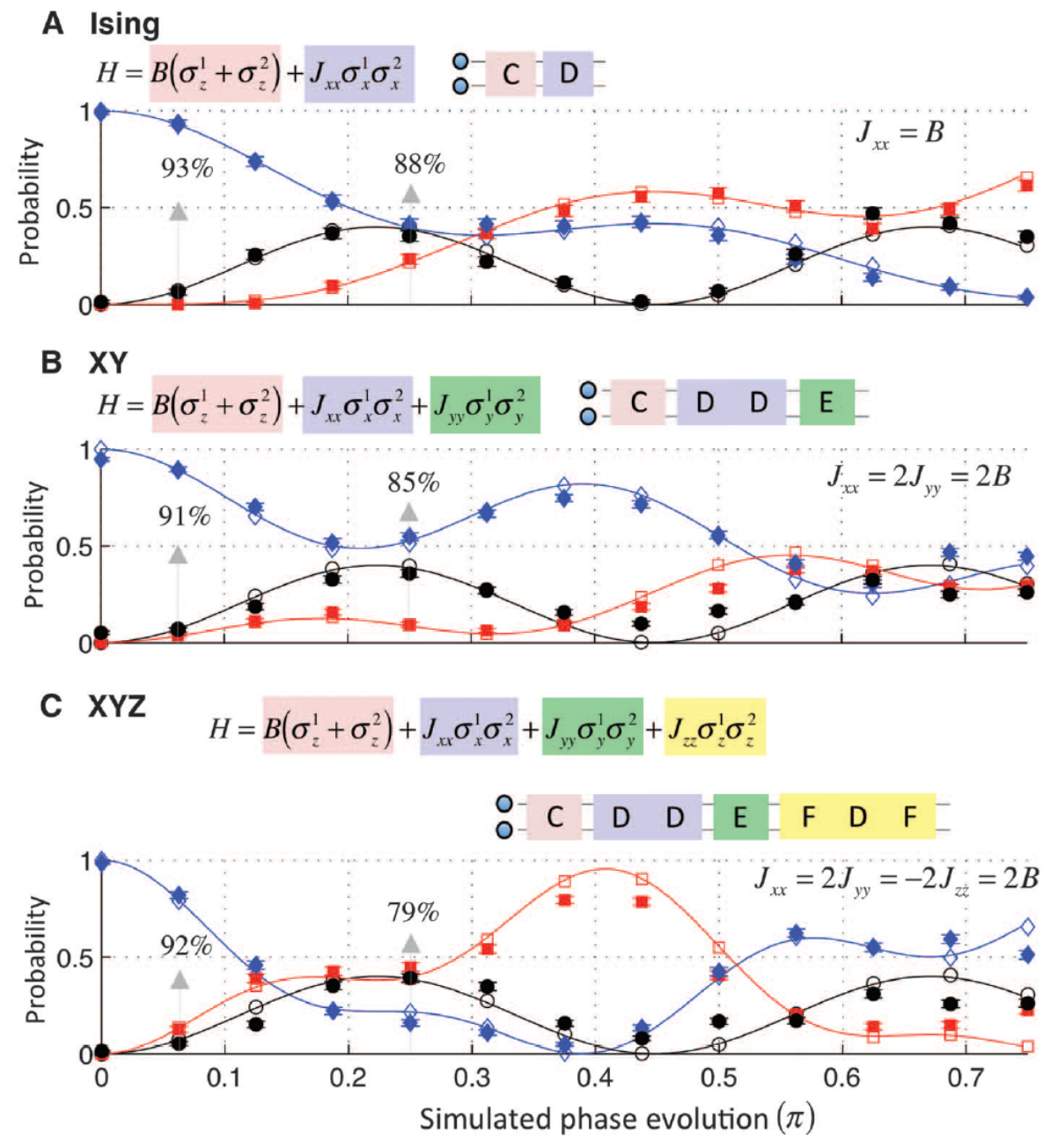}
\caption{Experimental quantum simulation of two-spin models of increasing complexity is shown in the bottom panels: the digital resolution was kept fixed as $\theta/n = \pi/16$, i.e. up to $n=12$ Trotter steps for the data shown in the figure, and each panel displays the corresponding sequence of unitary operations in each digital step; the lines correspond to the exact evolution, the empty symbols correspond to the ideal digitized evolution, and filled symbols are the quantum simulator results for the evolution of the different eigenstates (reprinted with permission from Ref.~\onlinecite{lanyon_universal_2011}). Copyright 2011, American Association for the Advancement of Science.}
\label{fig:ExpIons}
\end{figure}

\blu{After reviewing the basic aspects of trapped ion quantum hardware, and the up-to-date  figures of merit currently achieved in state-of-art experimental platforms, we hereby show a few examples of digital quantum simulation performed on one such a NISQ device. An example of results showing the first universal quantum simulations performed on NISQ processor based on optical qubits is given in Fig.~\ref{fig:ExpIons},} as reported back in 2011 in Ref.~\onlinecite{lanyon_universal_2011} from Lanyon et al. This represents a seminal work since it shows that the same quantum hardware can be experimentally reprogrammed to \blu{run the digital quantum simulation of} different spin models and interaction terms, \blu{even for Hamiltonian terms not natively implemented on the simulator}. Each spin-1/2 is directly mapped onto a single ionic qubit, and unitary operations ($C$, $D$, $E$, and $F$) are defined in terms of the universal set of gates in  Eqs.~\ref{eq:T1gate}-\ref{eq:T4gate}, which is native on this hardware (see original reference for details). The time evolution is quantified by a dimensionless phase $\theta= Et/\hbar$, which is reported \blu{on the horizontal axis}. The initial state is chosen as an eigenstate of $\sum_i \sigma_x^{(i)} $, and the population in each of the eigenstates is monitored as a function of $\theta$. The same work reported digital quantum simulation of up to 6 spins and multi-spin interaction terms, allowing to envision the potentialities of digital quantum simulators for fundamental physics studies. Hence, the work from Lanyon et al. sets a reference standard for all the following demonstrations of digital quantum simulation on NISQ processors. 
\blu{At time of writing, we are not aware of any digital quantum simulation of spin models performed on quantum processors based on hyperfine qubits such as the one presented, e.g., in Ref.~\onlinecite{wright_benchmarking_2019}, although it would be interesting to test them with some of the algorithms presented in this review, indeed. }

The quantum hardware employed to obtain the results of Fig.~\ref{fig:ExpIons} has more recently been applied to the quantum simulation of the real time dynamics underlying particle-antiparticle pair creation in lattice gauge field theories~\cite{martinez_real-time_2016}, where up to 4 qubits were employed to run the related quantum circuit. There, digital quantum simulation is obtained after mapping the fermionic degrees of freedom into Pauli spin operators applying the Jordan-Wigner transformation, as outlined before. Given the size of the quantum register, a toy model successfully simulating the electron-positron spontaneous creation from vacuum fluctuations and the persistence of their entanglement was reported, which creates a bridge between digital quantum simulators and elementary particle physics. \\
Without restricting to the main object of the present review, it is worth mentioning recent results obtained on trapped ion quantum technology. First, a 20 qubits register has been shown to reliably allow for the creation of multi-qubit entangled states \cite{friis_observation_2018}, thus opening the door to quantum simulations of larger spin systems. The same quantum hardware has been used to show a hybrid quantum-classical approach to the simulation of the Schwinger model~\cite{kokail_self-verifying_2019}. \blu{The latter belongs to the class of variational optimization algorithms commonly employed in quantum chemistry \cite{mcclean_theory_2016}, an approach now commonly defined Variational Quantum Eigensolver (VQE). In fact, the latter methods have been recently applied to accurately calculate the ground state energy of simple molecules \cite{hempel_quantum-chemistry_2018}. } 
Conceptually similar approaches have also been applied to the quantum simulation of effective field theories in nuclear physics, such as calculating the deuteron nucleus binding  energy with percent accuracy and \blu{using record depth} quantum circuit on a ion trap quantum processor~\cite{shehab_toward_2019}. \blu{A VQE algorithm has been recently described and applied to obtain the ground state energy of a large chemical complex, such as the water molecule, by using a hyperfine qubits-based quantum processor~\cite{nam_water_2019}.}  \\ 
%OTHER EXAMPLES

\blu{In summary, despite considerable progress and the maturity reached, many challenges still need to be addressed before practical quantum computers based on trapped ion technology may allow  universal quantum simulations outperforming classical computations, in perspective. It currently appears a significant challenge to operate a quantum computer with a number of fully controlled qubits larger than a few tens, still preserving the required figures of merit in terms of gates fidelity and coherence properties, which might still need a few years of intense research and development.}

\subsection{UQS with superconducting circuits }
\label{sec:SCcircuits}

Superconducting quantum circuits have lately emerged as a practical quantum computing technology after a fast development in the last decade \cite{blais_cavity_2004,wallraff_strong_2004,koch_charge-insensitive_2007,gambetta_protocols_2007,majer_coupling_2007,gambetta_quantum_2008,mariantoni_implementing_2011,rigetti_superconducting_2012,barends_superconducting_2014,nersisyan_manufacturing_2019}. In fact, this platform has reached the level of reliability typical of trapped ions quantum processors in only few years of continuous improvement \cite{linke_experimental_2017}. It is worth mentioning that some of the worldwide leading high-tech companies that are currently targeting the  realization of NISQ hardware are concentrating their efforts on this technology. \blu{As for the trapped ion case, restricted cloud access is also provided to superconducting quantum processors of up to 20 qubits from IBM (see https://www.research.ibm.com/ibm-q/, accessed 2019-09-01) and up to 16 qubits from Rigetti Computing (see https://www.rigetti.com/, accessed 2019-09-01), respectively.} These devices work with cryogenic set-up in a $^3$He/$^4$He  dilution refrigerator with 10-15 mK base temperature, in which qubits can be efficiently encoded into the anharmonic energy spectrum of the lowest collective charge/current excitations in a micro-LC resonator, with a nanostructured Josephson junction playing the role of a nonlinear inducting element. \blu{With respect to quantum hardware based on trapped atomic species, these architectures certainly bring along the advantages of a solid state chip-based microelectronic technology, although working at low temperature and with intrinsic limitations in inter-qubit connectivity. }
\blu{Several families of superconducting qubits have been realized (i.e., phase qubits, rf-sQUID, flux qubits, charge qubits, transmons), characterized by different ratios between the characteristic parameters of their quantized Hamiltonian, namely charging energy of a Cooper pair, inductive and Josephson energy. These systems are multi-level quantum oscillators whose energy spectrum is made sufficiently anharmonic to enable selective addressing of only the two lowest energy levels, which then become the $|0\rangle$ and $|1\rangle$ qubit states.
While additional levels may become a resource to implement two-qubit gates \cite{Nature_DiCarlo_2009, mariantoni_implementing_2011} or quantum-error correction codes \cite{hussain_coherent_2018}, they also represent a possible source of leakage errors that must necessarily be limited by using, e.g., long control pulses for qubit transitions.
We refer to Ref.~\onlinecite{wendin_quantum_2017} for a detailed description and a comparison between the different types of superconducting qubits. Here we only note that the evolution of these devices led to the realization of the \textit{transmon} \cite{koch_charge-insensitive_2007}, which has allowed to reach coherence times in the 100 $\mu$s range for single qubits \cite{rigetti_superconducting_2012}. This is a development of the Cooper pair box into a circuit less sensitive to charge noise but still characterized by a sufficiently anharmonic energy spectrum. The frequency of the transmon can be tuned by varying the Josephson energy using a SQUID. The 2D transmon is nowadays the elementary unit of several scalable architectures \cite{corcoles_demonstration_2015}, but other transmon-like qubits have been realized, such as the Xmon \cite{barends_superconducting_2014}, which has already shown remarkable fidelities even in a setup consisting of 9 qubits \cite{Nature_Kelly_2015,Nature_Barends_2016}. }\\
Different qubits in these platforms are typically interconnected through superconducting transmission line resonators, and they can be individually addressed through other transmission lines wired at the edges of the chip board. The latter allow to perform single-qubit initialization, manipulation, and read out through microwave pulses. Current superconducting quantum circuits allow for single-qubit gate fidelities above $99.9\%$ \cite{barends_superconducting_2014}.
\blu{Each qubit is dispersively coupled to a resonator to mediate an effective qubit-qubit interaction, employed to implement two-qubit gates. These are obtained either (i) by tuning the qubit transition frequency by a local magnetic field, or (ii) by using a cross-resonant (CR) drive, in which microwaves resonant with a target qubit are applied on another control
qubit. While the first realizations of scheme (i) were obtained by bringing the two qubits into mutual resonance to get a virtual photon exchange (thus yielding an effective XY interaction \cite{salathe_digital_2015}), the most promising implementations are currently based on tuning one qubit along a ``fast adiabatic trajectory"\cite{barends_superconducting_2014} that moves the $|11\rangle$ component of the wave-function close to the avoided level crossing with state $|02\rangle$, leading to a state-dependent phase and hence to the implementation of the controlled-phase gate \cite{Nature_DiCarlo_2009}.
This approach results in very fast (40 ns) and high-fidelity two-qubit gates, compared to relaxation and coherence times in the 20-40 $\mu$s range on average~\cite{barends_superconducting_2014}. 
Conversely, scheme (ii) uses fixed-frequency qubits in order to avoid frequency crowding and implements a CNOT gate by activating the resonator-mediated interaction via a cross-resonant drive \cite{chow_simple_2011}.
This requires more selective pulses and hence results in slower two-qubit gates, taking $\sim 200-300$ ns  \cite{chow_simple_2011,gambetta_building_2017} with only slightly longer coherence times (50 $\mu$s on average quality devices) and average fidelity of $96\%$. 
These results make platforms based on scheme (i) more promising in view of concatenating several gates to perform quantum simulation algorithms. Furthermore, the possibility to tune the phase of $C\Phi$ gates 
makes architecture (i) more flexible than (ii), and it reduces the circuit depth for the simulation of many model Hamiltonians of physical interest. In this respect, a novel proposal to directly implement exchange-type gates with tunable amplitude and phase on fixed frequency qubits (thus making also this architecture much more flexible for quantum simulations) has been recently reported \cite{PRApplied_Ganzhorn_2019}. \\
Readout still represents one of the main limitations of superconducting platforms. Indeed, accurate readout based on transmission measurements of the frequency shift of auxiliary resonators is usually slow, taking few  hundreds of ns, with average accuracy in excess of $96\%$ \cite{gambetta_building_2017}. However, high-fidelity (i.e., more than $99\%$) readout was demonstrated in 140 ns on a four-qubit setup \cite{PRL_Sank_2014}.\\
As already noted for trapped-ion architectures, we finally point out that while remarkable results have been obtained in samples specifically aimed at testing the basic operations on few qubits circuits, thus reaching fidelities no more limited by unitary errors \cite{PRA_Gambetta_2016}, the realization of a scalable platform able to implement fast and high fidelity single- and two-qubit gates, as well as efficient readout, still represents a key challenge. Indeed, recent studies have demonstrated that systematic coherent errors due to an  imperfect implementation of the elementary gates are still one of the leading error sources when several qubits are operated together in a quantum simulation \cite{chiesa_quantum_2019}. This will also become evident from the examples of digital quantum simulations reported below.}

\begin{figure}[t]
\centering
\includegraphics[width=\columnwidth]{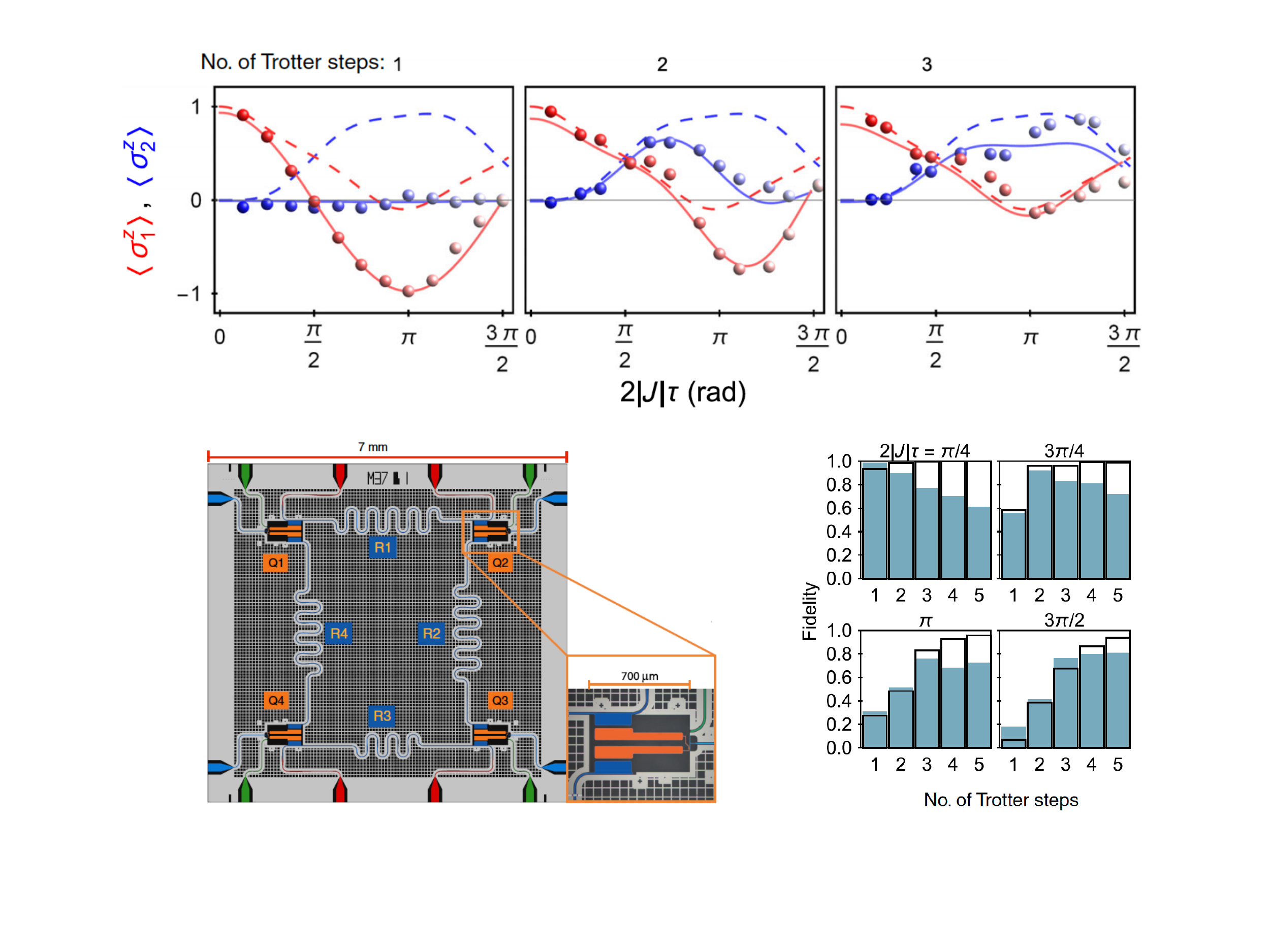}
\caption{Experimental quantum simulation of the Ising model in a transverse homogeneous field for two spins with increasing number of Trotter steps, performed on a superconducting quantum processor with 4 Niobium qubits interconnected through Aluminum transmission line resonators (shown in the picture, with false color images, input and output ports and single qubit flux bias lines are also highlighted). Dependence of final state fidelity on the number of digital steps used in the quantum simulation is also shown, for different phase angles (color bars), as compared to ideal unitary evolution for the given Trotter step (reprinted under Creative Commons Attribution 3.0 License from Ref.~\onlinecite{salathe_digital_2015}, published by the American Physical Society).}
\label{fig:ExpSCcircuits}
\end{figure}

The first digital quantum simulations of spin models on superconducting quantum hardware were experimentally reported in 2015~\cite{salathe_digital_2015}. Here, the evolution of the spin magnetization in Heisenberg and Ising models was systematically studied for 2-spin type Hamiltonians on a 4-qubits quantum processor, as a function of the number of ST steps. Superconducting processors with tunable frequency qubits (through external flux bias lines) naturally implement a XY-type interacting spin Hamiltonian, which can be used as the basis to digitally program a full Heisenberg or Ising type evolution through a circuit model, as outlined in the previous section and explained in detail in the original references~\cite{las_heras_digital_2014,salathe_digital_2015}. An example is reported in Fig.~\ref{fig:ExpSCcircuits}, where the digital evolution is explicitly shown for the two spins projections along the magnetic field direction, $z$, with up to 3 digital time steps, for an initial state prepared in $\left|\uparrow\right\rangle\left(\left|\uparrow\right\rangle-i\left|\downarrow\right\rangle\right)/\sqrt{2}$ that evolves non trivially in time. A summary of the fidelity obtained from these quantum simulations on the same quantum hardware with up to 5 Trotter steps is also reported from the original reference~\cite{salathe_digital_2015}. While the ideal fidelity of the simulated quantum state with respect to the exact evolution increases against the number of Trotter steps, the experimental one starts to decrease after about 2 or 3 digitized steps, depending on the phase. It is evident that 5 digitized steps in the simulated time evolution still presented limited fidelities, due to the short coherence times and to systematic circuit errors. Nevertheless, such results have set a milestone as a proof of concept demonstration of universal quantum simulations in superconducting quantum circuits. 

\begin{figure}[t]
\centering
\includegraphics[width=\columnwidth]{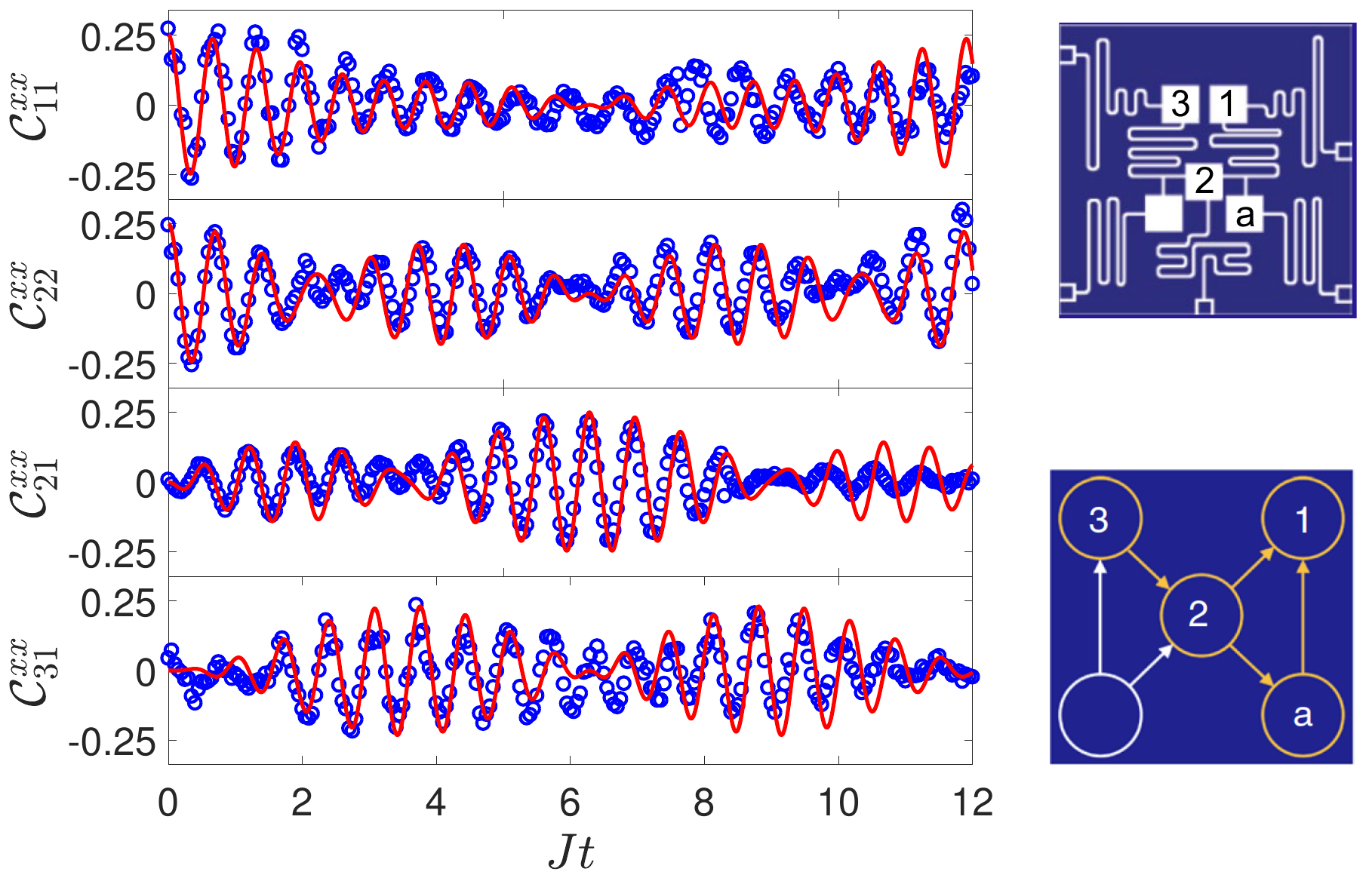}
\caption{Experimental quantum simulation of the dynamical correlations for the Heisenberg model of three spins in a chain (symbols) in external magnetic field along $z$, compared to the ideal evolution for the given number of digital time steps ($n=2$, lines), including autocorrelations as well as nearest neighbors and next-to-nearest neighbors cross correlations. Only the real part of the correlation functions are plotted; the agreement on the imaginary part is analogously good. Digital quantum simulations have been performed on IBM Quantum Experience, chips ibmqx4 (a sketch of the chip layout and inter-qubits connectivities is explicitly shown), ibmqx5, and ibmqx20, respectively, accessed online for cloud quantum computing (data replotted  from the original Ref.~\onlinecite{chiesa_quantum_2019}).} %; the corresponding quantum circuit run on the actual quantum hardware is explicitly shown
\label{fig:ExpCorr}
\end{figure}

Correlation functions represent some of the most useful and informative quantities to be calculated in quantum many body physics. A first attempt at simulating the digital time evolution of two-point correlations was already reported in~\cite{salathe_digital_2015}. More recently, dynamical correlation functions have been experimentally simulated on the quantum processors made freely available by IBM, through their IBM Quantum Experience (see quantumexperience.ng.bluemix.net/qx/experience) \cite{chiesa_quantum_2019,smith_simulating-manybody_2019}.
\blu{These processors are based on fixed frequency superconducting qubits allowing the experimenter to implement arbitary single-qubit rotations and CNOT two-qubit gates, along the lines of scheme (ii) described above (see qiskit.org). These elementary operations are then combined as in Fig. \ref{fig:circuitsHeis2} to obtain complex expectation values for the correlation functions. 
The results reported} in Ref. \onlinecite{chiesa_quantum_2019} are encouraging in view of scalability: dynamical correlations were digitally simulated for various basic spin models, ranging from Ising to isotropic and anisotropic exchange Hamiltonians, both for spin dimers and trimers~\cite{chiesa_quantum_2019}. The largest number of ST steps that could be reliably simulated on the IBM quantum processor was $n=4$ for spin dimers, and $n=2$ for trimers (due to the larger depth of the corresponding quantum circuit). An example of the digital quantum simulation of  time-dependent two-body correlation functions, as defined in the previous section, is reported in Fig.~\ref{fig:ExpCorr} for a three spin-1/2 Heisenberg model in external magnetic field along $z$ for an initial state $\left|\downarrow\downarrow\downarrow\right\rangle$, as compared to the ideal digitized evolution with $n=2$ ST steps, showing truly remarkable agreement. A sketch of the available 5-qubit quantum hardware is also shown, with an outline of chip connectivity. In Ref.~\onlinecite{chiesa_quantum_2019}, the largest quantum simulation reported on the actual IBM quantum hardware actually employed 5 qubits (4 encoding the target system, plus 1 ancilla for correlations readout), showing good agreement with the expected behavior despite the noisy nature of the quantum processors. 
\blu{It is important to note that the accuracy of these results was largely improved by systematic post-processing  corrections based on general properties of the extracted correlation functions, a problem-specific procedure which could be applied to many other quantum simulations, after analysis of the symmetries of the target Hamiltonian \cite{mcardle_error_2019}. }
Fitting of such digitally simulated correlations allows one to extract four-dimensional inelastic neutron scattering spectra, a crucial experimental tool to characterize magnetic molecules \cite{baker_spin_2012,garlatti_portraying_2017}. 
The speedup of a quantum processor in simulating the dynamical correlations needed to compute the inelastic neutron cross-section could allow for an efficient and real time interpretation of experiments on complex molecules, a task which is nowadays infeasible with classical computer simulations.
\blu{This study also reports a systematic analysis of the errors propagating on the quantum hardware, highlighting that one of the main limitations to circuit depth is currently represented by systematic off-resonant driving errors \cite{mckay_efficient_2017}. This could be a direct consequence of the many transitions that must be addressed on a chip consisting of several qubits with only slightly different transition frequency (the so-called {\it frequency crowding}). In fact, recent quantum simulations of similar models performed on a 20-qubit quantum hardware show that a proper accounting of such errors is essential to obtain a reasonable agreement with expected results \cite{smith_simulating-manybody_2019}.}  \\
\blu{Going beyond spin models, the superconducting quantum hardware has been tested as a UQS of the Fermi-Hubbard model~\cite{las_heras_fermionic_2015}, which has been experimentally achieved with up to 4 modes~\cite{barends_digital_2015}.} \\
\blu{A hybrid approach combining adiabatic to digital quantum computation was reported in Ref.~\cite{Nature_Barends_2016}, where a superconducting circuit made of frequency-tunable qubits and implementing two-qubit controlled-phase gates according to scheme (i) was used to probe the dynamics of a chain of 9 interacting spins, starting in a factorized state and evolving along an adiabatic trajectory by switching on an anisotropic exchange interaction term. 
Recent results in hybrid quantum-classical approaches, such as the VQE method already recalled at the end of the previous Section, have been efficiently employed to show quantum chemistry simulations on superconducting NISQ processors~\cite{omalley_scalable_2016,moll_optimizing_2016,kandala_hardware-efficient_2017}, in which the ground state energy of multi-atomic molecules was calculated with precision approaching the chemical accuracy limits \cite{PRApplied_Ganzhorn_2019}.} Using an evolution of the VQE algorithm, nuclear physics quantum simulations have also been reported in superconducting quantum hardware, with the cloud computing of the deuteron binding energy~\cite{dumitrescu_cloud_2018}. Along the same lines of trapped ions quantum simulators~\cite{martinez_real-time_2016,kokail_self-verifying_2019}, a similar quantum classical algorithm has been used to solve for the Schwinger model dynamics on a superconducting quantum hardware~\cite{klco_quantum-classical_2018}. 

Interesting comparisons between trapped ions and superconducting quantum processors in applying such hybrid quantum-classical approaches have been reported, in which the same algorithm was simulated on different platforms, showing a substantial equivalence of the two leading architectures when the same number of qubits could be used, but trapped ions processors allowing for a larger system size to be simulated~\cite{shehab_toward_2019}. 
\blu{Compared to trapped-ion technologies, superconducting circuits show much larger gates speed, even if the ratio of coherence time to gate operation remains smaller. This can be a remarkable advantage when comparing the performance of quantum and classical devices in terms of absolute execution time of a given algorithm.
While progress in the last few years has led to a substantial increase of the coherence times in transmon qubits, the leading errors in multi-qubit architectures are still coherent~\cite{chiesa_quantum_2019}. Indeed, the need to address several qubits with only slightly different transition frequencies, as well as the small anharmonicities required to keep long coherence times, require long and often not completely resonant driving pulses \cite{PRA_Gambetta_2016}. In addition, inter-qubit interactions are never completely switched off, even when qubits are in the idle phase. This leads to unwanted evolution of the multi-qubit wave-function ({\it cross-talk}), an effect increasing with the system size and strongly depending also on the chip connectivity. 
For quantum simulation purposes, the latter should be as close as possible to that of the target system, in order to reduce the circuit depth and avoid cumbersome SWAP gates. In fact, trapped ions are more promising for establishing entanglement between distant pairs of qubits, and hence also to simulate fermionic Hamiltonians involving multi-qubit interaction terms, although the first attempts to go  beyond nearest-neighbors coupling have been reported on three-qubit quantum processors \cite{roy_programmable_2018}.
Finally, compared to trapped ion-based technologies in which all qubits are identical, superconducting qubits are all characterized by different parameters (such as transition frequencies and mutual couplings), which are also affected by thermal cycling and hence require a detailed and frequent characterization to accurately calibrate the control pulses~\cite{klimov_thermal-cycling_2018}.}\\
As mentioned above, considerable work is currently focusing on understanding the main sources of error, and on developing error mitigation techniques to enhance the overall quantum simulation fidelities~\cite{temme_error_2017,mcardle_error_2019,chiesa_quantum_2019}. \blu{These theoretical proposals have been recently and successfully applied to improve the accuracy of the observables extracted from VQE calculations on a superconducting chip \cite{kandala_error_2019}. Although demonstrated on elementary single- and two-qubit gates in a VQE experiment, this protocol can be applied to mitigate errors of any quantum algorithm since it is problem-independent and does not lead to any hardware overhead. However, its application requires the experimenter to control the amount of noise on the hardware (which in Ref. \cite{kandala_error_2019} is mapped on the evolution under a scaled drive) and it is limited to incoherent errors.}\\
In the ongoing effort to develop a fault tolerant quantum computing architecture, these results pave the way towards reaching the quantum advantage, possibly already within the NISQ time frame.
\blu{As a summary of all the previous results, IBM tested the concept of quantum volume on three of its devices, finding that it doubled each year (from 2017 to 2019), from 4 to 8 to 16, a trend similar to Moore's Law for classical computers and  promising for future perspectives. Nevertheless, similar conclusions apply here as already given for the trapped ion quantum hardware: despite considerable progress and the first attempts at performing full digital quantum simulations, several challenges need to be addressed before quantitative and not only qualitative accuracy be reached, especially on larger system sizes and running deeper quantum circuits. Specifically, a deeper understanding of error and noise sources, as well as suitable strategies to mitigate them on hardware with larger number of qubits is crucial at this NISQ stage \cite{smith_simulating-manybody_2019}, which will be the focus of intense research and development in the coming years. }

%\vspace{-0.8cm}

\subsection{Prospective technologies for UQS}
\label{sec:Prospect}

While the two leading technologies outlined above are currently the mainstream in practical quantum computing, it is still unclear if they will be able to overcome the challenges to reach a large number ($N>100$) of logical qubits and a significantly larger amount of error correcting ones. \blu{Alternative technologies might also start playing a role in the meantime. In this respect, it is interesting to follow recent progress in semiconductor-based technologies.} After the widespread success of semiconductors in microelectronics applications, they have been a little behind the scene in the quest for practical quantum computing devices despite the early proposals~\cite{loss_quantum_1998}. In particular, semiconductor based quantum dots have long been considered as potential spin qubits. Single-spin read-out and manipulation has been shown quite early~\cite{elzerman_single-shot_2004,petta_coherent_2005}, but scalability has been hindered so far, mostly due to coherence times being limited by nuclear spin dephasing and spin-orbit coupling~\cite{hanson_spins_2007}. However, recent advances in silicon-based quantum dots have renewed interest in the actual possibilities of these technologies: CNOT gates between two quantum dots have been shown with about $78\%$ fidelity~\cite{zajac_resonantly_2018}, and two-qubit gates with fidelities in the order of $80\%$ have also been reported, with single qubit rotation precision of $\sim 99\%$~\cite{veldhorst_two-qubit_2015,watson_programmable_2018,noiri_fast_2018,huang_fidelity_2019}. Gating times are below 100 ns, and dephasing times about 200 ns~\cite{watson_programmable_2018}. These results, \blu{although still far from the required performances achieved, e.g., in the trapped ion or superconducting circuits architectures}, are \blu{quite} promising and set a stepping stone on the development of a fully semiconductor-based quantum technology, evidently interesting for a number of reasons (from cheap costs to the potential for mass scale manufacturing). \\ 
In parallel to research on quantum dots, controlled impurities and defect ions in silicon have been lately considered for a potentially low-cost quantum technologies~\cite{awschalom_quantum_2013}. After the demonstration of single-spin read-out and manipulation of localized donor impurities in silicon~\cite{morello_single-shot_2010,pla_single-atom_2012}, two-qubit quantum gates have been proposed for this prospective platform~\cite{kalra_robust_2014,tosi_silicon_2017}. \blu{Very recently, the first $\sqrt{\mathrm{SWAP}}$ operation between Phosphorus donor electrons in silicon, with gating time of $800\,$ps and fidelity around $94$\% has been reported~\cite{he_two-qubits-Phosphorus_2019}.} These results hold promise that further developments might be seen in the near future, if challenges related to scalability will be overcome. \\
While electronic states in engineered potentials or in impurity states are naturally emerging as reasonable candidates for a qubit-based architecture, it is less obvious that photons, in particular photonic integrated circuits, could play a significant role as UQS. Photonic circuits have been largely explored as analog quantum simulators~\cite{aspuru-guzik_photonic_2012}. The main limitation to exploit photonic states for quantum computing lies in their weak interactions, due to intrinsically small material nonlinearities. While it may be argued that suitable electromagnetic confinement in nonlinear materials might lead to single-photon sensitivity \cite{ferretti_single-photon_2012,flayac_all-silicon_2015}, no such effect has been measured at time of writing. Mixed radiation-matter excitations in semiconductors, also called exciton-polaritons, have been shown to be sensitive at the single quantum level \cite{munoz-matutano_emergence_2019,delteil_towards_2019}, which might play a role in realizing analog simulators of strongly interacting photonic lattices \cite{gerace_quantum-optical_2009,carusotto_fermionized_2009}, but their effective use as qubits is still immature. On the other hand, a few companies are investing in a photonic-based quantum computer, which could then be used as a UQS employing continuous variable cluster states \cite{menicucci_universal_2006}, but we are not aware of any proof-of-principle demonstration at the moment. \\
Magnetic molecules manipulated through electromagnetic pulses have also been proposed as a potential platform for quantum information processing~\cite{santini_molecular_2011}, thanks to their long coherence times and high degree of chemical tunability. This allows to engineer suitable structures of elecronic \cite{ferrando-soria_modular_2016} or nuclear \cite{atzori_two-qubit_2018} spin qubits in which the qubit-qubit interaction is effectively switched on and off by electromagnetic pulses. Furthermore, the richness of the molecular Hilbert space can be exploited to directly encode logical qubits with embedded quantum error correction in single molecules \cite{hussain_coherent_2018}. \\
\blu{Finally, also arrays of optically or magnetically trapped neutral atoms, typically excited in Rydberg states, have been proposed as a platform for quantum simulations \cite{weimer_rydberg_2010}, although several challenges need to be addressed for quantum computing applications, particularly related to the short coherence times as compared to trapped ions \cite{saffman_rydberg_2016}. While analog simulators with more than 50 Rydberg atoms have been reported \cite{bernien_probing_2017}, and  two-qubit entanglement has been recently shown with $\sim 97\%$ fidelity \cite{levine_rydberg-qubits_2018}, it is still premature to expect digital quantum simulators based on fully controlled Rydberg-atomic qubits. }

%existing technologies, under development
%?FIGURE: CNOT in SC quantum dots and photonic circuits?
\begin{figure}[t]
\centering
\includegraphics[width=\columnwidth]{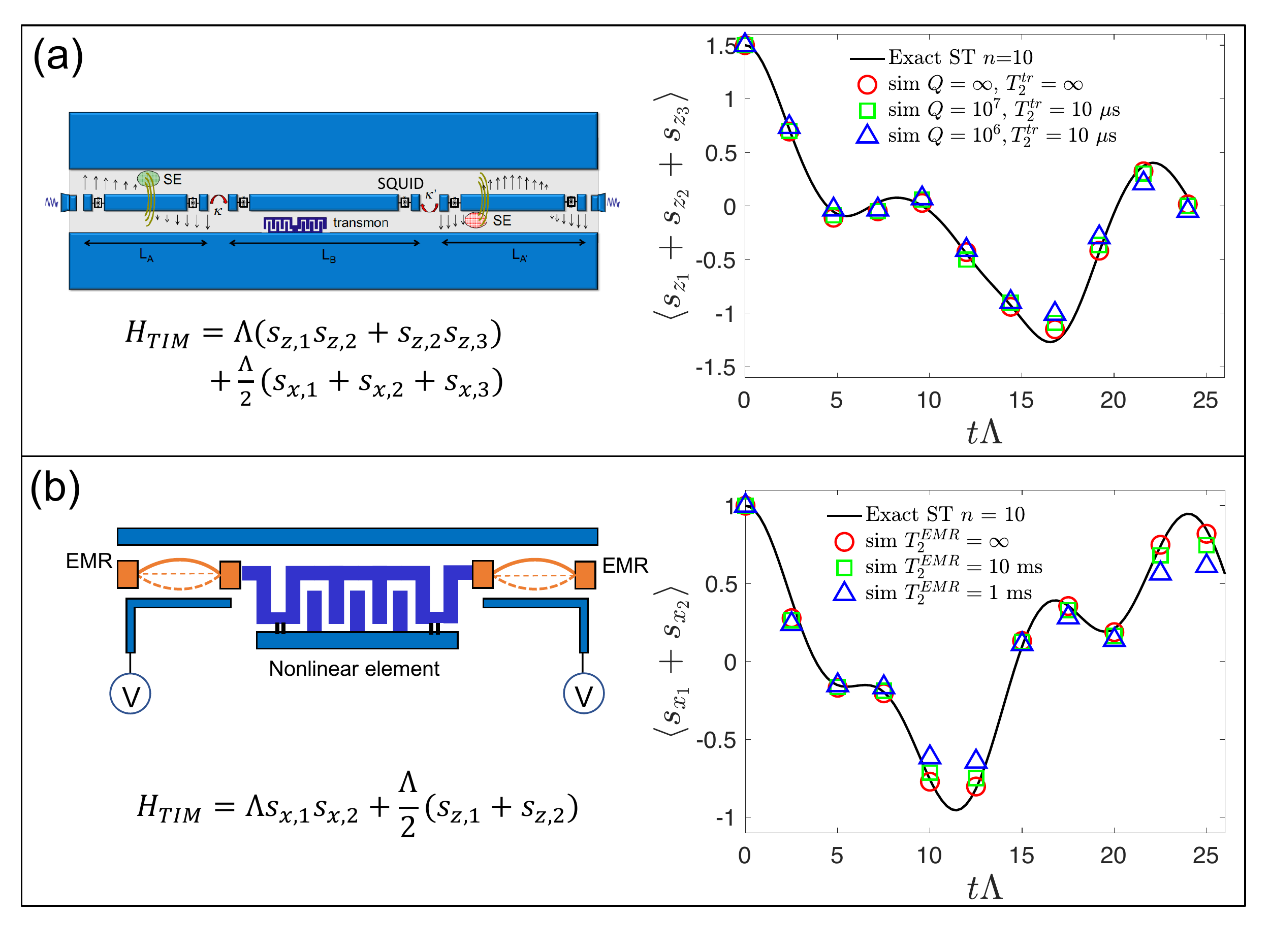}
\caption{Prospective hybrid platforms for UQS: (a) hybrid spin-photon qubits encoded in a superconducting resonators array with spin ensembles in each resonator, inter-connected through transmon qubits playing the role of nonlinear elements employed for two-qubits gating, and resulting theoretical quantum simulation of a 3-spin \blu{transverse field Ising model} for different values of photon and spin dissipation rates (see original Ref. \onlinecite{chiesa_digital_2015}); (b) electromechanical nanoresonators (EMR) mutually coupled through a superconducting nonlinear element represent the building block of a scalable UQS architecture, in which qubits are encoded in mechanical degrees of freedom, and the corresponding test of a digital quantum simulation of a 2-spin \blu{transverse field Ising model} taking all the sources of error and dissipation into account (see original Ref. \onlinecite{tacchino_electromechanical_2018}).}
\label{fig:Prospective}
\end{figure}

Together with existing technologies that are moving their first steps into actual quantum computing applications, it is worth concluding this brief overview by mentioning a few potentially promising hybrid technologies, which are usually aimed at merging the best characteristics of two or more existing approaches \cite{xiang_hybrid_2013,kurizki_quantum_2015}. The philosophy behind these proposals is simple: it is quite likely that a hybrid technology will be in the best position to simultaneously meet all the requirements in terms of scalability (possibly in multi-dimensional arrays), chip-scale integration, and high operational reliability (i.e., long qubits coherence and short gating times). A number of proposals for prospective quantum technologies have been reported, which we will hereby summarize briefly and refer to the original references for further details.\\ 
For instance, spin ensembles coherently coupled to superconducting microwave resonators have been proposed as a backbone of a novel hybrid quantum technology~\cite{ping_measurement_2012,carretta_quantum_2013,chiesa_robustness_2014}. This hybrid architecture would exploit the long coherence times of spin ensembles and the easy manipulation of photons in tunable resonators. A full digital quantum computing architecture has been devised \cite{chiesa_digital_2015,chiesa_long-lasting_2016}, for which we report an example in Fig.~\ref{fig:Prospective}(a), where the \blu{transverse field Ising model} of 3 spins (see Hamiltonian model in the inset) is theoretically shown to be simulated with a  large overall fidelity ($\sim 95\%$ on average) when realistic dissipation parameters are assumed~\cite{chiesa_digital_2015}. This kind of architecture could be even built with single magnetic molecules strongly coupled to the quantized resonator field \cite{jenkins_coupling_2013,jenkins_scalable_2016}. \\
Along similar lines, hybrid architectures based on Nitrogen Vacancy (NV) centers coupled to Carbon nanotubes have also been proposed \cite{li_hybrid_2016}. Mechanical degrees of freedom have also been considered to be part of hybrid platforms, due to their intrinsically low dephasing rates. In particular, quantum information processing has been theoretically shown in optical devices in which qubits are encoded in the lowest lying mechanical levels~\cite{rips_quantum_2013}, as well as in NV centers coupled to mechanical resonators and superconducting waveguides \cite{li_hybrid_2015}.
Recently, mechanical qubits encoding has been considered in a hybrid set up coupling vibrating nanoresonators to superconducting circuits~\cite{tacchino_electromechanical_2018}. An example of such an electromechanical quantum computing architecture is reported in  Fig.~\ref{fig:Prospective}(b), with theoretical simulations of the digitized evolution of the \blu{transverse field Ising model} of 2 spins (see Hamiltonian model in the inset) performed on such hypothetical platform, with very interesting fidelities in the order of $99.9\%$ for the overall quantum simulation if the EMR dephasing is neglected, which reduces to about  $99\%$ for realistic pure dephasing rates. These numbers are extremely promising, and could motivate further experimental efforts towards realization of the required building blocks.

\section{Outlook and perspectives}

We have given a brief summary of the current status on quantum simulators, restricting our overview to the use of quantum computers as general purpose machines that can be programmed to solve for the exact time evolution of an arbitrary Hamiltonian model. The only prescription for such a universal quantum simulator is that the physical model under analysis be mapped onto an effective \blu{local} Hamiltonian obeying the algebra of Pauli matrices, which is then encoded directly in a qubit-based quantum computer through a quantum circuit model. A number of observables, such as spectra and correlation functions, can be  accessed upon measurement in the computational basis. After giving a pedagogic introduction to the theoretical background allowing to translate the digitized unitary evolution in discrete time steps into the corresponding quantum algorithm made of a sequence of one- and two-qubits gates, we have reviewed recent experimental results on the two leading quantum technology platforms. Finally, we have outlined a few existing technologies that might develop into quantum computing hardware, and hence be useful for quantum simulation, as well as prospective hybrid approaches that might eventually be tested. 

Since most Hamiltonian models of physical interest can be expressed in terms of locally interacting spin terms, we have focused this review on the most widespread spin-type models, such as the Heisenberg and Ising models in an external magnetic field. Getting acquainted and applying the basic techniques \blu{developed} for such models allows to quickly grasp the quantum simulation of more general quantum many body systems that are typically intractable with classical simulations, mostly due to exponential scaling of the required resources with the system size, such as the Fermi-Hubbard model. The road to quantum advantage is an exciting targeted goal to be fulfilled in the coming years, following the availability of NISQ quantum hardware with few tens of non-error corrected qubits. While the advent of quantum error correction will most probably have a transformative impact allowing to realize universal quantum simulations with arbitrary digital precision, \blu{in the intervening years} error mitigation techniques and further technological improvement will bring interesting results also from current experiments, which are limited by noisy gates and qubits coherence times. In this respect, increasing gate fidelities and reducing gate duration are some of the technological challenges to be faced in the near term. In fact, these developments should all lead to an increase in ``quantum volume'' \cite{moll_quantum_2018,cross_validating_2018,wei_verifying_2019}, i.e.\ a larger number of actual qubits usefully participating in a given quantum computation, which is currently limited to less than 10 in essentially all of the available platforms. When such a number will actually be on the order of 30 to 40, quantum advantage will finally be within reach of such NISQ digital quantum simulators, at least for some targeted applications or models.

In the meantime, a great deal of work is ongoing to devise new potential use cases and algorithms to be run on these machines, for which learning techniques of quantum circuit programming might turn being useful. A brief, non-exhaustive list, with a bit of personal taste, is given in the following. Restricting to problems of academic interest, the dynamical localization of quantum Hamiltonians that have a classical chaotic behavior~\cite{benenti_dynamical_2003,montangero_dynamically_2004}, requiring simulation of the quantum Fourier transform, could be run on a universal quantum computer. More recently, universal quantum computers have been receiving attention from machine learning applications, in particular to develop quantum neural networks, with the aim of processing an exponentially large amount of data with polynomial resources \cite{biamonte_quantum_2017}. The first attempts in this direction have been reported \cite{havlicek_supervised_2019,schuld_quantum_2019,tacchino_artificial_2019}. Universal quantum simulators might also help solving problems in open quantum system dynamics, for which novel numerical approaches \blu{have already been developed} \cite{finazzi_corner-space_2015,mascarenhas_matrix-product-operator_2015}. Simulating the digitized non-unitary evolution of an open quantum system on a quantum computer is a topic of current interest \cite{lloyd_engineering_2001,wang_quantum_2011,muller_simulating_2011,schindler_quantum_2013-1,di_candia_quantum_2015,cleve_efficient_2017,lamm_simulation_2018}. Finally, the huge body of knowledge accumulated in the past half a century to classically simulate the many body dynamics of quantum systems of increasing complexity, such as quantum Monte-Carlo, molecular dynamics, and density matrix renormalization group could be integrated into quantum algorithms to be run on digital quantum computers, with far-reaching and still unknown consequences.

\section{Acknowledgments}
This review is based on a series of lectures on Universal Quantum Simulators DG has given at the Ph.D. school in Physics, University of Pavia, in January 2019. 
A number of collaborators and friends are acknowledged for useful discussions and inputs over the years: G. Amoretti, L. C. Andreani, A. Auff\`{e}ves, D. Bajoni, P. Barkoutsos, G. Benenti, S.-Y. Chang, M. Grossi, M. D. LaHaye, C. Macchiavello, S. Montangero, P. Santini, M. F. Santos, V. Savona, I. Tavernelli, F. Troiani, H. E. Tureci. 
This work was partly funded by the Italian Ministry of Education and Research (MIUR) through PRIN Project 2015 HYFSRT ``Quantum Coherence in Nanostructures of Molecular Spin Qubits,'' PRIN Project 2017 INPhoPOL ``Interacting Photons in Polariton Circuits'', and by EU H2020 funded QuantERA ERA-NET Cofund in Quantum Technologies through projects CUSPIDOR and SUMO, both cofunded by Italian MIUR.

\bibliography{biblioReview}{}

\begin{thebibliography}{100}

\bibitem{gubernatis_quantum_2016}
J.~Gubernatis, N.~Kawashima, P.~Werner, {\it Quantum Monte Carlo Methods\/},
  Cambridge University Press, Cambridge, UK {\bf 2016}.

\bibitem{haile_molecular_1992}
J.~M. Haile, {\it Molecular Dynamics Simulation: Elementary Methods\/},
  Wiley-Interscience, New York, USA {\bf 1992}.

\bibitem{montangero_introduction_2018}
S.~Montangero, {\it Introduction to Tensor Network Methods\/}, Springer Nature
  Switzerland AG, Cham, CH {\bf 2018}.

\bibitem{manin_computable_1980}
Y.~I. Manin, {\it Sov. Radio\/} {\bf 1980}, p.~13.

\bibitem{benioff_computer_1980}
P.~Benioff, {\it J. Stat. Phys.\/} {\bf 1980}, {\it 22\/}, 563.

\bibitem{feynman_simulating_1982}
R.~P. Feynman, {\it Int. J. Theor. Phys.\/} {\bf 1982}, {\it 21\/}, 467.

\bibitem{fisher_boson_1989}
M.~Fisher, P.~B. Weichman, G.~Grinstein, D.~S. Fisher, {\it Phys. Rev. B\/}
  {\bf 1989}, {\it 40\/}, 546.

\bibitem{jaksch_cold_1998}
D.~Jaksch, C.~Bruder, J.~I. Cirac, C.~W. Gardiner, P.~Zoller, {\it Phys. Rev.
  Lett.\/} {\bf 1998}, {\it 81\/}, 3108.

\bibitem{hartmann_strongly_2006}
M.~J. Hartmann, F.~G. S.~L. Brandao, M.~B. Plenio, {\it Nat. Phys.\/} {\bf
  2006}, {\it 2\/}, 848.

\bibitem{greentree_quantum_2006}
A.~D. Greentree, C.~Tahan, J.~H. Cole, L.~C.~L. Hollenberg, {\it Nat. Phys.\/}
  {\bf 2006}, {\it 2\/}, 856.

\bibitem{angelakis_photon-blockade-induced_2007}
D.~G. Angelakis, M.~F. Santos, S.~Bose, {\it Phys. Rev. A\/} {\bf 2007}, {\it
  76\/}, 031805.

\bibitem{carusotto_numerical_2008}
I.~Carusotto, S.~Fagnocchi, A.~Recati, R.~Balbinot, A.~Fabbri, {\it New J.
  Phys.\/} {\bf 2008}, {\it 10\/}, 103001.

\bibitem{carusotto_fermionized_2009}
I.~Carusotto, D.~Gerace, H.~E. Tureci, S.~De~Liberato, C.~Ciuti, A.~Imamo{\v
  g}lu, {\it Phys. Rev. Lett.\/} {\bf 2009}, {\it 103\/}, 033601.

\bibitem{tian_circuit_2010}
L.~Tian, {\it Phys. Rev. Lett.\/} {\bf 2010}, {\it 105\/}, 167001.

\bibitem{gerace_analog_2012}
D.~Gerace, I.~Carusotto, {\it Phys. Rev. B\/} {\bf 2012}, {\it 86\/}, 144505.

\bibitem{viehmann_observing_2013}
O.~Viehmann, J.~von Delft, F.~Marquardt, {\it Phys. Rev. Lett.\/} {\bf 2013},
  {\it 110\/}, 030601.

\bibitem{du_superconducting_2015}
L.-H. Du, J.~Q. You, L.~Tian, {\it Phys. Rev. A\/} {\bf 2015}, {\it 92\/},
  012330.

\bibitem{reiner_emulating_2016}
J.-M. Reiner, M.~Marthaler, J.~Braum{\"u}ller, M.~Weides, G.~Sch{\"o}n, {\it
  Phys. Rev. A\/} {\bf 2016}, {\it 94\/}, 032338.

\bibitem{greiner_quantum_2002}
M.~Greiner, O.~Mandel, T.~Esslinger, T.~W. H{\"a}nsch, I.~Bloch, {\it Nature\/}
  {\bf 2002}, {\it 415\/}, 39.

\bibitem{friedenauer_simulating_2008}
A.~Friedenauer, H.~Schmitz, J.~T. Glueckert, D.~Porras, T.~Schaetz, {\it Nat.
  Phys.\/} {\bf 2008}, {\it 4\/}, 757.

\bibitem{gerritsma_quantum_2010}
R.~Gerritsma, G.~Kirchmair, F.~Z{\"a}hringer, E.~Solano, R.~Blatt, C.~F. Roos,
  {\it Nature\/} {\bf 2010}, {\it 463\/}, 68.

\bibitem{kim_quantum_2010}
K.~Kim, M.-S. Chang, S.~Korenblit, R.~Islam, E.~E. Edwards, J.~K. Freericks,
  G.-D. Lin, L.-M. Duan, C.~Monroe, {\it Nature\/} {\bf 2010}, {\it 465\/},
  590.

\bibitem{islam_onset_2011}
R.~Islam, E.~E. Edwards, K.~Kim, S.~Korenblit, C.~Noh, H.~Carmichael, G.-D.
  Lin, L.-M. Duan, C.-C. Joseph~Wang, J.~K. Freericks, C.~Monroe, {\it Nat.
  Commun.\/} {\bf 2011}, {\it 2\/}, 377.

\bibitem{nguyen_acoustic_2015}
H.~S. Nguyen, D.~Gerace, I.~Carusotto, D.~Sanvitto, E.~Galopin,
  A.~Lema{\^i}tre, I.~Sagnes, J.~Bloch, A.~Amo, {\it Phys. Rev. Lett.\/} {\bf
  2015}, {\it 114\/}, 036402.

\bibitem{steinhauer_observation_2016}
J.~Steinhauer, {\it Nat. Phys.\/} {\bf 2016}, {\it 12\/}, 959.

\bibitem{labuhn_tunable_2016}
H.~Labuhn, D.~Barredo, S.~Ravets, S.~de~L{\'e}s{\'e}leuc, T.~Macr{\`i},
  T.~Lahaye, A.~Browaeys, {\it Nature\/} {\bf 2016}, {\it 534\/}, 667.

\bibitem{bernien_probing_2017}
H.~Bernien, S.~Schwartz, A.~Keesling, H.~Levine, A.~Omran, H.~Pichler, S.~Choi,
  A.~S. Zibrov, M.~Endres, M.~Greiner, V.~Vuleti{\'c}, M.~D. Lukin, {\it
  Nature\/} {\bf 2017}, {\it 551\/}, 579.

\bibitem{roushan_spectroscopic_2017}
P.~Roushan, C.~Neill, J.~Tangpanitanon, V.~M. Bastidas, A.~Megrant, R.~Barends,
  Y.~Chen, Z.~Chen, B.~Chiaro, A.~Dunsworth, A.~Fowler, B.~Foxen, M.~Giustina,
  E.~Jeffrey, J.~Kelly, E.~Lucero, J.~Mutus, M.~Neeley, C.~Quintana, D.~Sank,
  A.~Vainsencher, J.~Wenner, T.~White, H.~Neven, D.~G. Angelakis, J.~Martinis,
  {\it Science\/} {\bf 2017}, {\it 358\/}, 1175.

\bibitem{zhang_observation_2017}
J.~Zhang, G.~Pagano, P.~W. Hess, A.~Kyprianidis, P.~Becker, H.~Kaplan, A.~V.
  Gorshkov, Z.-X. Gong, C.~Monroe, {\it Nature\/} {\bf 2017}, {\it 551\/}, 601.

\bibitem{zhang_experimental_2018}
X.~Zhang, K.~Zhang, Y.~Shen, S.~Zhang, J.-N. Zhang, M.-H. Yung, J.~Casanova,
  J.~S. Pedernales, L.~Lamata, E.~Solano, K.~Kim, {\it Nat. Commun.\/} {\bf
  2018}, {\it 9\/}, 195.

\bibitem{lloyd_universal_1996}
S.~Lloyd, {\it Science\/} {\bf 1996}, {\it 273\/}, 1073.

\bibitem{ortiz_quantum_2001}
G.~Ortiz, J.~E. Gubernatis, E.~Knill, R.~Laflamme, {\it Phys. Rev. A\/} {\bf
  2001}, {\it 64\/}, 022319.

\bibitem{somma_simulating_2002}
R.~Somma, G.~Ortiz, J.~E. Gubernatis, E.~Knill, R.~Laflamme, {\it Phys. Rev.
  A\/} {\bf 2002}, {\it 65\/}, 042323.

\bibitem{verstraete_quantum_2009}
F.~Verstraete, J.~I. Cirac, J.~I. Latorre, {\it Phys. Rev. A\/} {\bf 2009},
  {\it 79\/}, 032316.

\bibitem{weimer_rydberg_2010}
H.~Weimer, M.~M{\"u}ller, I.~Lesanovsky, P.~Zoller, H.~P. B{\"u}chler, {\it
  Nat. Phys.\/} {\bf 2010}, {\it 6\/}, 382.

\bibitem{santini_molecular_2011}
P.~Santini, S.~Carretta, F.~Troiani, G.~Amoretti, {\it Phys. Rev. Lett.\/} {\bf
  2011}, {\it 107\/}, 230502.

\bibitem{casanova_quantum_2012}
J.~Casanova, A.~Mezzacapo, L.~Lamata, E.~Solano, {\it Phys. Rev. Lett.\/} {\bf
  2012}, {\it 108\/}, 190502.

\bibitem{mezzacapo_digital_2012}
A.~Mezzacapo, J.~Casanova, L.~Lamata, E.~Solano, {\it Phys. Rev. Lett.\/} {\bf
  2012}, {\it 109\/}, 200501.

\bibitem{jordan_quantum_2012}
S.~P. Jordan, K.~S.~M. Lee, J.~Preskill, {\it Science\/} {\bf 2012}, {\it
  336\/}, 1130.

\bibitem{raeisi_quantum-circuit_2012}
S.~Raeisi, N.~Wiebe, B.~C. Sanders, {\it New J. of Phys.\/} {\bf 2012}, {\it
  14\/}, 103017.

\bibitem{hauke_quantum_2013}
P.~Hauke, D.~Marcos, M.~Dalmonte, P.~Zoller, {\it Phys. Rev. X\/} {\bf 2013},
  {\it 3\/}, 041018.

\bibitem{las_heras_digital_2014}
U.~Las~Heras, A.~Mezzacapo, L.~Lamata, S.~Filipp, A.~Wallraff, E.~Solano, {\it
  Phys. Rev. Lett.\/} {\bf 2014}, {\it 112\/}, 200501.

\bibitem{chiesa_digital_2015}
A.~Chiesa, P.~Santini, D.~Gerace, J.~Raftery, A.~A. Houck, S.~Carretta, {\it
  Sci. Rep.\/} {\bf 2015}, {\it 5\/}, 16036.

\bibitem{garcia-alvarez_digital_2017}
L.~Garc{\'i}a-{\'A}lvarez, I.~L. Egusquiza, L.~Lamata, A.~del Campo, J.~Sonner,
  E.~Solano, {\it Phys. Rev. Lett.\/} {\bf 2017}, {\it 119\/}, 040501.

\bibitem{jiang_quantum_2018}
Z.~Jiang, K.~Sung, K.~Kechedzhi, V.~N. Smelyanskiy, S.~Boixo, {\it Phys. Rev.
  Appl.\/} {\bf 2018}, {\it 9\/}, 044036.

\bibitem{kivlichan_quantum_2018}
I.~D. Kivlichan, J.~McClean, N.~Wiebe, C.~Gidney, A.~Aspuru-Guzik, G.~K.-L.
  Chan, R.~Babbush, {\it Phys. Rev. Lett.\/} {\bf 2018}, {\it 120\/}, 110501.

\bibitem{divincenzo_physical_2000}
D.~P. DiVincenzo, {\it Fortschr. Phys.\/} {\bf 2000}, {\it 48\/}, 771.

\bibitem{nielsen_quantum_2000}
M.~A. Nielsen, I.~L. Chuang, {\it Quantum computation and quantum
  information\/}, Cambridge University Press, Cambridge, UK {\bf 2000}.

\bibitem{mezzacapo_digital_2015}
A.~Mezzacapo, U.~Las~Heras, J.~S. Pedernales, L.~DiCarlo, E.~Solano, L.~Lamata,
  {\it Sci. Rep.\/} {\bf 2015}, {\it 4\/}, 7482.

\bibitem{buluta_quantum_2009}
I.~Buluta, F.~Nori, {\it Science\/} {\bf 2009}, {\it 326\/}, 108.

\bibitem{georgescu_quantum_2014}
I.~M. Georgescu, S.~Ashhab, F.~Nori, {\it Rev. Mod. Phys.\/} {\bf 2014}, {\it
  86\/}, 153.

\bibitem{sanders_efficient_2013}
B.~C. Sanders, in {\it Reversible Computation\/} (Eds. G.~W. Dueck, D.~M.
  Miller), Springer Berlin Heidelberg, Berlin, Heidelberg {\bf 2013}, p.~1.

\bibitem{preskill_quantum_2018}
J.~Preskill, {\it Quantum\/} {\bf 2018}, {\it 2\/}, 79.

\bibitem{blatt_quantum_2012}
R.~Blatt, C.~F. Roos, {\it Nat. Phys.\/} {\bf 2012}, {\it 8\/}, 277.

\bibitem{bloch_quantum_2012}
I.~Bloch, J.~Dalibard, S.~Nascimb{\`e}ne, {\it Nat. Phys.\/} {\bf 2012}, {\it
  8\/}, 267.

\bibitem{aspuru-guzik_photonic_2012}
A.~Aspuru-Guzik, P.~Walther, {\it Nat. Phys.\/} {\bf 2012}, {\it 8\/}, 285.

\bibitem{houck_-chip_2012}
A.~A. Houck, H.~E. T{\"u}reci, J.~Koch, {\it Nat. Phys.\/} {\bf 2012}, {\it
  8\/}, 292.

\bibitem{wendin_quantum_2017}
G.~Wendin, {\it Rep. Prog. Phys.\/} {\bf 2017}, {\it 80\/}, 106001.

\bibitem{lamata_digital-analog_2018}
L.~Lamata, A.~Parra-Rodriguez, M.~Sanz, E.~Solano, {\it Adv. Phys.: X\/} {\bf
  2018}, {\it 3\/}, 1457981.

\bibitem{pednault_breaking_2017}
E.~Pednault, J.~A. Gunnels, G.~Nannicini, L.~Horesh, T.~Magerlein,
  E.~Solomonik, E.~W. Draeger, E.~T. Holland, R.~Wisnieff, {\it
  arXiv:1710.05867 [quant-ph]\/} {\bf 2017}.

\bibitem{boixo_characterizing_2018}
S.~Boixo, S.~V. Isakov, V.~N. Smelyanskiy, R.~Babbush, N.~Ding, Z.~Jiang,
  M.~Bremner, J.~M. Martinis, H.~Neven, {\it Nat. Phys.\/} {\bf 2018}, {\it
  14\/}, 595.

\bibitem{schindler_experimental_2011}
P.~Schindler, J.~T. Barreiro, T.~Monz, V.~Nebendahl, D.~Nigg, M.~Chwalla,
  M.~Hennrich, R.~Blatt, {\it Science\/} {\bf 2011}, {\it 332\/}, 1059.

\bibitem{you_simulating_2013}
H.~You, M.~R. Geller, P.~C. Stancil, {\it Phys. Rev. A\/} {\bf 2013}, {\it
  87\/}, 032341.

\bibitem{barends_superconducting_2014}
R.~Barends, J.~Kelly, A.~Megrant, A.~Veitia, D.~Sank, E.~Jeffrey, T.~C. White,
  J.~Mutus, A.~G. Fowler, B.~Campbell, Y.~Chen, Z.~Chen, B.~Chiaro,
  A.~Dunsworth, C.~Neill, P.~O{\textquoteright}Malley, P.~Roushan,
  A.~Vainsencher, J.~Wenner, A.~N. Korotkov, A.~N. Cleland, J.~M. Martinis,
  {\it Nature\/} {\bf 2014}, {\it 508\/}, 500.

\bibitem{corcoles_demonstration_2015}
A.~C{\'o}rcoles, E.~Magesan, S.~J. Srinivasan, A.~W. Cross, M.~Steffen, J.~M.
  Gambetta, J.~M. Chow, {\it Nat. Commun.\/} {\bf 2015}, {\it 6\/}, 6979.

\bibitem{troyer_computational_2005}
M.~Troyer, U.-J. Wiese, {\it Phys. Rev. Lett.\/} {\bf 2005}, {\it 94\/},
  170201.

\bibitem{martinez_real-time_2016}
E.~A. Martinez, C.~A. Muschik, P.~Schindler, D.~Nigg, A.~Erhard, M.~Heyl,
  P.~Hauke, M.~Dalmonte, T.~Monz, P.~Zoller, R.~Blatt, {\it Nature\/} {\bf
  2016}, {\it 534\/}, 516.

\bibitem{ladd_quantum_2010}
T.~D. Ladd, F.~Jelezko, R.~Laflamme, Y.~Nakamura, C.~Monroe, J.~L.
  O{\textquoteright}Brien, {\it Nature\/} {\bf 2010}, {\it 464\/}, 45.

\bibitem{awschalom_quantum_2013}
D.~D. Awschalom, L.~C. Bassett, A.~S. Dzurak, E.~L. Hu, J.~R. Petta, {\it
  Science\/} {\bf 2013}, {\it 339\/}, 1174.

\bibitem{monroe_scaling_2013}
C.~Monroe, J.~Kim, {\it Science\/} {\bf 2013}, {\it 339\/}, 1164.

\bibitem{schindler_quantum_2013}
P.~Schindler, D.~Nigg, T.~Monz, J.~T. Barreiro, E.~Martinez, S.~X. Wang,
  S.~Quint, M.~F. Brandl, V.~Nebendahl, C.~F. Roos, M.~Chwalla, M.~Hennrich,
  R.~Blatt, {\it New J. Phys.\/} {\bf 2013}, {\it 15\/}, 123012.

\bibitem{bruzewicz_review-trap-ion_2019}
C.~D. Bruzewicz, J.~Chiaverini, R.~McConnell, J.~M. Sage, {\it Applied Physics
  Reviews\/} {\bf 2019}, {\it 6\/}, 021314.

\bibitem{clarke_superconducting_2008}
J.~Clarke, F.~K. Wilhelm, {\it Nature\/} {\bf 2008}, {\it 453\/}, 1031.

\bibitem{schoelkopf_wiring_2008}
R.~J. Schoelkopf, S.~M. Girvin, {\it Nature\/} {\bf 2008}, {\it 451\/}, 664.

\bibitem{devoret_superconducting_2013}
M.~H. Devoret, R.~J. Schoelkopf, {\it Science\/} {\bf 2013}, {\it 339\/}, 1169.

\bibitem{Gu_review_SCcircuits_2017}
X.~Gu, A.~F. Kockum, A.~Miranowicz, Y.~xi~Liu, F.~Nori, {\it Physics Reports\/}
  {\bf 2017}, {\it 718-719\/}, 1 , microwave photonics with superconducting
  quantum circuits.

\bibitem{woerner_quantum_2019}
S.~Woerner, D.~J. Egger, {\it npj Quantum Inf.\/} {\bf 2019}, {\it 5\/}, 15.

\bibitem{martin_towards_2019}
A.~Martin, B.~Candelas, {\'A}.~Rodr{\'i}guez-Rozas, J.~D. Mart{\'i}n-Guerrero,
  X.~Chen, L.~Lamata, R.~Or{\'u}s, E.~Solano, M.~Sanz, {\it arXiv:1904.05803
  [cond-mat, physics:quant-ph]\/} {\bf 2019}.

\bibitem{biamonte_quantum_2017}
J.~Biamonte, P.~Wittek, N.~Pancotti, P.~Rebentrost, N.~Wiebe, S.~Lloyd, {\it
  Nature\/} {\bf 2017}, {\it 549\/}, 195.

\bibitem{acin_quantum_2018}
A.~Ac{\'i}n, I.~Bloch, H.~Buhrman, T.~Calarco, C.~Eichler, J.~Eisert,
  D.~Esteve, N.~Gisin, S.~J. Glaser, F.~Jelezko, S.~Kuhr, M.~Lewenstein, M.~F.
  Riedel, P.~O. Schmidt, R.~Thew, A.~Wallraff, I.~Walmsley, F.~K. Wilhelm, {\it
  New J. Phys.\/} {\bf 2018}, {\it 20\/}, 080201.

\bibitem{barenco_elementary_1995}
A.~Barenco, C.~H. Bennett, R.~Cleve, D.~P. DiVincenzo, N.~Margolus, P.~Shor,
  T.~Sleator, J.~A. Smolin, H.~Weinfurter, {\it Phys. Rev. A\/} {\bf 1995},
  {\it 52\/}, 3457.

\bibitem{jordan_uber_1928}
P.~Jordan, E.~Wigner, {\it Zeitschrift f{\"u}r Physik\/} {\bf 1928}, {\it
  47\/}, 631.

\bibitem{cross_validating_2018}
A.~W. Cross, L.~S. Bishop, S.~Sheldon, P.~D. Nation, J.~M. Gambetta, {\it
  arXiv:1811.12926 [quant-ph]\/} {\bf 2018}.

\bibitem{tacchino_electromechanical_2018}
F.~Tacchino, A.~Chiesa, M.~D. LaHaye, S.~Carretta, D.~Gerace, {\it Phys. Rev.
  B\/} {\bf 2018}, {\it 97\/}, 214302.

\bibitem{barends_digital_2015}
R.~Barends, L.~Lamata, J.~Kelly, L.~Garc{\'i}a-{\'A}lvarez, A.~G. Fowler,
  A.~Megrant, E.~Jeffrey, T.~C. White, D.~Sank, J.~Y. Mutus, B.~Campbell,
  Y.~Chen, Z.~Chen, B.~Chiaro, A.~Dunsworth, I.-C. Hoi, C.~Neill, P.~J.~J.
  O{\textquoteright}Malley, C.~Quintana, P.~Roushan, A.~Vainsencher, J.~Wenner,
  E.~Solano, J.~M. Martinis, {\it Nat. Commun.\/} {\bf 2015}, {\it 6\/}, 7654.

\bibitem{kandala_hardware-efficient_2017}
A.~Kandala, A.~Mezzacapo, K.~Temme, M.~Takita, M.~Brink, J.~M. Chow, J.~M.
  Gambetta, {\it Nature\/} {\bf 2017}, {\it 549\/}, 242.

\bibitem{dumitrescu_cloud_2018}
E.~F. Dumitrescu, A.~J. McCaskey, G.~Hagen, G.~R. Jansen, T.~D. Morris,
  T.~Papenbrock, R.~C. Pooser, D.~J. Dean, P.~Lougovski, {\it Phys. Rev.
  Lett.\/} {\bf 2018}, {\it 120\/}, 210501.

\bibitem{klco_quantum-classical_2018}
N.~Klco, E.~F. Dumitrescu, A.~J. McCaskey, T.~D. Morris, R.~C. Pooser, M.~Sanz,
  E.~Solano, P.~Lougovski, M.~J. Savage, {\it Phys. Rev. A\/} {\bf 2018}, {\it
  98\/}, 032331.

\bibitem{chow_universal_2012}
J.~M. Chow, J.~M. Gambetta, A.~D. C{\'o}rcoles, S.~T. Merkel, J.~A. Smolin,
  C.~Rigetti, S.~Poletto, G.~A. Keefe, M.~B. Rothwell, J.~R. Rozen, M.~B.
  Ketchen, M.~Steffen, {\it Phys. Rev. Lett.\/} {\bf 2012}, {\it 109\/},
  060501.

\bibitem{rigetti_fully_2010}
C.~Rigetti, M.~Devoret, {\it Phys. Rev. B\/} {\bf 2010}, {\it 81\/}, 134507.

\bibitem{chow_simple_2011}
J.~M. Chow, A.~D. C{\'o}rcoles, J.~M. Gambetta, C.~Rigetti, B.~R. Johnson,
  J.~A. Smolin, J.~R. Rozen, G.~A. Keefe, M.~B. Rothwell, M.~B. Ketchen,
  M.~Steffen, {\it Phys. Rev. Lett.\/} {\bf 2011}, {\it 107\/}, 080502.

\bibitem{sheldon_procedure_2016}
S.~Sheldon, E.~Magesan, J.~M. Chow, J.~M. Gambetta, {\it Phys. Rev. A\/} {\bf
  2016}, {\it 93\/}, 060302.

\bibitem{salathe_digital_2015}
Y.~Salath{\'e}, M.~Mondal, M.~Oppliger, J.~Heinsoo, P.~Kurpiers, A.~Poto{\v
  c}nik, A.~Mezzacapo, U.~Las~Heras, L.~Lamata, E.~Solano, S.~Filipp,
  A.~Wallraff, {\it Phys. Rev. X\/} {\bf 2015}, {\it 5\/}, 021027.

\bibitem{mckay_universal_2016}
D.~C. McKay, S.~Filipp, A.~Mezzacapo, E.~Magesan, J.~M. Chow, J.~M. Gambetta,
  {\it Phys. Rev. Appl.\/} {\bf 2016}, {\it 6\/}, 064007.

\bibitem{carretta_quantum_2013}
S.~Carretta, A.~Chiesa, F.~Troiani, D.~Gerace, G.~Amoretti, P.~Santini, {\it
  Phys. Rev. Lett.\/} {\bf 2013}, {\it 111\/}, 110501.

\bibitem{chiesa_robustness_2014}
A.~Chiesa, D.~Gerace, F.~Troiani, G.~Amoretti, P.~Santini, S.~Carretta, {\it
  Phys. Rev. A\/} {\bf 2014}, {\it 89\/}, 052308.

\bibitem{reagor_demonstration_2018}
M.~Reagor, C.~B. Osborn, N.~Tezak, A.~Staley, G.~Prawiroatmodjo, M.~Scheer,
  N.~Alidoust, E.~A. Sete, N.~Didier, M.~P. da~Silva, E.~Acala, J.~Angeles,
  A.~Bestwick, M.~Block, B.~Bloom, A.~Bradley, C.~Bui, S.~Caldwell,
  L.~Capelluto, R.~Chilcott, J.~Cordova, G.~Crossman, M.~Curtis, S.~Deshpande,
  T.~El~Bouayadi, D.~Girshovich, S.~Hong, A.~Hudson, P.~Karalekas, K.~Kuang,
  M.~Lenihan, R.~Manenti, T.~Manning, J.~Marshall, Y.~Mohan,
  W.~O{\textquoteright}Brien, J.~Otterbach, A.~Papageorge, J.-P. Paquette,
  M.~Pelstring, A.~Polloreno, V.~Rawat, C.~A. Ryan, R.~Renzas, N.~Rubin,
  D.~Russel, M.~Rust, D.~Scarabelli, M.~Selvanayagam, R.~Sinclair, R.~Smith,
  M.~Suska, T.-W. To, .~Vahidpour, N.~Vodrahalli, T.~Whyland, K.~Yadav,
  W.~Zeng, C.~T. Rigetti, {\it Sci. Adv.\/} {\bf 2018}, {\it 4\/}, eaao3603.

\bibitem{jones_robust_2003}
J.~A. Jones, {\it Phys. Rev. A\/} {\bf 2003}, {\it 67\/}, 012317.

\bibitem{schindler_quantum_2013-1}
P.~Schindler, M.~M{\"u}ller, D.~Nigg, J.~T. Barreiro, E.~A. Martinez,
  M.~Hennrich, T.~Monz, S.~Diehl, P.~Zoller, R.~Blatt, {\it Nat. Phys.\/} {\bf
  2013}, {\it 9\/}, 361.

\bibitem{lanyon_universal_2011}
B.~P. Lanyon, C.~Hempel, D.~Nigg, M.~Muller, R.~Gerritsma, F.~Zahringer,
  P.~Schindler, J.~T. Barreiro, M.~Rambach, G.~Kirchmair, M.~Hennrich,
  P.~Zoller, R.~Blatt, C.~F. Roos, {\it Science\/} {\bf 2011}, {\it 334\/}, 57.

\bibitem{molmer_multiparticle_1999}
K.~M{\o}lmer, A.~S{\o}rensen, {\it Phys. Rev. Lett.\/} {\bf 1999}, {\it 82\/},
  1835.

\bibitem{vidal_universal_2004}
G.~Vidal, C.~M. Dawson, {\it Phys. Rev. A\/} {\bf 2004}, {\it 69\/}, 010301.

\bibitem{chiesa_quantum_2019}
A.~Chiesa, F.~Tacchino, M.~Grossi, P.~Santini, I.~Tavernelli, D.~Gerace,
  S.~Carretta, {\it Nat. Phys.\/} {\bf 2019}, {\it 15\/}, 455.

\bibitem{muller_simulating_2011}
M.~M{\"u}ller, K.~Hammerer, Y.~L. Zhou, C.~F. Roos, P.~Zoller, {\it New J.
  Phys.\/} {\bf 2011}, {\it 13\/}, 085007.

\bibitem{hatano_finding_2005}
N.~Hatano, M.~Suzuki, in {\it Quantum {Annealing} and {Other} {Optimization}
  {Methods}\/} (Eds. A.~Das, B.~K.~Chakrabarti), pp. 37--68, Springer Berlin
  Heidelberg, Berlin, Heidelberg, {\bf 2005}.

\bibitem{ferrando-soria_switchable_2016}
J.~Ferrando-Soria, S.~A. Magee, A.~Chiesa, S.~Carretta, P.~Santini, I.~J.
  Vitorica-Yrezabal, F.~Tuna, G.~F. Whitehead, S.~Sproules, K.~M. Lancaster,
  A.-L. Barra, G.~A. Timco, E.~J. McInnes, R.~E. Winpenny, {\it Chem\/} {\bf
  2016}, {\it 1\/}, 727.

\bibitem{pedernales_efficient_2014}
J.~S. Pedernales, R.~Di~Candia, I.~L. Egusquiza, J.~Casanova, E.~Solano, {\it
  Phys. Rev. Lett.\/} {\bf 2014}, {\it 113\/}, 020505.

\bibitem{bari_classical_1973}
R.~A. Bari, {\it Phys. Rev. B\/} {\bf 1973}, {\it 7\/}, 4318.

\bibitem{moll_quantum_2018}
N.~Moll, P.~Barkoutsos, L.~S. Bishop, J.~M. Chow, A.~Cross, D.~J. Egger,
  S.~Filipp, A.~Fuhrer, J.~M. Gambetta, M.~Ganzhorn, A.~Kandala, A.~Mezzacapo,
  P.~M{\"u}ller, W.~Riess, G.~Salis, J.~Smolin, I.~Tavernelli, K.~Temme, {\it
  Quantum Sci. Tech.\/} {\bf 2018}, {\it 3\/}, 030503.

\bibitem{linke_experimental_2017}
N.~M. Linke, D.~Maslov, M.~Roetteler, S.~Debnath, C.~Figgatt, K.~A. Landsman,
  K.~Wright, C.~Monroe, {\it Proceedings of the National Academy of Sciences\/}
  {\bf 2017}, {\it 114\/}, 3305.

\bibitem{mcardle_error_2019}
S.~McArdle, X.~Yuan, S.~Benjamin, {\it Phys. Rev. Lett.\/} {\bf 2019}, {\it
  122\/}, 180501.

\bibitem{PRX_Benjamin_2017}
Y.~Li, C.~Benjamin, {\it Phys. Rev. X\/} {\bf 2017}, {\it 7\/}, 021050.

\bibitem{PRX_Benjamin_2018}
S.~Endo, C.~Benjamin, {\it Phys. Rev. X\/} {\bf 2018}, {\it 8\/}, 031027.

\bibitem{kandala_error_2019}
A.~Kandala, K.~Temme, A.~D. C{\'o}rcoles, A.~Mezzacapo, J.~M. Chow, J.~M.
  Gambetta, {\it Nature\/} {\bf 2019}, {\it 567\/}, 491.

\bibitem{blatt_entangled_2008}
R.~Blatt, D.~Wineland, {\it Nature\/} {\bf 2008}, {\it 453\/}, 1008.

\bibitem{wright_benchmarking_2019}
K.~Wright, K.~M. Beck, S.~Debnath, J.~M. Amini, Y.~Nam, N.~Grzesiak, J.-S.
  Chen, N.~C. Pisenti, M.~Chmielewski, C.~Collins, K.~M. Hudek, J.~Mizrahi,
  J.~D. Wong-Campos, S.~Allen, J.~Apisdorf, P.~Solomon, M.~Williams, A.~M.
  Ducore, A.~Blinov, S.~M. Kreikemeier, V.~Chaplin, M.~Keesan, C.~Monroe, ,
  J.~Kim, {\it arxiv:1903.08181 [quant-ph]\/} {\bf 2019}.

\bibitem{leibfried_review-trap-ion_2003}
D.~Leibfried, R.~Blatt, C.~Monroe, D.~Wineland, {\it Rev. Mod. Phys.\/} {\bf
  2003}, {\it 75\/}, 281.

\bibitem{bautista_multilayer-iontraps_2019}
A.~Bautista-Salvador, G.~Zarantonello, H.~Hahn, A.~Preciado-Grijalva,
  J.~Morgner, M.~Wahnschaffe, C.~Ospelkaus, {\it New Journal of Physics\/} {\bf
  2019}, {\it 21\/}, 043011.

\bibitem{bruzewicz_2Darray_ions_2016}
C.~D. Bruzewicz, R.~McConnell, J.~Chiaverini, J.~M. Sage, {\it Nature
  Communications\/} {\bf 2016}, {\it 7\/}, 13005.

\bibitem{friis_observation_2018}
N.~Friis, O.~Marty, C.~Maier, C.~Hempel, M.~Holz{\"a}pfel, P.~Jurcevic, M.~B.
  Plenio, M.~Huber, C.~Roos, R.~Blatt, B.~Lanyon, {\it Phys. Rev. X\/} {\bf
  2018}, {\it 8\/}, 021012.

\bibitem{hempel_quantum-chemistry_2018}
C.~Hempel, C.~Maier, J.~Romero, J.~McClean, T.~Monz, H.~Shen, P.~Jurcevic,
  B.~P. Lanyon, P.~Love, R.~Babbush, A.~Aspuru-Guzik, R.~Blatt, C.~F. Roos,
  {\it Phys. Rev. X\/} {\bf 2018}, {\it 8\/}, 031022.

\bibitem{ballance_fidelity-gates_2016}
C.~J. Ballance, T.~P. Harty, N.~M. Linke, M.~A. Sepiol, D.~M. Lucas, {\it Phys.
  Rev. Lett.\/} {\bf 2016}, {\it 117\/}, 060504.

\bibitem{debnath_demonstration_2016}
S.~Debnath, N.~M. Linke, C.~Figgatt, K.~A. Landsman, K.~Wright, C.~Monroe, {\it
  Nature\/} {\bf 2016}, {\it 536\/}, 63.

\bibitem{cirac_quantum_1995}
J.~I. Cirac, P.~Zoller, {\it Phys. Rev. Lett.\/} {\bf 1995}, {\it 74\/}, 4091.

\bibitem{shapira_robust_2018}
Y.~Shapira, R.~Shaniv, T.~Manovitz, N.~Akerman, R.~Ozeri, {\it Phys. Rev.
  Lett.\/} {\bf 2018}, {\it 121\/}, 180502.

\bibitem{webb_resilient_2018}
A.~E. Webb, S.~C. Webster, S.~Collingbourne, D.~Bretaud, A.~M. Lawrence,
  S.~Weidt, F.~Mintert, W.~K. Hensinger, {\it Phys. Rev. Lett.\/} {\bf 2018},
  {\it 121\/}, 180501.

\bibitem{harty_high-fidelity_2014}
T.~P. Harty, D.~T.~C. Allcock, C.~J. Ballance, L.~Guidoni, H.~A. Janacek, N.~M.
  Linke, D.~N. Stacey, D.~M. Lucas, {\it Phys. Rev. Lett.\/} {\bf 2014}, {\it
  113\/}, 220501.

\bibitem{gaebler_fidelity-gate-set_2016}
J.~P. Gaebler, T.~R. Tan, Y.~Lin, Y.~Wan, R.~Bowler, A.~C. Keith, S.~Glancy,
  K.~Coakley, E.~Knill, D.~Leibfried, D.~J. Wineland, {\it Phys. Rev. Lett.\/}
  {\bf 2016}, {\it 117\/}, 060505.

\bibitem{schafer_fast_2018}
V.~M. Sch{\"a}fer, C.~J. Ballance, K.~Thirumalai, L.~J. Stephenson, T.~G.
  Ballance, A.~M. Steane, D.~M. Lucas, {\it Nature\/} {\bf 2018}, {\it 555\/},
  75.

\bibitem{meyerson_readout_2008}
A.~H. Myerson, D.~J. Szwer, S.~C. Webster, D.~T.~C. Allcock, M.~J. Curtis,
  G.~Imreh, J.~A. Sherman, D.~N. Stacey, A.~M. Steane, D.~M. Lucas, {\it Phys.
  Rev. Lett.\/} {\bf 2008}, {\it 100\/}, 200502.

\bibitem{bermudez_progress_2017}
A.~Bermudez, X.~Xu, R.~Nigmatullin, J.~O'Gorman, V.~Negnevitsky, P.~Schindler,
  T.~Monz, U.~G. Poschinger, C.~Hempel, J.~Home, F.~Schmidt-Kaler, M.~Biercuk,
  R.~Blatt, S.~Benjamin, M.~M\"uller, {\it Phys. Rev. X\/} {\bf 2017}, {\it
  7\/}, 041061.

\bibitem{wang_memory_2017}
Y.~Wang, M.~Um, J.~Zhang, S.~An, M.~Lyu, J.-N. Zhang, L.~M. Duan, D.~Yum,
  K.~Kim, {\it Nature Photonics\/} {\bf 2017}, {\it 11\/}, 646.

\bibitem{erhard_characterizing_2019}
A.~Erhard, J.~J. Wallman, L.~Postler, M.~Meth, R.~Stricker, E.~A. Martinez,
  P.~Schindler, T.~Monz, J.~Emerson, R.~Blatt, {\it arxiv:1902.08543
  [quant-ph]\/} {\bf 2019}.

\bibitem{kokail_self-verifying_2019}
C.~Kokail, C.~Maier, R.~van Bijnen, T.~Brydges, M.~K. Joshi, P.~Jurcevic, C.~A.
  Muschik, P.~Silvi, R.~Blatt, C.~F. Roos, P.~Zoller, {\it Nature\/} {\bf
  2019}, {\it 569\/}, 355.

\bibitem{mcclean_theory_2016}
J.~R. McClean, J.~Romero, R.~Babbush, A.~Aspuru-Guzik, {\it New J. Phys.\/}
  {\bf 2016}, {\it 18\/}, 023023.

\bibitem{shehab_toward_2019}
O.~Shehab, K.~A. Landsman, Y.~Nam, D.~Zhu, N.~M. Linke, M.~J. Keesan, R.~C.
  Pooser, C.~R. Monroe, {\it arXiv:1904.04338 [nucl-th, physics:quant-ph]\/}
  {\bf 2019}.

\bibitem{nam_water_2019}
Y.~Nam, J.-S. Chen, N.~C. Pisenti, K.~Wright, C.~Delaney, D.~Maslov, K.~R.
  Brown, S.~Allen, J.~M. Amini, J.~Apisdorf, K.~M. Beck, A.~Blinov, V.~Chaplin,
  M.~Chmielewski, C.~Collins, S.~Debnath, A.~M. Ducore, K.~M. Hudek, M.~Keesan,
  S.~M. Kreikemeier, J.~Mizrahi, P.~Solomon, M.~Williams, J.~D. Wong-Campos,
  C.~Monroe, J.~Kim, {\it arxiv:1902.10171 [quant-ph]\/} {\bf 2019}.

\bibitem{blais_cavity_2004}
A.~Blais, R.-S. Huang, A.~Wallraff, S.~M. Girvin, R.~J. Schoelkopf, {\it Phys.
  Rev. A\/} {\bf 2004}, {\it 69\/}, 062320.

\bibitem{wallraff_strong_2004}
A.~Wallraff, D.~I. Schuster, A.~Blais, L.~Frunzio, R.-S. Huang, J.~Majer,
  S.~Kumar, S.~M. Girvin, R.~J. Schoelkopf, {\it Nature\/} {\bf 2004}, {\it
  431\/}, 162.

\bibitem{koch_charge-insensitive_2007}
J.~Koch, T.~M. Yu, J.~M. Gambetta, A.~A. Houck, D.~I. Schuster, J.~Majer,
  A.~Blais, M.~H. Devoret, S.~M. Girvin, R.~J. Schoelkopf, {\it Phys. Rev. A\/}
  {\bf 2007}, {\it 76\/}, 042319.

\bibitem{gambetta_protocols_2007}
J.~Gambetta, W.~A. Braff, A.~Wallraff, S.~M. Girvin, R.~J. Schoelkopf, {\it
  Phys. Rev. A\/} {\bf 2007}, {\it 76\/}, 012325.

\bibitem{majer_coupling_2007}
J.~Majer, J.~M. Chow, J.~M. Gambetta, J.~Koch, B.~R. Johnson, J.~A. Schreier,
  L.~Frunzio, D.~I. Schuster, A.~A. Houck, A.~Wallraff, A.~Blais, M.~H.
  Devoret, S.~M. Girvin, R.~J. Schoelkopf, {\it Nature\/} {\bf 2007}, {\it
  449\/}, 443.

\bibitem{gambetta_quantum_2008}
J.~Gambetta, A.~Blais, M.~Boissonneault, A.~A. Houck, D.~I. Schuster, S.~M.
  Girvin, {\it Phys. Rev. A\/} {\bf 2008}, {\it 77\/}, 012112.

\bibitem{mariantoni_implementing_2011}
M.~Mariantoni, H.~Wang, T.~Yamamoto, M.~Neeley, R.~C. Bialczak, Y.~Chen,
  M.~Lenander, E.~Lucero, A.~D. O'Connell, D.~Sank, M.~Weides, J.~Wenner,
  Y.~Yin, J.~Zhao, A.~N. Korotkov, A.~N. Cleland, J.~M. Martinis, {\it
  Science\/} {\bf 2011}, {\it 334\/}, 61.

\bibitem{rigetti_superconducting_2012}
C.~Rigetti, J.~M. Gambetta, S.~Poletto, B.~L.~T. Plourde, J.~M. Chow, A.~D.
  C{\'o}rcoles, J.~A. Smolin, S.~T. Merkel, J.~R. Rozen, G.~A. Keefe, M.~B.
  Rothwell, M.~B. Ketchen, M.~Steffen, {\it Phys. Rev. B\/} {\bf 2012}, {\it
  86\/}, 100506(R).

\bibitem{nersisyan_manufacturing_2019}
A.~Nersisyan, S.~Poletto, N.~Alidoust, R.~Manenti, R.~Renzas, C.-V. Bui, K.~Vu,
  T.~Whyland, Y.~Mohan, E.~A. Sete, S.~Stanwyck, A.~Bestwick, M.~Reagor, {\it
  arXiv:1901.08042 [physics, physics:quant-ph]\/} {\bf 2019}.

\bibitem{Nature_DiCarlo_2009}
L.~DiCarlo, J.~M. Chow, J.~Gambetta, L.~S. Bishop, B.~R. Johnson, D.~I.
  Schuster, J.~Majer, A.~Blais, F.~A., S.~M. Girvin, R.~J. Schoelkopf, {\it
  Nature\/} {\bf 2009}, {\it 460\/}, 240.

\bibitem{hussain_coherent_2018}
R.~Hussain, G.~Allodi, A.~Chiesa, E.~Garlatti, D.~Mitcov, A.~Konstantatos,
  K.~S. Pedersen, R.~De~Renzi, S.~Piligkos, S.~Carretta, {\it J. Am. Chem.
  Soc.\/} {\bf 2018}, {\it 140\/}, 9814.

\bibitem{Nature_Kelly_2015}
J.~Kelly, R.~Barends, A.~G. Fowler, A.~Megrant, E.~Jeffrey, T.~C. White,
  D.~Sank, J.~Y. Mutus, B.~Campbell, Y.~Chen, Z.~Chen, B.~Chiaro, A.~Dunsworth,
  I.-C.Hoi, C.~Neill, P.~J.~J. O'Malley, C.~Quintana, P.~Roushan,
  A.Vainsencher, J.~Wenner, A.~N. Cleland, J.~M. Martinis, {\it Nature\/} {\bf
  2015}, {\it 519\/}, 66.

\bibitem{Nature_Barends_2016}
R.~Barends, A.~Shabani, L.~Lamata, J.~Kelly, A.~Mezzacapo, U.~L. Heras,
  R.~Babbush, A.~G. Fowler, B.~Campbell, Y.~Chen, Z.~Chen, B.~Chiaro,
  A.~Dunsworth, E.~Jeffrey, E.~Lucero, A.~Megrant, J.~Y. Mutus, M.~Neeley,
  C.~Neill, P.~J.~J. O'Malley, C.~Quintana, P.~Roushan, D.~Sank,
  A.~Vainsencher, J.~Wenner, T.~C. White, E.~Solano, H.~Neven, J.~M. Martinis,
  {\it Nature\/} {\bf 2016}, {\it 534\/}, 222.

\bibitem{gambetta_building_2017}
J.~M. Gambetta, J.~M. Chow, M.~Steffen, {\it npj Quantum Inf.\/} {\bf 2017},
  {\it 3\/}, 2.

\bibitem{PRApplied_Ganzhorn_2019}
M.~Ganzhorn, D.~Egger, P.~Barkoutsos, P.~Ollitrault, G.~Salis, N.~Moll,
  M.~Roth, A.~Fuhrer, P.~Mueller, S.~Woerner, I.~Tavernelli, S.~Filipp, {\it
  Phys. Rev. A\/} {\bf 2019}, {\it 11\/}, 044092.

\bibitem{PRL_Sank_2014}
D.~Sank, E.~Jeffrey, J.~Mutus, T.~White, J.~Kelly, R.~Barends, Y.~Chen,
  Z.~Chen, B.~Chiaro, A.~Dunsworth, A.~Megrant, P.~J.~J. O'Malley, C.~Neill,
  P.~Roushan, A.~Vainsencher, J.~Wenner, A.~N. Cleland, J.~M. Martinis, {\it
  Phys. Rev. Lett.\/} {\bf 2014}, {\it 112\/}, 190504.

\bibitem{PRA_Gambetta_2016}
S.~Sheldon, L.~S. Bishop, E.~Magesan, S.~Filipp, J.~M. Chow, J.~M. Gambetta,
  {\it Phys. Rev. A\/} {\bf 2016}, {\it 93\/}, 012301.

\bibitem{smith_simulating-manybody_2019}
A.~Smith, M.~S. Kim, F.~Pollmann, J.~Knolle, {\it arxiv:1906.06343
  [quant-ph]\/} {\bf 2019}.

\bibitem{baker_spin_2012}
M.~L. Baker, T.~Guidi, S.~Carretta, J.~Ollivier, H.~Mutka, H.~U. G{\"u}del,
  G.~A. Timco, E.~J.~L. McInnes, G.~Amoretti, R.~E.~P. Winpenny, P.~Santini,
  {\it Nat. Phys.\/} {\bf 2012}, {\it 8\/}, 906.

\bibitem{garlatti_portraying_2017}
E.~Garlatti, T.~Guidi, S.~Ansbro, P.~Santini, G.~Amoretti, J.~Ollivier,
  H.~Mutka, G.~Timco, I.~J. Vitorica-Yrezabal, G.~F.~S. Whitehead, R.~E.~P.
  Winpenny, S.~Carretta, {\it Nat. Commun.\/} {\bf 2017}, {\it 8\/}, 14543.

\bibitem{mckay_efficient_2017}
D.~C. McKay, C.~J. Wood, S.~Sheldon, J.~M. Chow, J.~M. Gambetta, {\it Phys.
  Rev. A\/} {\bf 2017}, {\it 96\/}, 022330.

\bibitem{las_heras_fermionic_2015}
U.~Las~Heras, L.~Garc{\'i}a-{\'A}lvarez, A.~Mezzacapo, E.~Solano, L.~Lamata,
  {\it EPJ Quantum Tech.\/} {\bf 2015}, {\it 2\/}, 8.

\bibitem{omalley_scalable_2016}
P.~J.~J. O{\textquoteright}Malley, R.~Babbush, I.~D. Kivlichan, J.~Romero,
  J.~R. McClean, R.~Barends, J.~Kelly, P.~Roushan, A.~Tranter, N.~Ding,
  B.~Campbell, Y.~Chen, Z.~Chen, B.~Chiaro, A.~Dunsworth, A.~G. Fowler,
  E.~Jeffrey, E.~Lucero, A.~Megrant, J.~Y. Mutus, M.~Neeley, C.~Neill,
  C.~Quintana, D.~Sank, A.~Vainsencher, J.~Wenner, T.~C. White, P.~V. Coveney,
  P.~J. Love, H.~Neven, A.~Aspuru-Guzik, J.~M. Martinis, {\it Phys. Rev. X\/}
  {\bf 2016}, {\it 6\/}, 031007.

\bibitem{moll_optimizing_2016}
N.~Moll, A.~Fuhrer, P.~Staar, I.~Tavernelli, {\it J. Phys. A: Math. Theor.\/}
  {\bf 2016}, {\it 49\/}, 295301.

\bibitem{roy_programmable_2018}
T.~Roy, S.~Hazra, S.~Kundu, M.~Chand, M.~P. Patankar, R.~Vijay, {\it
  arXiv:1809.00668 [quant-ph]\/} {\bf 2018}.

\bibitem{klimov_thermal-cycling_2018}
P.~V. Klimov, J.~Kelly, Z.~Chen, M.~Neeley, A.~Megrant, B.~Burkett, R.~Barends,
  K.~Arya, B.~Chiaro, Y.~Chen, A.~Dunsworth, A.~Fowler, B.~Foxen, C.~Gidney,
  M.~Giustina, R.~Graff, T.~Huang, E.~Jeffrey, E.~Lucero, J.~Y. Mutus,
  O.~Naaman, C.~Neill, C.~Quintana, P.~Roushan, D.~Sank, A.~Vainsencher,
  J.~Wenner, T.~C. White, S.~Boixo, R.~Babbush, V.~N. Smelyanskiy, H.~Neven,
  J.~M. Martinis, {\it Phys. Rev. Lett.\/} {\bf 2018}, {\it 121\/}, 090502.

\bibitem{temme_error_2017}
K.~Temme, S.~Bravyi, J.~M. Gambetta, {\it Phys. Rev. Lett.\/} {\bf 2017}, {\it
  119\/}, 180509.

\bibitem{loss_quantum_1998}
D.~Loss, D.~P. DiVincenzo, {\it Phys. Rev. A\/} {\bf 1998}, {\it 57\/}, 120.

\bibitem{elzerman_single-shot_2004}
J.~M. Elzerman, R.~Hanson, L.~H. Willems~van Beveren, B.~Witkamp, L.~M.~K.
  Vandersypen, L.~P. Kouwenhoven, {\it Nature\/} {\bf 2004}, {\it 430\/}, 431.

\bibitem{petta_coherent_2005}
J.~R. Petta, A.~C. Johnson, J.~M. Taylor, E.~A. Laird, A.~Yacoby, M.~D. Lukin,
  C.~M. Marcus, M.~P. Hanson, A.~C. Gossard, {\it Science\/} {\bf 2005}, {\it
  309\/}, 2180.

\bibitem{hanson_spins_2007}
R.~Hanson, L.~P. Kouwenhoven, J.~R. Petta, S.~Tarucha, L.~M.~K. Vandersypen,
  {\it Rev. Mod. Phys.\/} {\bf 2007}, {\it 79\/}, 1217.

\bibitem{zajac_resonantly_2018}
D.~M. Zajac, A.~J. Sigillito, M.~Russ, F.~Borjans, J.~M. Taylor, G.~Burkard,
  J.~R. Petta, {\it Science\/} {\bf 2018}, {\it 359\/}, 439.

\bibitem{veldhorst_two-qubit_2015}
M.~Veldhorst, C.~H. Yang, J.~C.~C. Hwang, W.~Huang, J.~P. Dehollain, J.~T.
  Muhonen, S.~Simmons, A.~Laucht, F.~E. Hudson, K.~M. Itoh, A.~Morello, A.~S.
  Dzurak, {\it Nature\/} {\bf 2015}, {\it 526\/}, 410.

\bibitem{watson_programmable_2018}
T.~F. Watson, S.~G.~J. Philips, E.~Kawakami, D.~R. Ward, P.~Scarlino,
  M.~Veldhorst, D.~E. Savage, M.~G. Lagally, M.~Friesen, S.~N. Coppersmith,
  M.~A. Eriksson, L.~M.~K. Vandersypen, {\it Nature\/} {\bf 2018}, {\it 555\/},
  633.

\bibitem{noiri_fast_2018}
A.~Noiri, T.~Nakajima, J.~Yoneda, M.~R. Delbecq, P.~Stano, T.~Otsuka,
  K.~Takeda, S.~Amaha, G.~Allison, K.~Kawasaki, Y.~Kojima, A.~Ludwig, A.~D.
  Wieck, D.~Loss, S.~Tarucha, {\it Nat. Commun.\/} {\bf 2018}, {\it 9\/}, 5066.

\bibitem{huang_fidelity_2019}
W.~Huang, C.~H. Yang, K.~W. Chan, T.~Tanttu, B.~Hensen, R.~C.~C. Leon, M.~A.
  Fogarty, J.~C.~C. Hwang, F.~E. Hudson, K.~M. Itoh, A.~Morello, A.~Laucht,
  A.~S. Dzurak, {\it Nature\/} {\bf 2019}.

\bibitem{morello_single-shot_2010}
A.~Morello, J.~J. Pla, F.~A. Zwanenburg, K.~W. Chan, K.~Y. Tan, H.~Huebl,
  M.~M{\"o}tt{\"o}nen, C.~D. Nugroho, C.~Yang, J.~A. van Donkelaar, A.~D.~C.
  Alves, D.~N. Jamieson, C.~C. Escott, L.~C.~L. Hollenberg, R.~G. Clark, A.~S.
  Dzurak, {\it Nature\/} {\bf 2010}, {\it 467\/}, 687.

\bibitem{pla_single-atom_2012}
J.~J. Pla, K.~Y. Tan, J.~P. Dehollain, W.~H. Lim, J.~J.~L. Morton, D.~N.
  Jamieson, A.~S. Dzurak, A.~Morello, {\it Nature\/} {\bf 2012}, {\it 489\/},
  541.

\bibitem{kalra_robust_2014}
R.~Kalra, A.~Laucht, C.~D. Hill, A.~Morello, {\it Phys. Rev. X\/} {\bf 2014},
  {\it 4\/}, 021044.

\bibitem{tosi_silicon_2017}
G.~Tosi, F.~A. Mohiyaddin, V.~Schmitt, S.~Tenberg, R.~Rahman, G.~Klimeck,
  A.~Morello, {\it Nat. Commun.\/} {\bf 2017}, {\it 8\/}, 450.

\bibitem{he_two-qubits-Phosphorus_2019}
Y.~He, S.~K. Gorman, D.~Keith, L.~Kranz, J.~G. Keizer, M.~Y. Simmons, {\it
  Nature\/} {\bf 2019}, {\it 571\/}, 371.

\bibitem{ferretti_single-photon_2012}
S.~Ferretti, D.~Gerace, {\it Phys. Rev. B\/} {\bf 2012}, {\it 85\/}, 033303.

\bibitem{flayac_all-silicon_2015}
H.~Flayac, D.~Gerace, V.~Savona, {\it Sci. Rep.\/} {\bf 2015}, {\it 5\/},
  11223.

\bibitem{munoz-matutano_emergence_2019}
G.~Mu{\~n}oz-Matutano, A.~Wood, M.~Johnsson, X.~Vidal, B.~Q. Baragiola,
  A.~Reinhard, A.~Lema{\^i}tre, J.~Bloch, A.~Amo, G.~Nogues, B.~Besga,
  M.~Richard, T.~Volz, {\it Nat. Mater.\/} {\bf 2019}, {\it 18\/}, 213.

\bibitem{delteil_towards_2019}
A.~Delteil, T.~Fink, A.~Schade, S.~H{\"o}fling, C.~Schneider, A.~{\.I}mamo{\u
  g}lu, {\it Nat. Mater.\/} {\bf 2019}, {\it 18\/}, 219.

\bibitem{gerace_quantum-optical_2009}
D.~Gerace, H.~E. T{\"u}reci, A.~Imamoglu, V.~Giovannetti, R.~Fazio, {\it Nat.
  Phys.\/} {\bf 2009}, {\it 5\/}, 281.

\bibitem{menicucci_universal_2006}
N.~C. Menicucci, P.~van Loock, M.~Gu, C.~Weedbrook, T.~C. Ralph, M.~A. Nielsen,
  {\it Phys. Rev. Lett.\/} {\bf 2006}, {\it 97\/}, 110501.

\bibitem{ferrando-soria_modular_2016}
J.~Ferrando-Soria, E.~Moreno~Pineda, A.~Chiesa, A.~Fernandez, S.~A. Magee,
  S.~Carretta, P.~Santini, I.~J. Vitorica-Yrezabal, F.~T., G.~A. Timco, E.~J.
  McInnes, R.~E. Winpenny, {\it Nat. Commun.\/} {\bf 2016}, {\it 7\/}, 11377.

\bibitem{atzori_two-qubit_2018}
M.~Atzori, A.~Chiesa, E.~Morra, M.~Chiesa, L.~Sorace, S.~Carretta, R.~Sessoli,
  {\it Chem. Sci.\/} {\bf 2018}, {\it 9\/}, 6183.

\bibitem{saffman_rydberg_2016}
M.~Saffman, {\it Journal of Physics B: Atomic, Molecular and Optical Physics\/}
  {\bf 2016}, {\it 49\/}, 202001.

\bibitem{levine_rydberg-qubits_2018}
H.~Levine, A.~Keesling, A.~Omran, H.~Bernien, S.~Schwartz, A.~S. Zibrov,
  M.~Endres, M.~Greiner, V.~Vuleti\ifmmode~\acute{c}\else \'{c}\fi{}, M.~D.
  Lukin, {\it Phys. Rev. Lett.\/} {\bf 2018}, {\it 121\/}, 123603.

\bibitem{xiang_hybrid_2013}
Z.-L. Xiang, S.~Ashhab, J.~Q. You, F.~Nori, {\it Rev. Mod. Phys.\/} {\bf 2013},
  {\it 85\/}, 623.

\bibitem{kurizki_quantum_2015}
G.~Kurizki, P.~Bertet, Y.~Kubo, K.~M{\o}lmer, D.~Petrosyan, P.~Rabl,
  J.~Schmiedmayer, {\it Proceedings of the National Academy of Sciences\/} {\bf
  2015}, {\it 112\/}, 3866.

\bibitem{ping_measurement_2012}
Y.~Ping, E.~M. Gauger, S.~C. Benjamin, {\it New J. Phys.\/} {\bf 2012}, {\it
  14\/}, 013030.

\bibitem{chiesa_long-lasting_2016}
A.~Chiesa, P.~Santini, D.~Gerace, S.~Carretta, {\it Phys. Rev. B\/} {\bf 2016},
  {\it 93\/}, 094432.

\bibitem{jenkins_coupling_2013}
.~Jenkins, T.~H{\"u}mmer, M.~J. Mart{\'i}nez-P{\'e}rez, J.~Garc{\'i}a-Ripoll,
  D.~Zueco, F.~Luis, {\it New J. Phys.\/} {\bf 2013}, {\it 15\/}, 095007.

\bibitem{jenkins_scalable_2016}
M.~D. Jenkins, D.~Zueco, O.~Roubeau, G.~Arom{\'i}, J.~Majer, F.~Luis, {\it
  Dalton Trans.\/} {\bf 2016}, {\it 45\/}, 16682.

\bibitem{li_hybrid_2016}
P.-B. Li, Z.-L. Xiang, P.~Rabl, F.~Nori, {\it Phys. Rev. Lett.\/} {\bf 2016},
  {\it 117\/}, 015502.

\bibitem{rips_quantum_2013}
S.~Rips, M.~J. Hartmann, {\it Phys. Rev. Lett.\/} {\bf 2013}, {\it 110\/},
  120503.

\bibitem{li_hybrid_2015}
P.-B. Li, Y.-C. Liu, S.-Y. Gao, Z.-L. Xiang, P.~Rabl, Y.-F. Xiao, F.-L. Li,
  {\it Phys. Rev. Appl.\/} {\bf 2015}, {\it 4\/}, 044003.

\bibitem{wei_verifying_2019}
K.~X. Wei, I.~Lauer, S.~Srinivasan, N.~Sundaresan, D.~T. McClure, D.~Toyli,
  D.~C. McKay, J.~M. Gambetta, S.~Sheldon, {\it arXiv:1905.05720 [quant-ph]\/}
  {\bf 2019}.

\bibitem{benenti_dynamical_2003}
G.~Benenti, G.~Casati, S.~Montangero, D.~L. Shepelyansky, {\it Phys. Rev. A\/}
  {\bf 2003}, {\it 67\/}, 052312.

\bibitem{montangero_dynamically_2004}
S.~Montangero, {\it Phys. Rev. A\/} {\bf 2004}, {\it 70\/}, 032311.

\bibitem{havlicek_supervised_2019}
V.~Havl{\'i}{\v c}ek, A.~D. C{\'o}rcoles, K.~Temme, A.~W. Harrow, A.~Kandala,
  J.~M. Chow, J.~M. Gambetta, {\it Nature\/} {\bf 2019}, {\it 567\/}, 209.

\bibitem{schuld_quantum_2019}
M.~Schuld, N.~Killoran, {\it Phys. Rev. Lett.\/} {\bf 2019}, {\it 122\/},
  040504.

\bibitem{tacchino_artificial_2019}
F.~Tacchino, C.~Macchiavello, D.~Gerace, D.~Bajoni, {\it npj Quantum Inf.\/}
  {\bf 2019}, {\it 5\/}, 26.

\bibitem{finazzi_corner-space_2015}
S.~Finazzi, A.~Le~Boit{\'e}, F.~Storme, A.~Baksic, C.~Ciuti, {\it Phys. Rev.
  Lett.\/} {\bf 2015}, {\it 115\/}, 080604.

\bibitem{mascarenhas_matrix-product-operator_2015}
E.~Mascarenhas, H.~Flayac, V.~Savona, {\it Phys. Rev. A\/} {\bf 2015}, {\it
  92\/}, 022116.

\bibitem{lloyd_engineering_2001}
S.~Lloyd, L.~Viola, {\it Phys. Rev. A\/} {\bf 2001}, {\it 65\/}, 010101.

\bibitem{wang_quantum_2011}
H.~Wang, S.~Ashhab, F.~Nori, {\it Phys. Rev. A\/} {\bf 2011}, {\it 83\/},
  062317.

\bibitem{di_candia_quantum_2015}
R.~Di~Candia, J.~S. Pedernales, A.~del Campo, E.~Solano, J.~Casanova, {\it Sci.
  Rep.\/} {\bf 2015}, {\it 5\/}, 09981.

\bibitem{cleve_efficient_2017}
R.~Cleve, C.~Wang, in {\it 44th International Colloquium on Automata,
  Languages, and Programming (ICALP 2017)\/}, {\it Leibniz International
  Proceedings in Informatics (LIPIcs)\/}, Vol.~80, Schloss
  Dagstuhl--Leibniz-Zentrum fuer Informatik, Dagstuhl, Germany {\bf 2017}, p.
  17:1.

\bibitem{lamm_simulation_2018}
H.~Lamm, S.~Lawrence, {\it Phys. Rev. Lett.\/} {\bf 2018}, {\it 121\/}, 170501.

\end{thebibliography}
\bibliographystyle{myadvquantumtech} 

\end{document}